%% file: arxiv.tex
\documentclass[twocolumn,astrosymb]{aastex702}
\graphicspath{{./}{Figures/}{Figures/Maps/}{Figures/C60_Fits/}{Figures/C70_Fits/}{Figures/Other_Fullerene_Fits/}}
\usepackage{xspace}
\usepackage{multirow}
\usepackage{threeparttable}
\usepackage{textcomp, gensymb}
\usepackage{comment}
\usepackage{rotating}

\usepackage[dvipsnames]{xcolor}

\definecolor{cbpurple}{rgb}{0.47, 0.37, 0.94}

\newcommand{\csixty}{\texorpdfstring{C$_{60}$}{C60}\xspace}
\newcommand{\um}{\texorpdfstring{$\mu$m}{micron}\xspace}
\newcommand{\wno}{\texorpdfstring{cm$^{-1}$}{wavenumbers}\xspace}
\newcommand{\cseventy}{\texorpdfstring{C$_{70}$}{C70}\xspace}
\newcommand{\csixtyplus}{C$_{60}^{+}$\xspace}
\newcommand{\overviewpaper}{Paper~I\xspace}
\newcommand{\dustpaper}{Paper~III\xspace}
\newcommand{\csixtyminus}{C$_{60}^{-}$\xspace}
\newcommand{\cseventyplus}{C$_{70}^{+}$\xspace}
\newcommand{\csixtyHplus}{C$_{60}$H$^{+}$\xspace}

\newcommand{\ArII}{\textrm{[Ar~{\textsc{ii}}] }}

\begin{document}

\title{Studying the Fullerene Nest: JWST Observations of the Planetary Nebula Tc~1. \\II. A Spectroscopic and Spatial Study of \csixty and \cseventy Emission}

\correspondingauthor{Jan Cami}
\email{jcami@uwo.ca}

\author[0009-0007-3950-335X]{Morgan~M. Giese}
\email{mgiese@uwo.ca}
\affiliation{Department of Physics \& Astronomy, The University of Western Ontario, London, Canada, N6A 3K7}
\affiliation{Institute for Earth and Space Exploration, The University of Western Ontario, London, Canada, N6A 3K7}

\author[0000-0002-2666-9234]{Jan Cami}
\email{jcami@uwo.ca}
\affiliation{Department of Physics \& Astronomy, The University of Western Ontario, London, Canada, N6A 3K7}
\affiliation{Institute for Earth and Space Exploration, The University of Western Ontario, London, Canada, N6A 3K7}

\author[0009-0004-9270-0019]{Simon Van Schuylenbergh}
\email{svanschu@uwo.ca}
\affiliation{Department of Physics \& Astronomy, The University of Western Ontario, London, Canada, N6A 3K7}
\affiliation{Institute for Earth and Space Exploration, The University of Western Ontario, London, Canada, N6A 3K7}

\author[0000-0002-2541-1602]{Els Peeters}
\email{epeeters@uwo.ca}
\affiliation{Department of Physics \& Astronomy, The University of Western Ontario, London, Canada, N6A 3K7}
\affiliation{Institute for Earth and Space Exploration, The University of Western Ontario, London, Canada, N6A 3K7}

\author[0009-0000-0191-6756]{Charmi Bhatt}
\email{cbhatt7@uwo.ca}
\affiliation{Department of Physics \& Astronomy, The University of Western Ontario, London, Canada, N6A 3K7}
\affiliation{Institute for Earth and Space Exploration, The University of Western Ontario, London, Canada, N6A 3K7}

\author[0000-0002-5895-8268]{Dries Van De Putte}
\email{dvandepu@uwo.ca}
\affiliation{Department of Physics \& Astronomy, The University of Western Ontario, London, Canada, N6A 3K7}
\affiliation{Institute for Earth and Space Exploration, The University of Western Ontario, London, Canada, N6A 3K7}

\author{Cedric Baerts}
\email{cbaerts@uwo.ca}
\affiliation{Department of Physics \& Astronomy, The University of Western Ontario, London, Canada, N6A 3K7}
\affiliation{Institute for Earth and Space Exploration, The University of Western Ontario, London, Canada, N6A 3K7}

\author[0000-0002-3875-1171]{Michael J. Barlow}
\affiliation{Department of Physics and Astronomy, University College London, Gower Street, London WC1E 6BT, United Kingdom}
\email{mjb@star.ucl.ac.uk}

\author[0000-0002-8452-8675]{Jeronimo Bernard-Salas}
\affiliation{ACRI-ST, Centre d'Etudes et de Recherche de Grasse (CERGA), 10Av. Nicolas Copernic, 06130 Grasse, France}
\affiliation{INCLASS Common Laboratory, 10 Av. Nicolas Copernic, 06130 Grasse, France}
\email{jeronimo.bernard-salas@acri-st.fr}

\author[0000-0002-5431-4449]{Alessandra Candian}
\affiliation{Anton Pannekoek Institute, University of Amsterdam, Science Park 904, 1098XH Am-
sterdam, The Netherlands}
\email{a.candian2@uva.nl}

\author[0000-0003-0304-9814]{P. Bryan Changala}
\email{changala@jila.colorado.edu}
\affiliation{JILA, University of Colorado Boulder and National Institute of Standards and Technology, Boulder, CO 80309, USA}
\affiliation{Department of Physics, University of Colorado Boulder, Boulder CO 80309}

\author[0000-0002-7926-4492]{Nick L.~J. Cox}
\affiliation{ACRI-ST, Centre d'Etudes et de Recherche de Grasse (CERGA), 10Av. Nicolas Copernic, 06130 Grasse, France}
\email{nljcox@gmail.com}

\author[0000-0002-4017-5572]{Harriet L. Dinerstein}
\affiliation{Department of Astronomy, University of Texas at Austin, Austin, TX 78712, USA }
\email{harriet@astro.as.utexas.edu}

\author[0000-0001-6035-3869]{Vincent J. Esposito}
\email{vesposito@chapman.edu}
\affiliation{Schmid College of Science and Technology, Chapman University, Orange, CA 92866, USA}

\author[0000-0002-1693-2721]{D.~A. Garc\'ia-Hern\'andez}
\affiliation{Instituto de Astrof\'isica de Canarias, C/ V\'ia L\'actea S/N, E-38205 La Laguna, Spain}
\affiliation{Departamento de Astrof\'isica, Universidad de La Laguna, E-38206 La Laguna, Spain}
\email{anibal.garcia.hernandez@iac.es}

\author[0000-0002-3938-4211]{Marco~A. G\'omez-Mu\~noz}
\email{magm@iac.es}
\affiliation{Departament de F\'isica Qu\`antica i Astrof\'isica (FQA), Universitat de Barcelona (UB), c. Mart\'i i Franqu\`es, 1, 08028 Barcelona, Spain}
\affiliation{Institut de Ci\`encies del Cosmos (ICCUB), Universitat de Barcelona (UB), c. Mart\'i i Franqu\`es, 1, 08028 Barcelona, Spain}

\author[0000-0003-1689-9201]{Kay Justtanont}
\affiliation{Department of Physics and Astronomy, Chalmers University of Technology, 412 96 Gothenburg, Sweden }
\email{kay.justtanont@chalmers.se}

\author[0000-0002-2626-7155]{Kathleen E. Kraemer}
\email{kathleen.kraemer@bc.edu}
\affiliation{Institute for Scientific Research, Boston College, 140 Commonwealth Avenue, Chestnut Hill, MA 02467, USA}

\author[0000-0002-1335-5623]{Eric Lagadec}
\affiliation{Universit\'e C\^{o}te d’Azur, Observatoire de la C\^{o}te d’Azur, CNRS, Lagrange,
96 Bd de l’Observatoire, 06300 Nice, France }
\email{Eric.Lagadec@oca.eu}

\author[0000-0002-3011-686X]{Arturo Manchado}
\affiliation{Instituto de Astrof\'isica de Canarias, C/ V\'ia L\'actea S/N, E-38205 La Laguna, Spain}
\affiliation{Departamento de Astrof\'isica, Universidad de La Laguna, E-38206 La Laguna, Spain}
\affiliation{Consejo Superior de Investigaciones Científicas (CSIC), Spain}
\email{arturo.manchado@iac.es}

\author[0000-0002-6455-2491]{Ana Monreal Ibero}
\affiliation{Leiden Observatory, Leiden University, PO Box 9513, 2300 RA Leiden, The Netherlands}
\email{monreal@strw.leidenuniv.nl}

\author[0000-0002-6858-5063]{Raghvendra Sahai}
\affiliation{Jet Propulsion Laboratory, MS 183-900, 4800 Oak Grove Dr., California Institute of Technology, Pasadena, CA 91109, USA}
\email{raghvendra.sahai@jpl.nasa.gov}

\author[0000-0003-3771-4990]{Ameek Sidhu}
\affiliation{Department of Physics \& Astronomy, The University of Western Ontario, London, Canada, N6A 3K7}
\affiliation{Institute for Earth and Space Exploration, The University of Western Ontario, London, Canada, N6A 3K7}
\email{asidhu92@uwo.ca}

\author[0000-0003-4520-1044]{G.~C.\ Sloan}
\affiliation{Space Telescope Science Institute, 3700 San Martin Drive, Baltimore, MD 21218, USA}
\affiliation{Department of Physics and Astronomy, University of North Carolina, Chapel Hill, NC 27599-3255, USA }
\email{gcsloan@stsci.edu}

\author[0000-0002-9604-1434]{N.~C.\ Sterling}
\affiliation{University of West Georgia, Carrollton, GA 30118, USA}
\email{nsterlin@westga.edu}

\author[0000-0003-0306-0028]{A.~G.~G.~M. Tielens}
\email{tielens@strw.leidenuniv.nl}
\affiliation{Astronomy Department, University of Maryland, College Park, MD
20742, USA}

\author[0000-0002-8008-910X]{Jeremy R. Walsh}
\affiliation{European Southern Observatory, Karl-Schwarzschild Strasse 2, 85748 Garching, Germany}
\email{jwalsh@eso.org}

\author[0000-0002-4000-4394]{Roger Wesson}
\email{rw@nebulousresearch.org}
\affiliation{Department of Physics and Astronomy, University College London, Gower Street, London WC1E 6BT, United Kingdom}
\affiliation{Cardiff Hub for Astrophysics Research and Technology (CHART),
School of Physics and Astronomy, \\Cardiff University, The Parade, Cardiff CF24 3AA, UK}
\email{rw@nebulousresearch.org}

\author{Joshua Cole Whitman}
\affiliation{Department of Physics, University of West Georgia, Carrollton, GA 30118, USA}
\email{jwhitma3@my.westga.edu}

\author[0000-0002-3171-5469]{Albert Zijlstra}
\affiliation{Jodrell Bank Centre for Astrophysics, Department of Physics \&\ Astronomy, The University of Manchester, \\Oxford Road, Manchester M13 9PL, UK}
\email{a.zijlstra@manchester.ac.uk}

\author[0009-0009-5675-7890]{Laura Finazzi}
\email{laura.finazzi@ru.nl}
\affiliation{HFML-FELIX, Toernooiveld 7, Nijmegen, 6525 ED, The Netherlands}
\affiliation{Institute for Molecules and Materials, Radboud University, Heyendaalseweg 135, Nijmegen, 6525 AJ, The Netherlands}

\author[0000-0001-9537-4117]{Jonathan Martens}
\email{jonathan.martens@ru.nl}
\affiliation{HFML-FELIX, Toernooiveld 7, Nijmegen, 6525 ED, The Netherlands}
\affiliation{Institute for Molecules and Materials, Radboud University, Heyendaalseweg 135, Nijmegen, 6525 AJ, The Netherlands}

\author[0000-0003-1500-922X]{Giel Berden}
\email{giel.berden@ru.nl}
\affiliation{HFML-FELIX, Toernooiveld 7, Nijmegen, 6525 ED, The Netherlands}
\affiliation{Institute for Molecules and Materials, Radboud University, Heyendaalseweg 135, Nijmegen, 6525 AJ, The Netherlands}

\author[0000-0002-2717-1278]{Jos Oomens}
\email{jos.oomens@ru.nl}
\affiliation{HFML-FELIX, Toernooiveld 7, Nijmegen, 6525 ED, The Netherlands}
\affiliation{Institute for Molecules and Materials, Radboud University, Heyendaalseweg 135, Nijmegen, 6525 AJ, The Netherlands}

\received{September 17, 2026}
\revised{October 11, 2026}
\accepted{October 16, 2026}

\begin{abstract}

We present a detailed analysis of the fullerene emission bands in JWST MIRI/MRS observations of the planetary nebula Tc~1. The observations reveal the emission features due to neutral \csixty and \cseventy in unprecedented detail. The combined fullerene emission represents $\sim$2.6\% of the total infrared output, suggesting that $\sim$0.6\% of the carbon budget is locked up in fullerenes. We find no clear evidence for \csixtyplus, \cseventyplus or \csixtyminus in the spectrum. The high spectral resolution allows us to separate the fullerene bands from atomic emission lines for the first time, allowing for a much better characterization of the band profiles and yielding more accurate band ratios than previous observations. We see no spatial variation in the individual \csixty and \cseventy band profile shapes across all pixels. The 17.4 and 18.9 \um \csixty bands show clear resolved ro-vibrational structure in their profiles, and similar structured emission is seen in the 12.5, 14.8, 17.7, and 21.8 \um \cseventy bands, further establishing that all emitting fullerene molecules are in the gas phase. We report a higher 7.0/18.9 \um \csixty band ratio ($\sim$1.4--1.9) than previously reported ($\sim$1.3). We note that all bands, except the 8.5 \um band, contain a significant contribution from \cseventy, complicating the analysis. The fullerene emission is spatially concentrated in a thin, asymmetric shell at a radius of 5–6\arcsec\ from the central star. The shell lies at the outer edge of the core of the nebula, which was ejected in a single, brief episode $\gtrsim$4,000 years ago.

\end{abstract}


\section{Introduction} \label{sec:intro}

The Universe is rich in carbonaceous compounds (molecules and dust grains) that appear in a bewildering variety of sizes, structures, shapes, and chemical compositions, depending on the environment in which they reside \citep[][]{2013RvMP...85.1021T}. About 15\% of all carbon in the Universe is locked up in the form of large gas-phase molecules: polycyclic aromatic hydrocarbons (PAHs), fullerenes, and related compounds. These are widespread and abundant in space and set the Universe aglow with their characteristic infrared (IR) emission. Indeed, the characteristic spectral features of PAHs have been seen out to cosmological distances, including by recent JWST observations \citep[e.g][]{2022ApJ...940L...8E, 2022A&A...666L...5G, 2023ApJ...942L..37A}. These molecules efficiently convert UV radiation to IR photons, drive key physical processes such as heating and cooling of the gas, and play an important role in interstellar chemistry. Consequently, they are major contributors to regulating large-scale processes in the Universe, including star and planet formation and the evolution of galaxies \citep{2008ARA&A..46..289T}. 

The detection and identification of the buckminsterfullerene \csixty \citep{Kroto1985} in diverse astrophysical environments offers a direct way to analyze how large individual species respond to differing physical conditions. The IR spectroscopic signatures of neutral \csixty were first recognized in the Spitzer-IRS spectrum of the planetary nebula (PN; PNe plural) Tc~1 \citep{Cami2010}, and have since been detected in a range of astronomical objects: PNe and other evolved stars \citep[e.g.,][]{Garcia-Hernandez:PN, Garcia-Hernandez:RcrB, Gielen:C60p-AGB, 2011AJ....142...54C}, reflection nebulae \citep{Sellgren2010}, young stellar objects \citep[]{Roberts:C60}, protoplanetary disks \citep{2019MNRAS.489.1509I,2023MNRAS.523.1601A}, meteorites \citep{Sabbah2022}, and the diffuse ISM \citep{Berne:C60_ISM}. Laboratory experiments have furthermore confirmed that several diffuse interstellar bands (DIBs) are due to C$_{60}^+$ (\citealp{Campbell:C60+DIBs} and others; see the review by \citealt{Harold:C60plusreview}). \csixty is abundant for an individual species, with estimated abundances between 10$^{-4}$ and up to 3\% of the available cosmic carbon \citep{2016A&A...590A..52O}. It is thus the only large carbonaceous species firmly identified in diverse astrophysical environments, it is widespread and abundant when detected, and it typically appears in the same environments where PAHs are seen \citep{Garcia-Hernandez:RcrB, Castellanos2014}. Indeed, PAHs and fullerenes are often observed together in the same spectra \citep[e.g.,][]{Garcia-Hernandez:PN, Sellgren2010}. 

What makes \csixty particularly valuable is that it shares
key physical properties with PAHs---both are large, carbon-based molecules with delocalized $\pi$-electron systems---and undergoes the same processes (e.g., excitation \& fluorescent cooling, ionization \& recombination, (de-)hydrogenation\dots) in astrophysical environments. Theoretical models developed for PAHs often include fullerenes as well \citep[sometimes with \csixty as the end product of UV photo-processing of large PAHs; see][]{BerneTielens2012,Berne:topdown,Ameek:theoryI}. Crucially, however, \csixty has recognizable and distinct spectroscopic fingerprints that are not subject to blending that affects PAH emission. Observations of \csixty in well-characterized environments thus provide direct tests of our understanding of the behavior of large molecules in astrophysical environments. 

Tc~1 remains one of the best targets for detailed fullerene studies. Its Spitzer-IRS spectrum exhibits the strongest known \csixty bands (relative to the mid-IR continuum), in addition to several weaker \cseventy features \citep{Cami2010}. It is known as one of the purest fullerene sources, and although some weak PAH features at 3.3, 6.2, and 11.2 \um have since been detected \citep{Bernard-Salas2012, Otsuka2014}, there is still little to no contamination by PAH bands (S. Van Schuylenbergh et al. 2026a, submitted). This greatly facilitates accurate fullerene measurements.

Tc~1 also offers excellent opportunities for spatially resolved fullerene studies. Optical images reveal a roughly spherical main nebula with a diameter of $\sim$12\arcsec, surrounded by a much fainter $\sim$55\arcsec\ halo \citep{Walsh_MUSE2026}. Most of the continuum dust emission appears to originate from a small ($\sim$3\arcsec\ diameter) horseshoe-shaped structure around the central star, which is surrounded by ionized gas following a more spherical distribution \citep{Cami:APN_formation}. Previous work has shown that the \csixty emission is prominent throughout the nebula but shows enhanced emission in an outer ring $\sim$5\arcsec\ from the central star, near the edge of the bright core \citep{Bernard-Salas2012,Cami:APN_formation}. Following J. Cami et al. (2026b, submitted), we refer to this main ionized nebula as the \textit{core}.

With JWST, we can now study the fullerene emission in Tc~1 at unprecedented spectral and spatial resolution. In this paper, we present MIRI/MRS observations that allow us to measure detailed fullerene spectral and spatial characteristics. An overview of the observing program, a spectral inventory, and a description of the morphology and physical conditions of the nebula are presented by J. Cami et al. (2026b, submitted; hereafter \overviewpaper); here, we focus on an in-depth analysis of the fullerene emission. The dust emission in Tc~1 is discussed by S. Van Schuylenbergh et al. (2026b, submitted; hereafter \dustpaper).

The paper is organized as follows: Section~\ref{Sec:Data} describes the observations and data reduction; Section~\ref{Sec:spectral_characteristics}  presents the detailed analysis of the fullerene spectral features; Section~\ref{sect:maps} describes the spatial distribution of the fullerene emission; Section~\ref{Sec:bandratios} describes the \csixty calculated band ratios; and Section~\ref{sec:discussion} discusses our findings in the context of fullerene abundances, excitation, and formation.

 \section{Observations and data reduction} 
\label{Sec:Data}

\begin{figure}[t]
    \centering
    \resizebox{\hsize}{!}{
    \includegraphics[trim={0.6cm 1.5cm 0cm 1.5cm},clip]{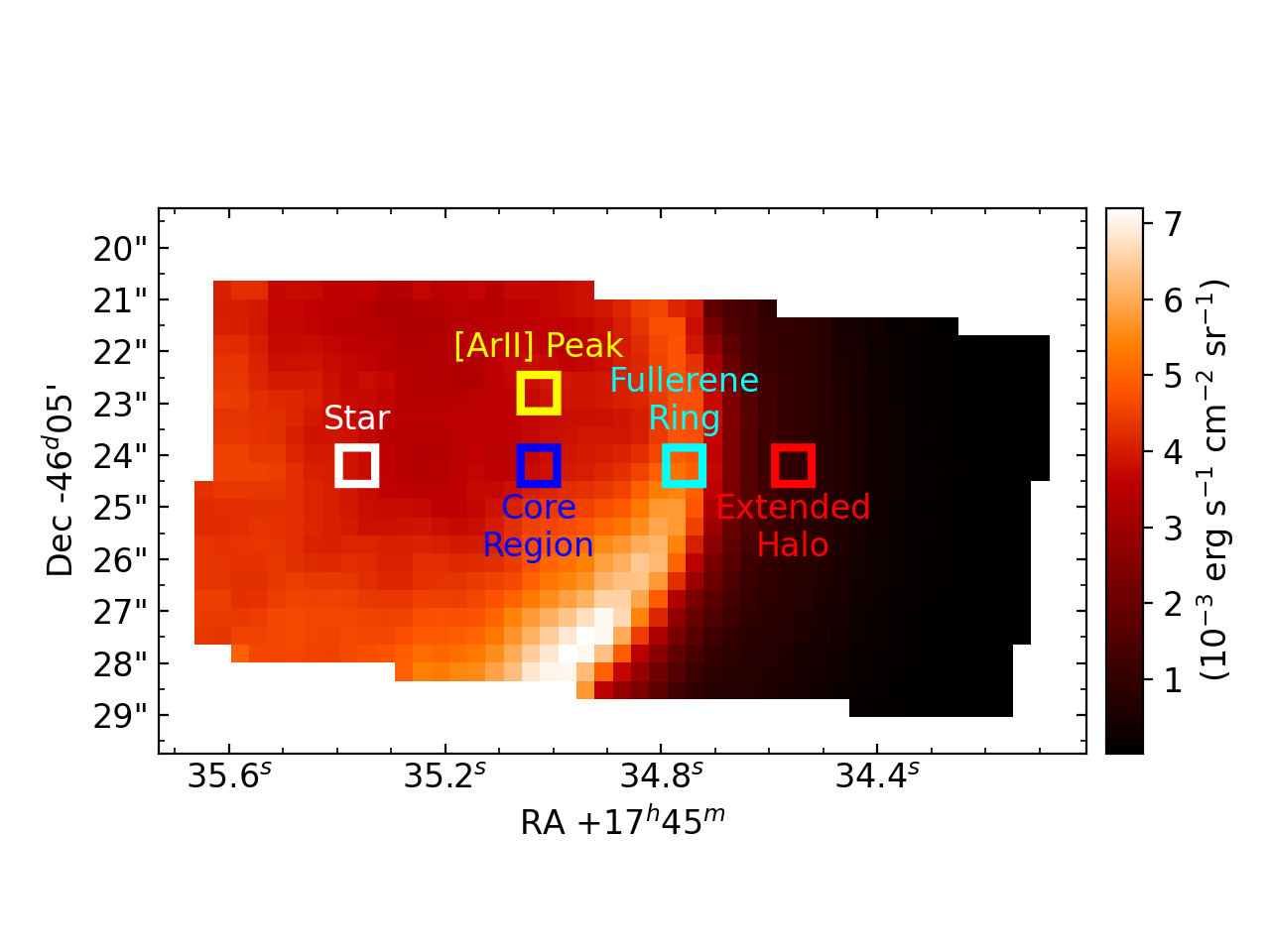}}
    \caption{Location of the five apertures used to describe notable areas of interest across Tc 1. The five apertures are plotted on top of the emission from the 18.9 \um fullerene band (further discussed in Section \ref{sect:maps}).}
    \label{fig:apertures}
    
\end{figure}

\begin{figure*}[t]
\resizebox{\hsize}{!}{\includegraphics[]{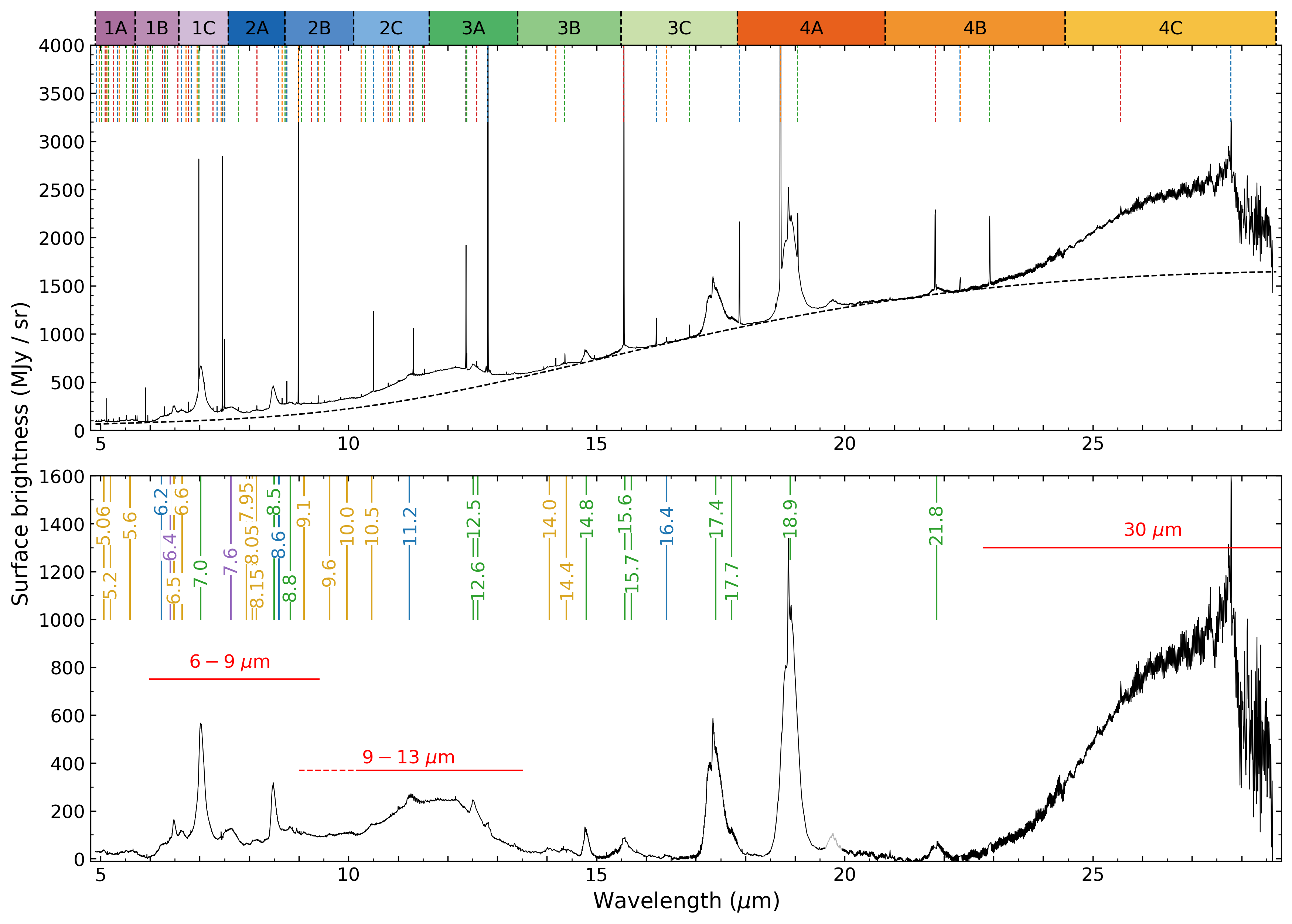}}
\caption{\label{Fig:overviewspectrum}Top: Composite JWST/MIRI spectrum of Tc~1 (black, solid) corresponding to the field of view of Channel 1. Unresolved lines are indicated using four different colors to aid in differentiating lines that are close together; the colors have no further meaning. The continuum is represented by the dashed line. The MIRI (sub)channels are indicated at the top. Bottom: The continuum-subtracted spectrum; note that all atomic emission lines have been removed from this spectrum as well. Features due to \csixty or \cseventy fundamental bands are indicated in green; PAH features in blue and unidentified features in purple. \csixty combination bands (see Section \ref{Sec:C60_spectral_characteristics}) are indicated in yellow. The plateaus discussed by \dustpaper are shown in red. The grey feature at 19.8~$\mu$m is due to an instrumental artifact. Reproduced from \overviewpaper.}
\end{figure*}

We observed Tc~1 with the JWST as part of JWST Cycle 3 program GO-4706 (P.I. J. Cami). An overview of the overall observational program as well as specific details on the data reduction process are described by \overviewpaper. The observations we discuss in this paper were taken on August 8, 2024 with the Mid-Infrared Instrument (MIRI) using Medium-Resolution Spectroscopy (MRS) mode \citep[][]{2015Wells, 2023Argyriou}. This provides a full spectrum at each pixel covering the 4.9--27.9 \um wavelength range at a spatial scale of 0.196--0.273\arcsec~per pixel, obtained through a $3\times1$ mosaic extending from the central star outward beyond the inner core and fullerene ring into the surrounding halo. 

The data were reduced using the standard JWST pipeline version 1.17.1 \citep{bushouse_2025_14597407} with the Calibration Reference Data System jwst\_1322.pmap context file. We refer to \overviewpaper for more details about the data reduction and calibration; for this paper, it is helpful to know that the spectral range is covered by four spectroscopic channels (1, 2, 3, and 4), with each channel subdivided into short (A), medium (B), and long (C) wavelength subchannels. The four channels sample different pixel scales and fields of view: channels 1 and 2 provide finer spatial sampling (pixel size 0.196\arcsec) over $3.2\arcsec\times3.7\arcsec$ and $4.0\arcsec\times4.8\arcsec$ fields respectively, while channels 3 and 4 cover larger fields at coarser resolution (respectively 0.245\arcsec ~pixels covering a FOV of $5.2\arcsec\times6.2\arcsec$ and 0.273\arcsec ~pixels covering a FOV of $6.6\arcsec\times7.7\arcsec$). The different pixel scales necessitate careful treatment when analyzing spatial variations as a function of wavelength, since a given physical structure may be well-resolved at shorter wavelengths but marginally resolved or blended at longer wavelengths.

To fully describe how these features change spatially across the entirety of the region, we therefore regridded all cubes into the spatial dimensions of channel 4 (longest wavelength and largest pixel size) while binning together pixels in each of the other three channels to form larger pixel sizes, averaging the individual spectra together. We furthermore defined five apertures corresponding to notable areas of interest: at the central star (designated ``Star"); within the bright inner core (designated ``Core Region"); at the edge of the bright region where the \csixty emission peaks (designated ``Fullerene Ring"; further discussed in Section \ref{sect:maps}); outside of the fullerene ring (designated ``Extended Halo"), and at the location of a low-ionization structure (LIS; see \overviewpaper) where the \ArII emission line peaks (designated ``\ArII Peak"). The location of each aperture can be seen in Figure \ref{fig:apertures}. Within each aperture, we binned spectra of each regridded cube, and then stitched the subchannels together to create five continuous spectra spanning the entire MIRI/MRS wavelength range covering the same regions. We stitched the spectra together by scaling the different subchannels with constant multiplicative factors so that the surface brightness in the overlapping wavelength range between each subchannel is the same. While these factors differ slightly for each pixel, they all generally lie between 0.8 and 1.2. While such regridding allows a direct comparison between the spectra of different channels, it does so at the cost of spatial resolution. For spatial studies of a single spectral feature that is fully contained within a single channel, we therefore also compared spectra on a pixel-by-pixel basis using the original data cubes.

\section{Spectral characteristics of the fullerene emission bands}
\label{Sec:spectral_characteristics}
In this section, we analyze the spectra of Tc~1 in detail. In particular, we describe the spectral characteristics of the neutral \csixty and \cseventy bands, and present the results of a search for other fullerenes, fullerene compounds and related carbonaceous species. 

\subsection{The \csixty emission bands}
\label{Sec:C60_spectral_characteristics}

\begin{figure*}[t]
      \centering
    \resizebox{\hsize}{!}{
    \hspace{.33\linewidth}
    \includegraphics[width=0.33\linewidth, clip, trim= 0.26cm 0.26cm 1.39cm 0cm]{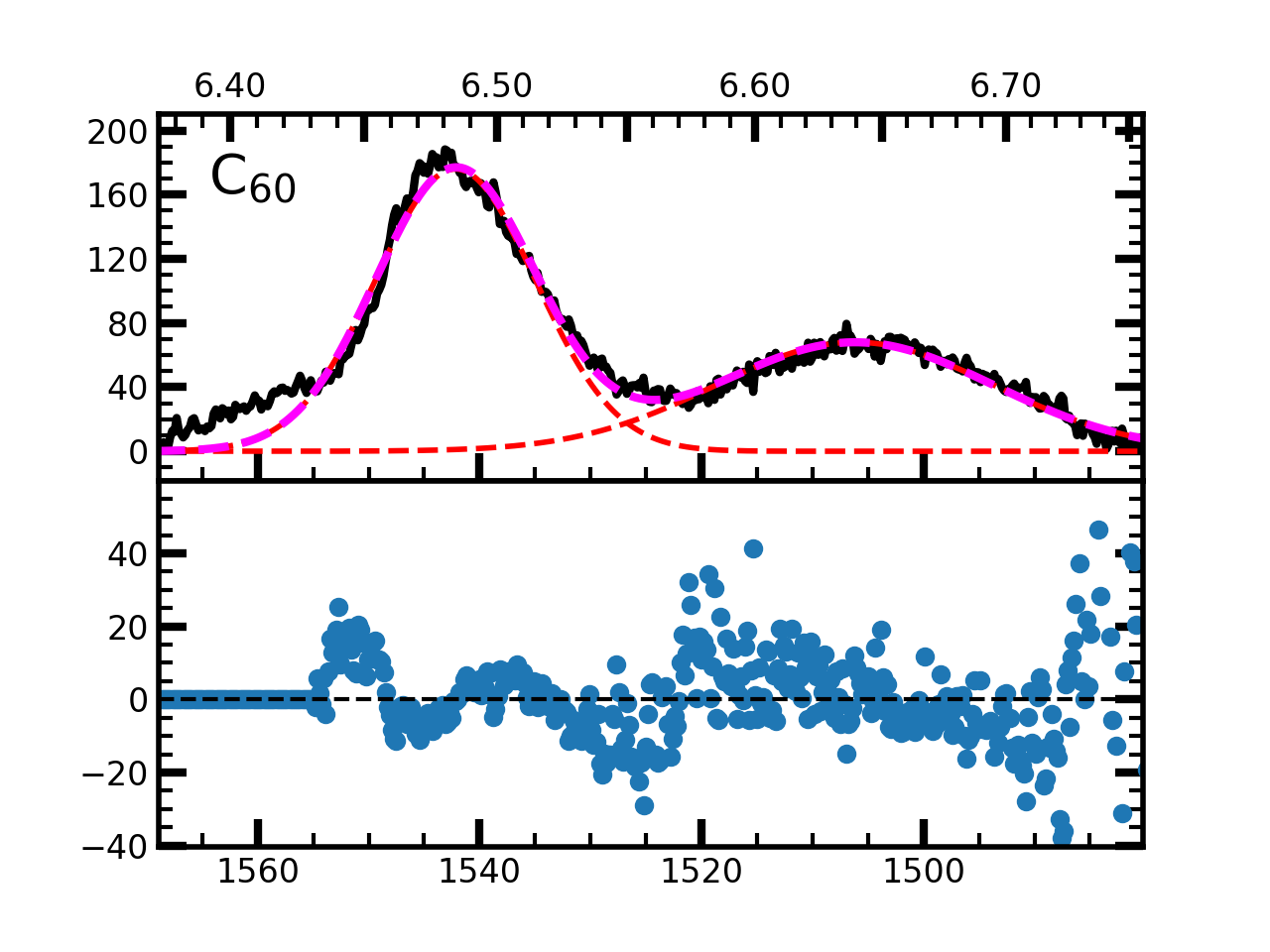}
    \hspace{.33\linewidth}
    }
    \resizebox{\hsize}{!}{
    \includegraphics[clip, trim= 0.26cm 0.26cm 1.39cm 0cm]{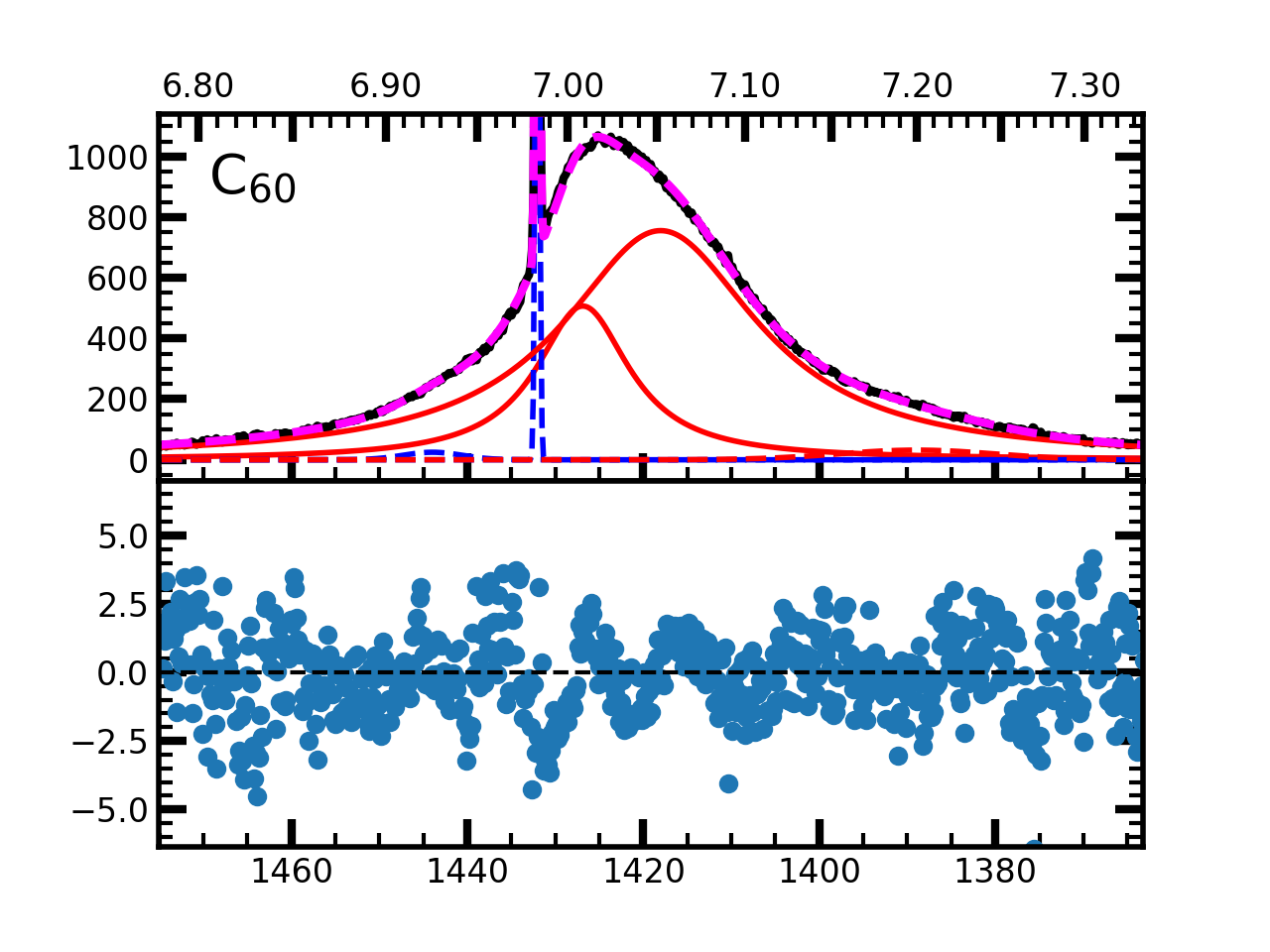}
    \includegraphics[clip, trim= 0.26cm 0.26cm 1.39cm 0cm]{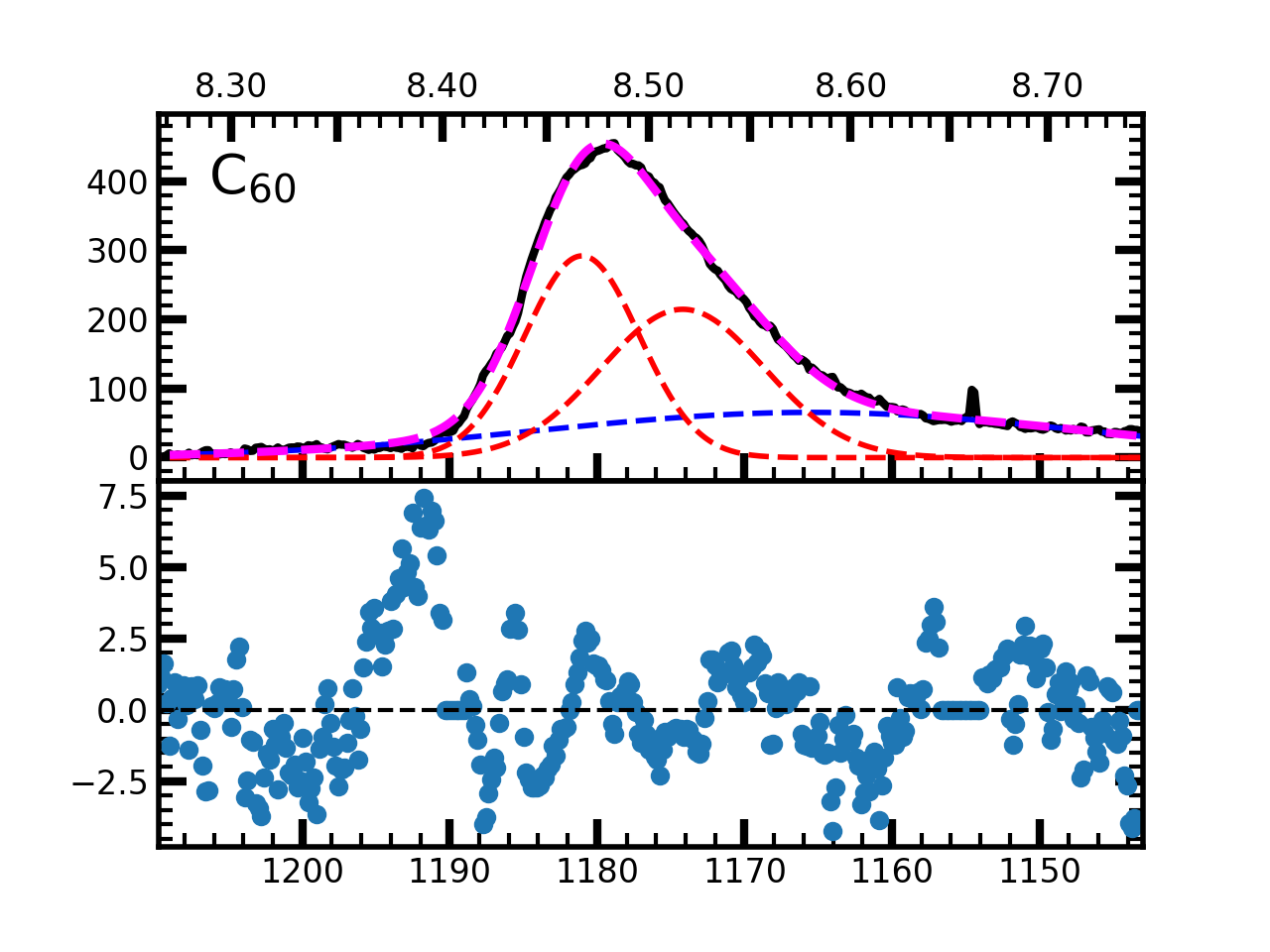}
    \includegraphics[clip, trim= 0.26cm 0.26cm 1.39cm 0cm]{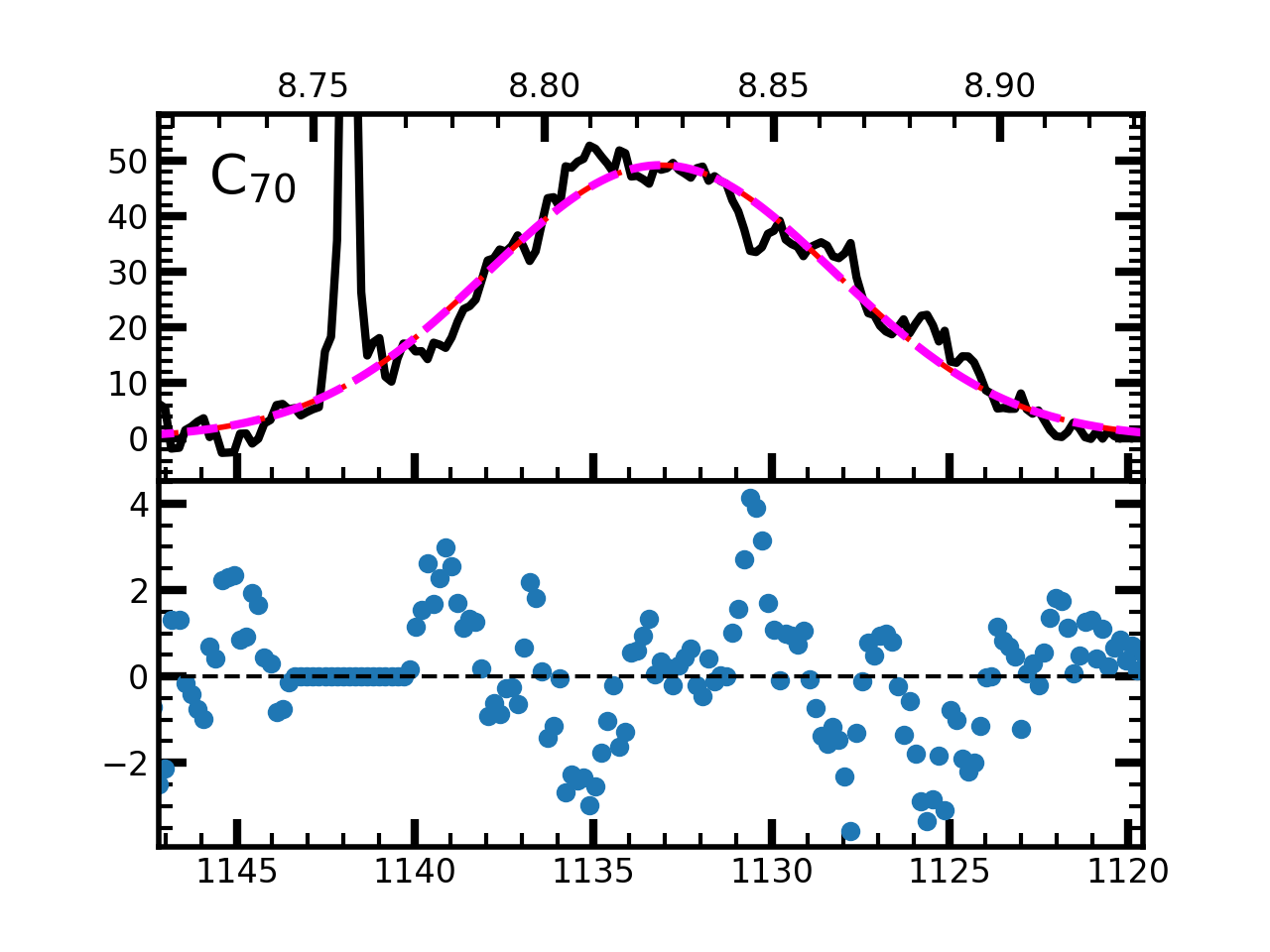}
    }
    \resizebox{\hsize}{!}{
    \includegraphics[clip, trim= 0.26cm 0.26cm 1.39cm 0cm]{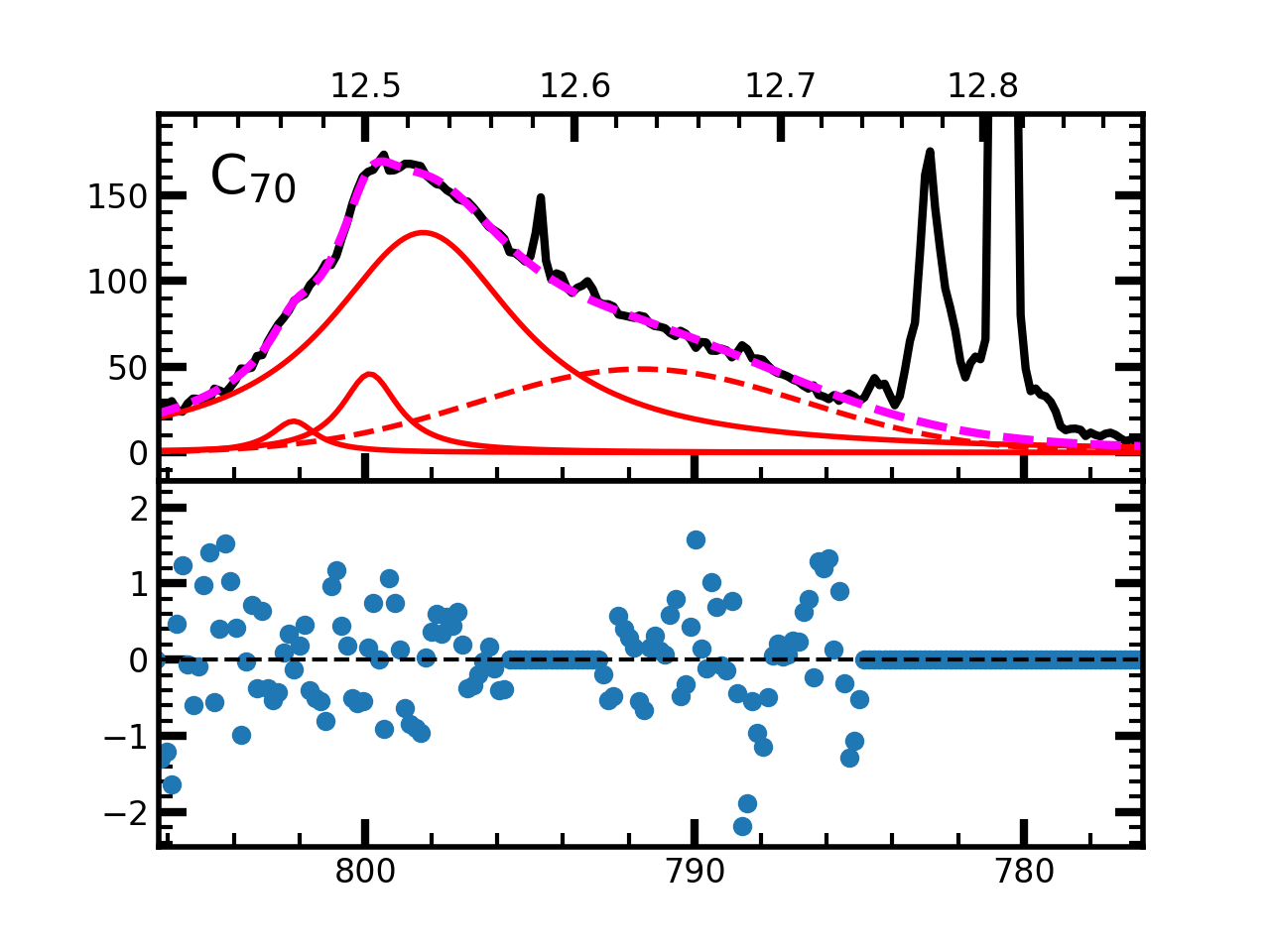}
    \includegraphics[clip, trim= 0.26cm 0.26cm 1.39cm 0cm]{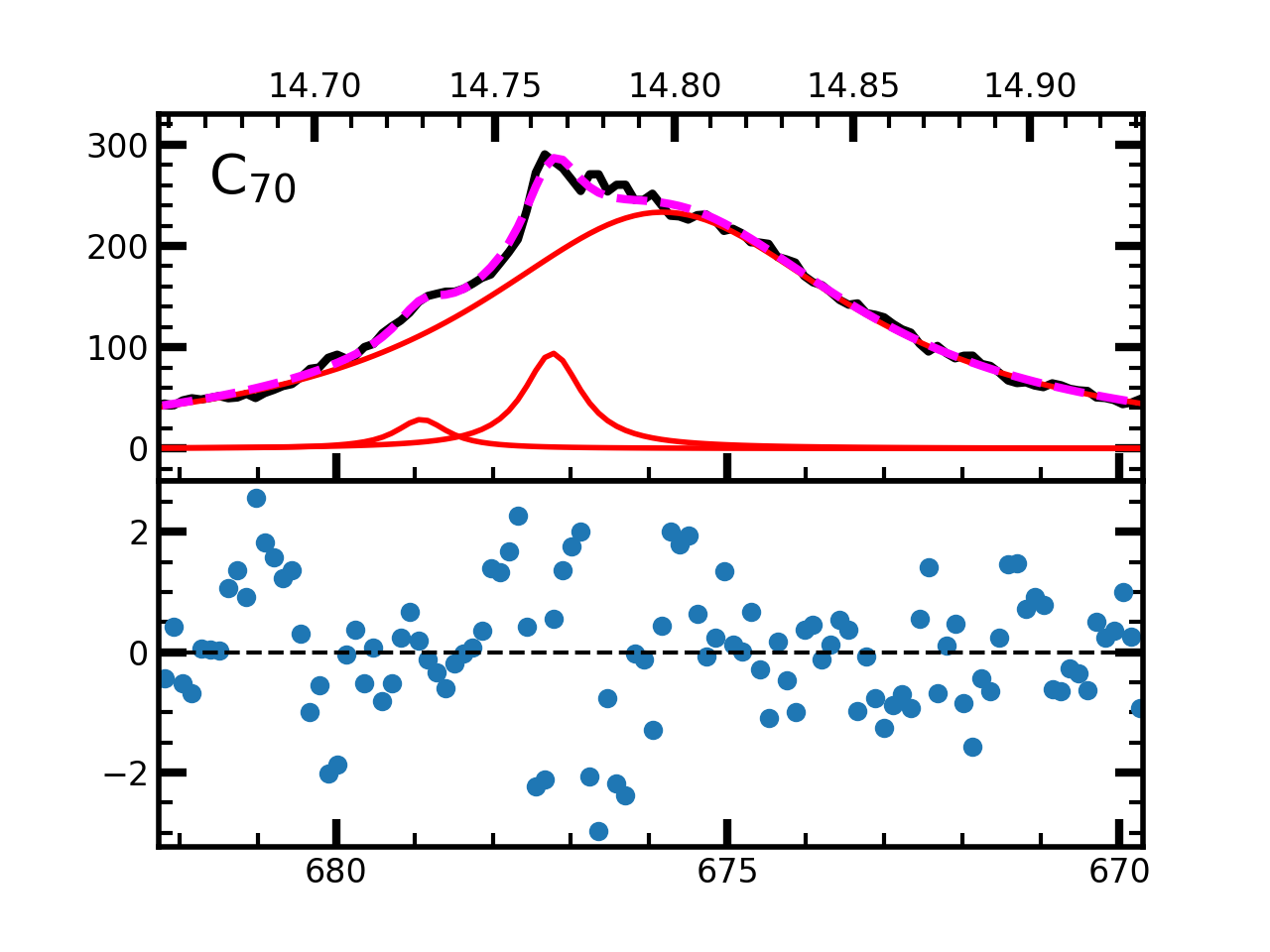}
    \includegraphics[clip, trim= 0.26cm 0.26cm 1.39cm 0cm]{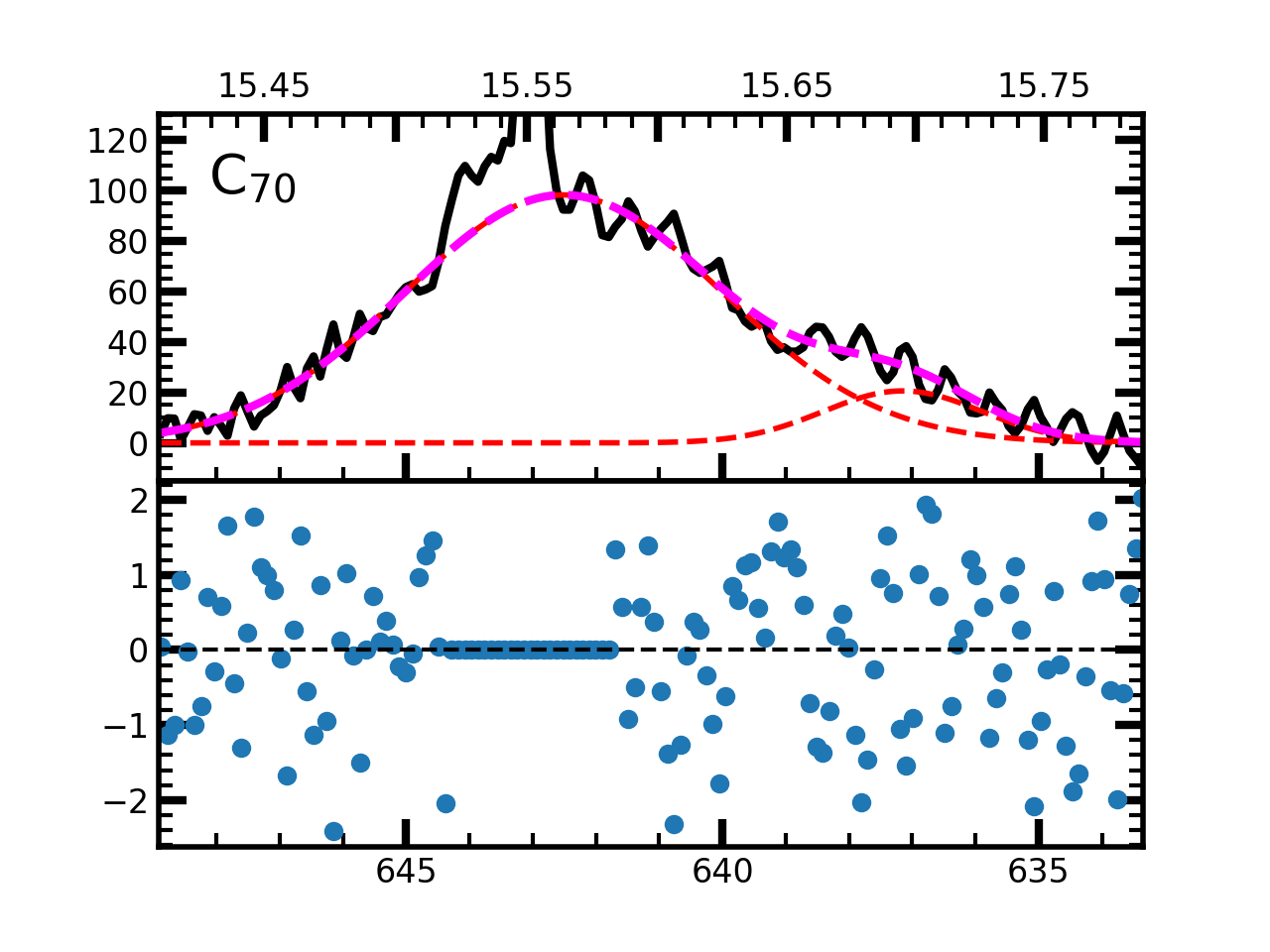}
    }
    \resizebox{\hsize}{!}{
    \includegraphics[clip, trim= 0.26cm 0.26cm 1.39cm 0cm]{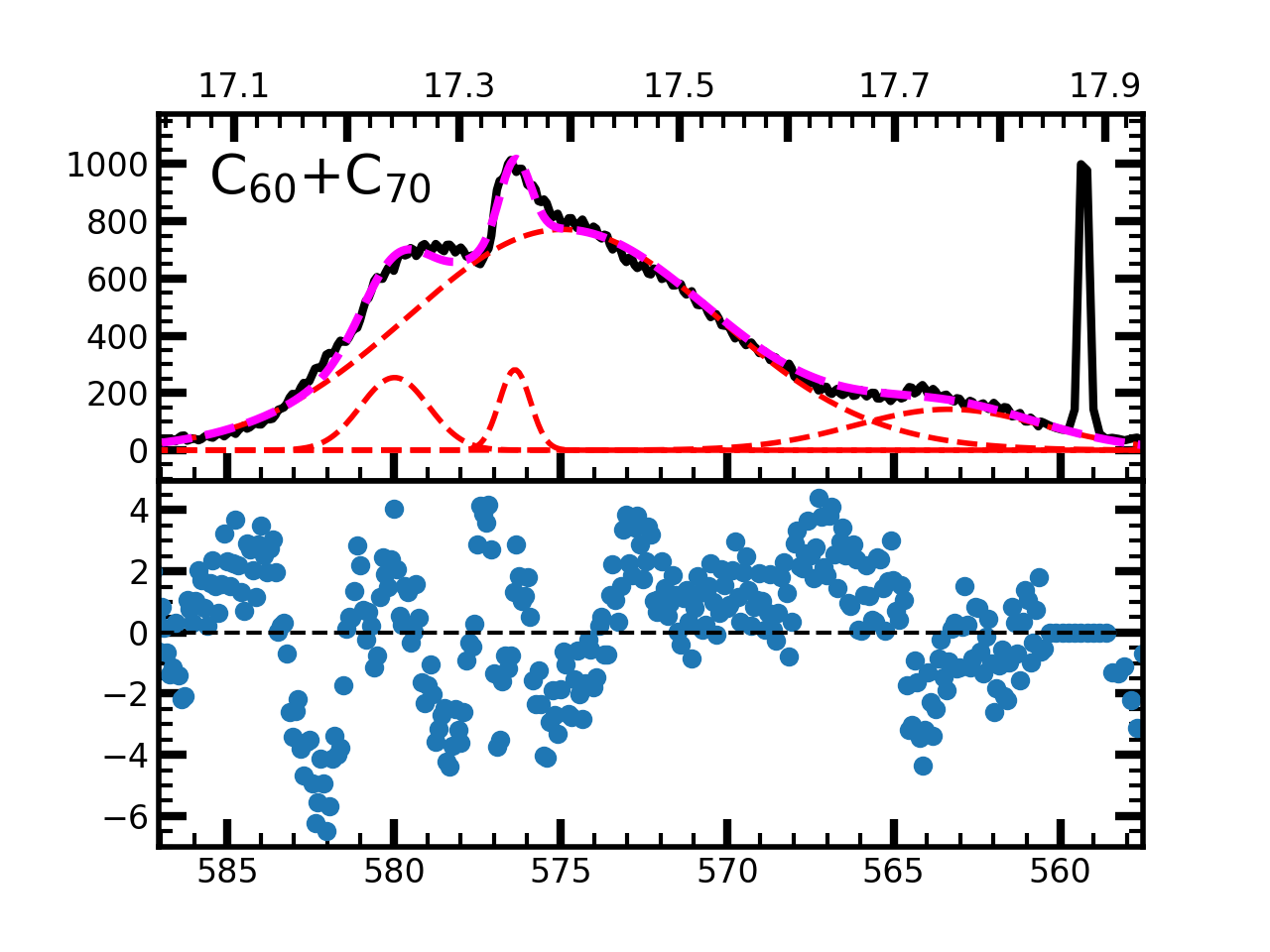}
    \includegraphics[clip, trim= 0.26cm 0.26cm 1.39cm 0cm]{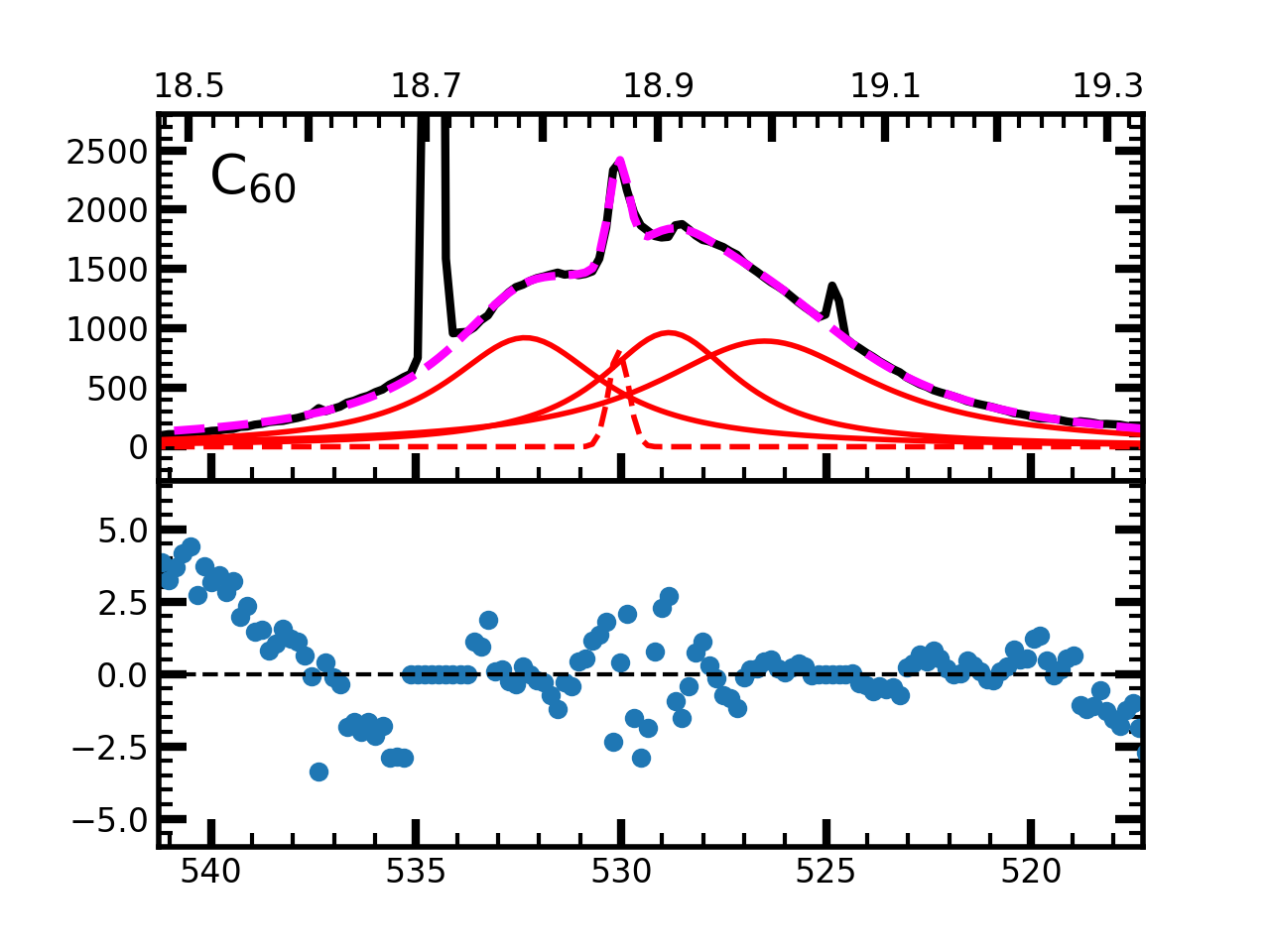}
    \includegraphics[clip, trim= 0.26cm 0.26cm 1.39cm 0cm]{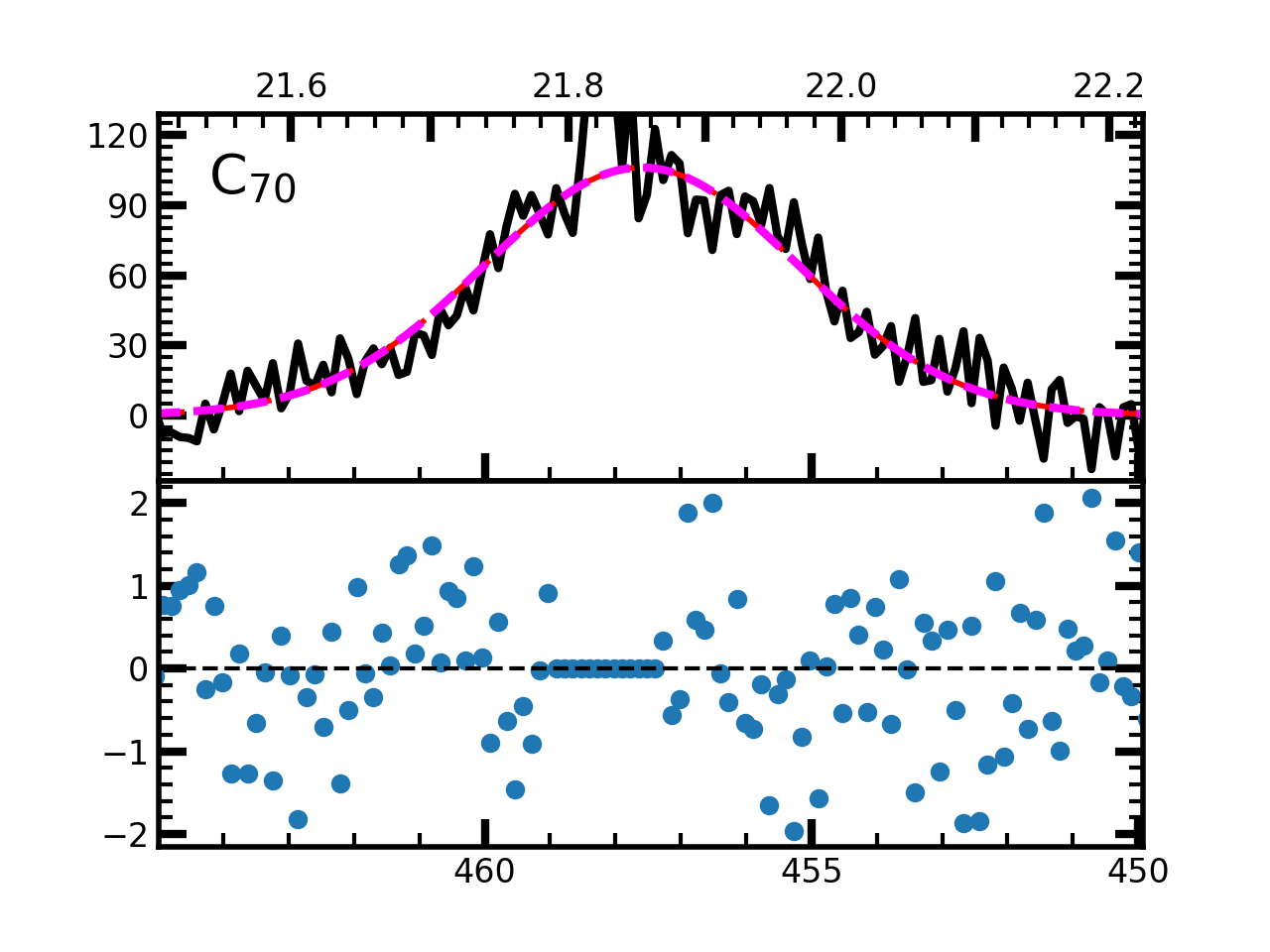}
    }

\caption{Example spectra showing the fullerene features (black) fit with multiple Gaussian and Lorentzian components. The red components correspond to fullerene emission, with dashed and solid lines styles denoting Gaussian and Lorentzian profiles, respectively. Blue components correspond to non-fullerene emission while the magenta line corresponds to the overall best fit. The bottom half of each panel shows the residuals displayed as percentage of the original, non-continuum subtracted spectrum. All spectra were taken from the ``Fullerene Ring" aperture. We subtracted a linear continuum from all spectra. Bands at 6.5, 6.6, 7.0, 8.5, 17.4, and 18.9 \um correspond to \csixty; bands at 8.8, 12.5, 14.8, 15.6, 17.7, and 21.8 \um correspond to \cseventy.}
\label{fig:c60_fits}
\end{figure*}

Figure~\ref{Fig:overviewspectrum} shows the composite spectrum of the JWST/MIRI observations of Tc~1. A detailed inventory of the numerous atomic emission lines is provided by \overviewpaper. After removal of the atomic emission lines, the most conspicuous spectral features in the continuum-subtracted spectrum are those corresponding to the four IR active $T_{1u}$ fundamental vibrational modes of \csixty at 7.0, 8.5, 17.4 and 18.9 \um. The high spectral resolution and sensitivity of JWST allow us to study these features in great detail. 

Many of the band profiles are asymmetric and appear to have subtle substructures superposed on the emission band. To better characterize the profiles and more easily describe variations in their overall emission across Tc~1, we decompose the bands into a combination of Gaussian and Lorentzian profiles. A more detailed discussion of this procedure can be found in Appendix \ref{App:fitting}, and an example fit spectrum for all bands is shown in Figure \ref{fig:c60_fits}. All four bands are quite broad and increase in width at shorter wavelengths. Additionally, each band is asymmetric with a broader wing on the red side than on the blue side. Both the 17.4 and 18.9 \um bands show clear evidence of ro-vibrational structure whereas the shorter wavelength bands have smooth and broad profiles (see J. Cami et al. 2026a, submitted for further discussion).

In addition to the four bands associated with the main fundamental vibrational modes of \csixty, we include a feature at 6.5 \um that has been seen in other evolved post-AGB objects \citep{Sloan:p-AGBdust} and is associated with a combination band of \csixty \citep{Brieva:C60ratios}. This assignment is further supported by the detection of more \csixty combination bands in the near-infrared spectrum of Tc~1 \citep{Giese:nirC60}. Many of these combination bands were first detected in laboratory measurements of \csixty \citep{Chase1992, Wang1993, Nemes:C60C70_temp, Hara1997}.

\begin{figure}
    \centering
    \resizebox{\hsize}{!}{
    \includegraphics[trim={0.6cm 0cm 0cm 1.4cm},clip]{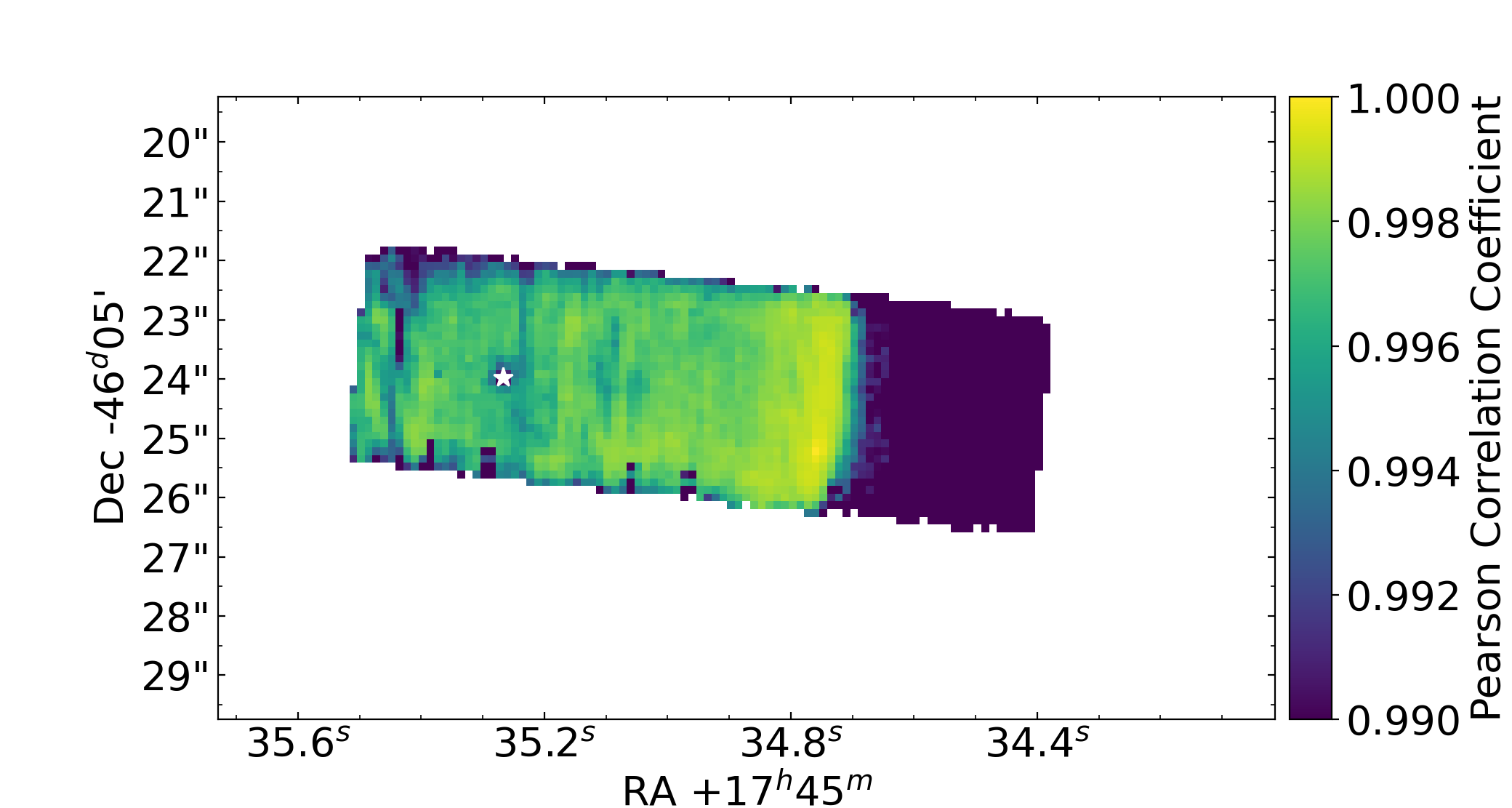}}
    \caption{Pearson correlation coefficients calculated for the 7.0 \um feature of each individual spectrum compared to a spectrum taken from the ring. The darker pixels near the bottom of the region are due to systematic artifacts in the data cubes that cannot be resolved by the pipeline, and do not affect the results. The location of the central star is denoted by a white star.}
    \label{fig:corr_coeff}
    
\end{figure}

These five bands show large changes in intensity across the full MIRI field of view (Section \ref{sect:maps}). To investigate if the profile shape of each feature changes significantly across the full region, we calculated the Pearson correlation coefficient (PCC) between each continuum-subtracted spectrum and a spectrum taken from the ring. All atomic lines were masked out for this calculation. Figure \ref{fig:corr_coeff} shows the PCC values for the 7 \um band. For all five bands, we find that there are overall no major changes to the profile shapes: the calculated PCC values are greater than 0.99 within the central core region for all bands. Apart from some small variations in the wings of the 17.4 and 18.9 \um bands (see Appendix \ref{App:profile}), we thus conclude that the individual \csixty band profile shapes are relatively identical at each pixel. We thus kept the same central wavelengths and widths for each of the components across all spectra when fitting the different \csixty bands, and only allowed the intensity of each component to vary. A full list of fit components and their derived parameters are given in Appendix \ref{App:fitting}; integrated surface brightness values for each band calculated from the decomposed fullerene emission in our five apertures are listed in in Table~\ref{tab:features}. Due to the high spectral resolution of JWST, these are the most accurate measurements of the cosmic fullerene bands to date.

Several of our band decompositions require weak components in addition to the dominant profiles (Figure~\ref{fig:c60_fits}), and our spectrum furthermore contains a number of weak, unassigned features. To decide whether such components are due to fullerenes, we use their spatial distribution as an additional criterion. As discussed in detail in Section~\ref{sect:maps}, the strong \csixty and \cseventy bands all share the same distinct spatial distribution: their emission peaks in a ring-like structure near the edge of the central core, which we interpret as a limb-brightened, geometrically thin shell. This morphology is very different from that of other emission components in our data, such as the hydrogen recombination lines (\overviewpaper) or the dust continuum, which mostly originates from the horseshoe-shaped structure around the central star (Section~\ref{sec:intro}). For each weak component, we therefore created a spatial map in the same way as for the strong bands, and considered it a fullerene component if the map shows the fullerene ring. For example, the weak Gaussian component at 7.2~\um\ in our decomposition of the 7.0~\um\ band clearly shows the ring and is included as a \csixty component, whereas the 6.9~\um\ component does not (Appendix~\ref{App:fitting}).

Our spectrum shows weak features at 5.6, 6.6, 7.95, 8.05, 8.15, 9.1, 9.6, 10.0, 10.5, 14.0, and 14.4~\um\ that have not been definitively assigned before in astronomical spectra. All of these features are seen in thin-film and gas-phase laboratory spectra of \csixty \citep{Chase1992, Wang1993, Nemes:C60C70_temp, Hara1997}; the 6.5 and 6.6~\um\ features in particular look remarkably similar to the high-temperature gas-phase spectrum of \citet[their Figure~2]{Nemes:C60C70_temp}. Moreover, their positions match those of weak \csixty combination bands in the same anharmonic calculations that we used to identify the near-IR combination bands in Tc~1 \citep{Giese:nirC60}. Finally, the spatial maps of all these features show the fullerene ring (Figure~\ref{fig:combination_maps}), although the maps of the weakest features are affected by noise and also show emission from other structures such as the horseshoe (Appendix~\ref{sec:append_combination}). We therefore tentatively assign all of these features to \csixty combination bands.

\input{fullerene_measurements_table}

\subsection{The \texorpdfstring{\cseventy}{C70} emission bands}
\label{Sec:C70_analysis}

\cseventy has 31 IR active fundamental modes \citep[e.g.,][]{vonCzarnowski1995321, Stratmann1998, Schettino2002}. \citet{Cami2010} detected 4 of those in the Spitzer-IRS spectrum of Tc~1 at 12.52, 14.79, 15.56 and 21.81 \um. We clearly detect these features in our JWST/MIRI spectra as well, and additionally confirm the presence of the 8.8 and 17.7 \um \cseventy features. \cseventy also has modes that are severely blended with the four fundamental \csixty bands described above. These blended features are located at similar wavelengths and are further discussed in Section \ref{sec:discussion}.

As with the five bands described above, we fit profiles to each of the isolated \cseventy features in the same manner with either Gaussian or Lorentzian components. The 12.5 and 14.8 \um features interestingly show similar fits to that of the 17.4 and 18.9 \um \csixty features. Indeed, the shape of the 14.8 \um band includes a weaker red wing followed by a sharp increase with a stronger blue wing, as in the two \csixty features (see Figure~\ref{fig:c60_fits}). Both the 17.7 and 21.8 \um \cseventy bands also show small features reminiscent of rovibrational structure, albeit at a much smaller scale. We interpret this similar shape across the features as evidence for rovibrational structure in \cseventy as well.

\begin{figure*}[t]
    \centering
    \includegraphics[width = \textwidth, clip, trim= 2cm 0cm 3cm 0cm]{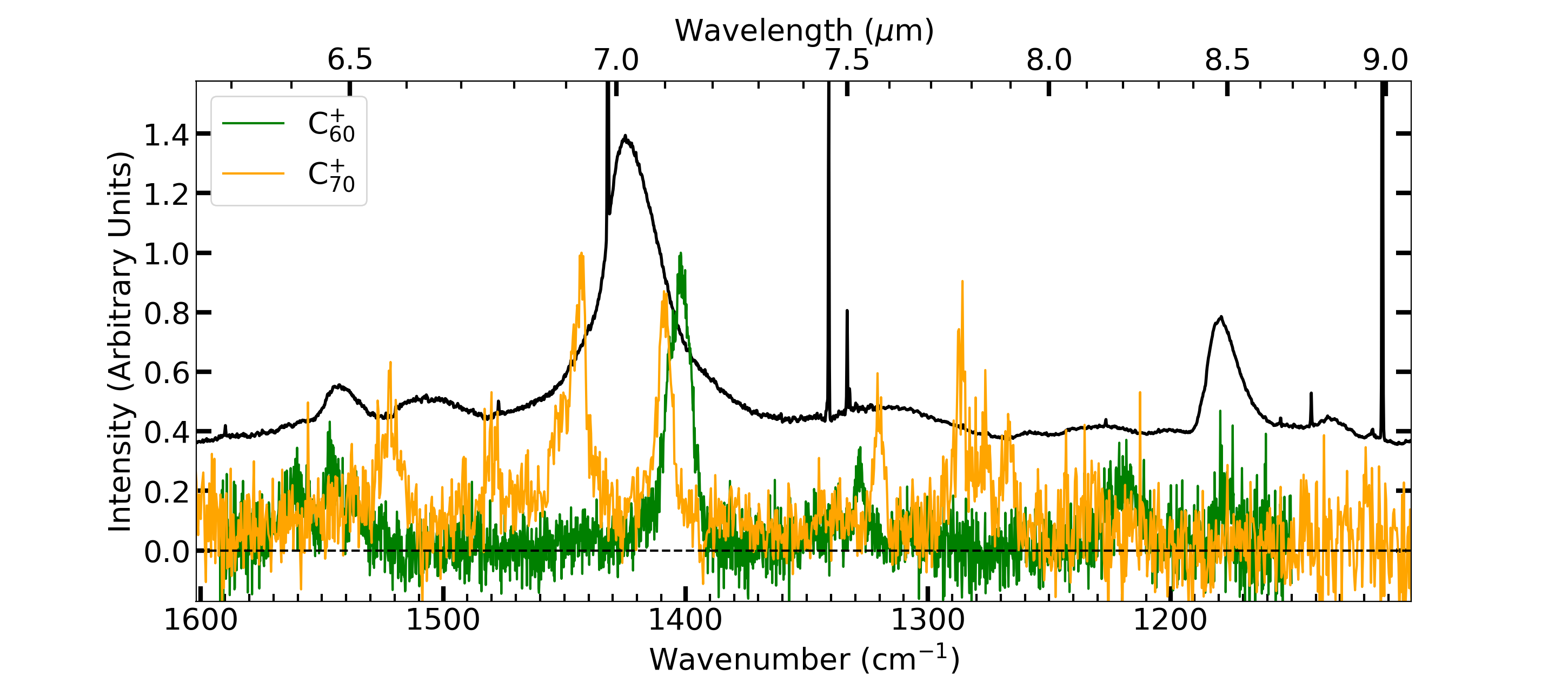}
    \caption{Laboratory spectra of \csixtyplus (green) and \cseventyplus (orange) \citep[priv. comm.,][]{2025ApJ...992..175D} plotted on top of the spectrum taken from the ``Fullerene Ring" aperture (black). Intensity units of all three spectra are normalized with the observed data offset.}
    \label{fig:plus_mols_overlay}
\end{figure*}

As was the case for \csixty, we find no major changes to the profile shapes for any of the \cseventy bands: the calculated PCC values are again all greater than 0.99 within the central core region, and thus all \cseventy band profiles are identical as well. An example fit spectrum for each \cseventy feature is also shown in Figure \ref{fig:c60_fits}. Details on fit parameters and individual components are listed in Appendix \ref{App:fitting}.

\subsection{A search for fullerene cations}

Across our MIRI field of view, the physical conditions change: there is dilution (and some attenuation) of the UV field as one moves away from the central star, and outside the bright core, the electron density decreases \citep{Walsh_MUSE2026}. One might thus expect changes in the charge balance across the object, with perhaps emission from fullerene cations in the inner regions. We therefore searched for evidence of \csixtyplus and \cseventyplus in our data as well. 

Recent experiments using helium-tagging messenger spectroscopy have obtained gas-phase infrared absorption spectra of both \csixtyplus and \cseventyplus in the 6--9 \um region \citep{2025ApJ...992..175D}. We reproduce those spectra in Figure~\ref{fig:plus_mols_overlay}, alongside the Tc~1 spectrum from the Fullerene Ring. Both \csixtyplus and \cseventyplus are dominated by absorption features around 7 \um.

The most intense \csixtyplus feature is at 1401 \wno (7.14 \um), with additional features at 1330, 1546, and 1561 \wno \citep{Kern2013, Strelnikov_c60+_2015, 2025ApJ...992..175D}. Neon matrix measurements covering a broader wavelength range \citep{Strelnikov_c60+_2015} show that \csixtyplus also has features in the 17–19 \um region, including a moderately strong $A_{2u}$ mode at 529.7 \wno (18.88 \um). However, the 7.14 \um feature is by far the strongest \csixtyplus band, with an integrated molar absorptivity approximately 3.5 times larger than the 18.88 \um feature. For \cseventyplus, the gas-phase spectrum from \citet{2025ApJ...992..175D} shows the strongest feature at 1448.5~\wno (6.90 \um), with several weaker features in the 6.5--8 \um range. Figure \ref{fig:plus_mols_overlay} compares the laboratory spectra of both cations to our observed spectrum.

On first sight, our data show no clear evidence for any contribution due to \csixtyplus or \cseventyplus. However, the laboratory spectra from \citet{2025ApJ...992..175D} are cold ($\sim$10 K) absorption spectra, whereas the emission in Tc 1 originates from vibrationally hot molecules. We should thus expect broader bands that are slightly redshifted from the cold absorption frequencies. If the cation bands would be as broad as the 7 \um band in our spectra, they would be much harder to recognize. We note that in our decomposition of the 7 \um band, we need weak emission features at about 6.9 and 7.2 \um. The 7.2 \um component shows the same spatial distribution as the neutral fullerene bands and is thus most likely due to \csixty itself (Section~\ref{Sec:C60_spectral_characteristics}), although we cannot exclude a contribution from \csixtyplus. The 6.9 \um component, on the other hand, does not show the fullerene ring, but its spatial map is too noisy to determine where this emission originates; it could perhaps be due to \cseventyplus, especially if this band is much broader than in the laboratory data. While we thus do not conclusively detect any cation emission in our data, we cannot exclude (weak) contributions to the observed spectra. 

We note that there may be an additional reason why fullerene ions are hard to detect in the IR. Their Jahn--Teller distorted ground states give rise to low-lying electronic states, and vibrationally hot ions could thus cool in part through recurrent (or Poincar\'e) fluorescence from these states \citep{1988PhRvL..60..921L} rather than through IR emission alone, which would reduce the strength of their IR bands. Recurrent fluorescence has indeed been shown to be highly competitive with vibrational emission for a wide range of carbon clusters containing 24--60 atoms \citep{Lacinbala2022_RF}, and could even account for the near-IR continuum emission observed in some reflection and planetary nebulae \citep{Lacinbala2023_RF}. For \csixtyplus, such emission would occur in the near-IR electronic transitions that are also responsible for the \csixtyplus DIBs at 9577 and 9632~\AA\ \citep{Campbell:C60+DIBs}. These wavelengths are covered by existing X-shooter spectra of Tc~1 \citep{Aleman:Tc1:XSHOOTER}, but no such emission has been reported. Emission from vibrationally hot ions would however result in broad bands that may be hard to recognize, and a targeted search for these features is warranted. At present, we thus have no evidence that recurrent fluorescence plays a significant role in the cooling of fullerene ions in Tc~1.

\subsection{A search for \texorpdfstring{\csixtyminus}{C60-}}\label{sec:c60minus}
\begin{figure}
\resizebox{\hsize}{!}{
\includegraphics[clip, trim= 0cm 0cm 1cm 0cm]{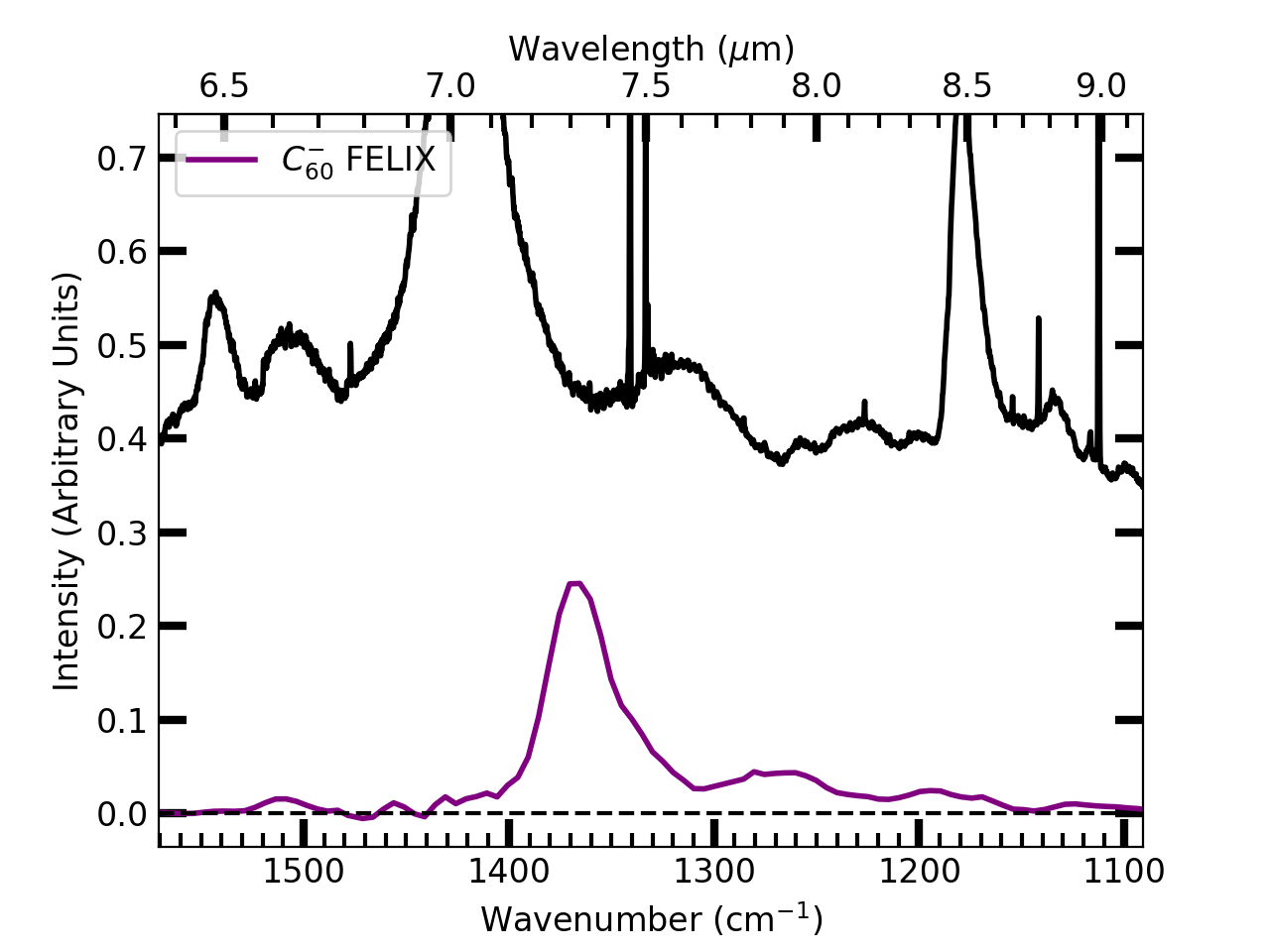}}
\resizebox{\hsize}{!}{\includegraphics[clip, trim= 0cm 0cm 1cm 0cm]{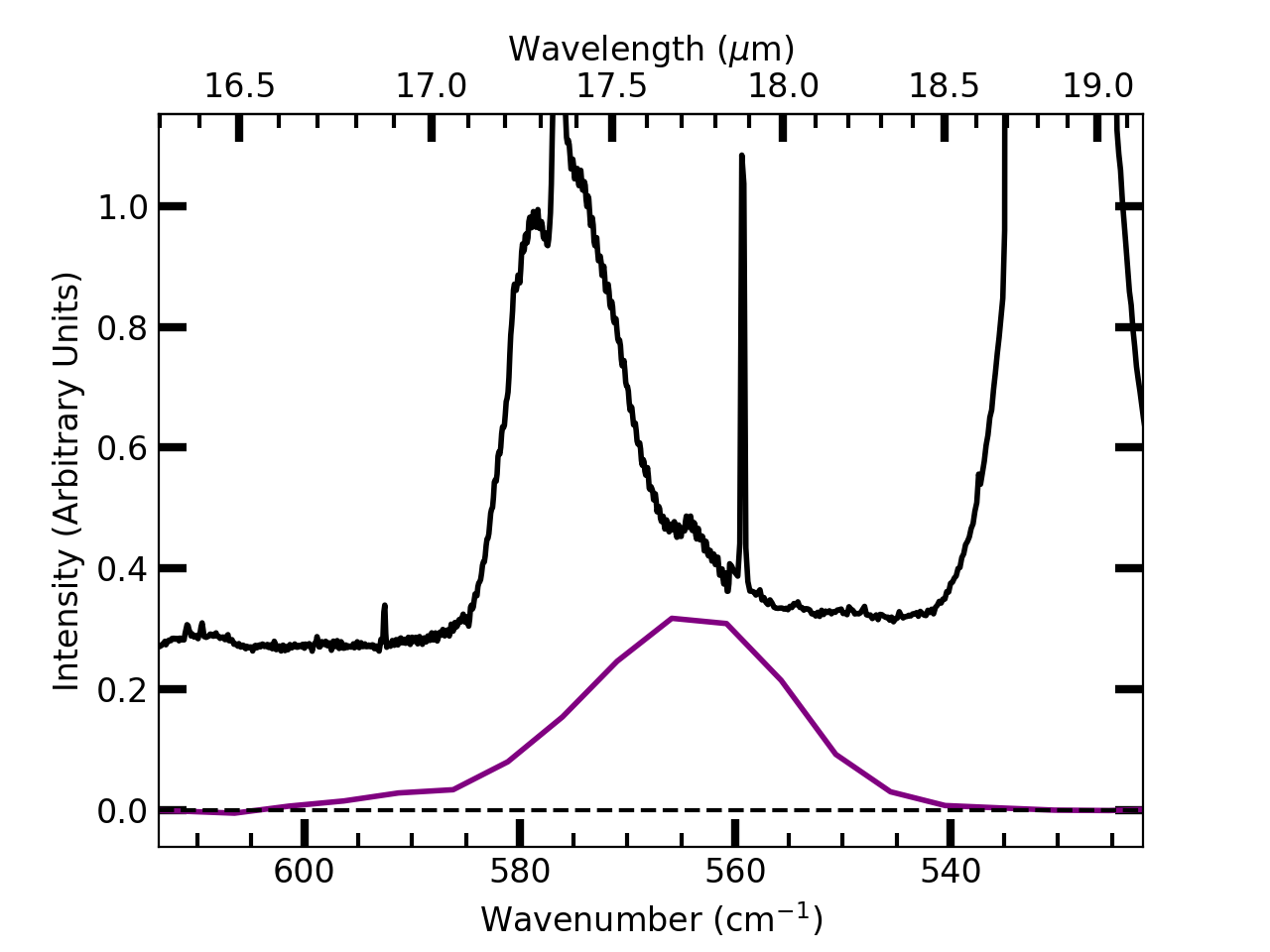}}
\caption{\label{Fig:C60anion}FELIX measurements of \csixtyminus (purple) plotted on top of the spectrum taken from the ``Fullerene Ring" aperture (black). Intensity units of both spectra are arbitrary. }
\end{figure}

\cite{Walsh_MUSE2026} determined the electron density of Tc~1 to be $n_e\sim2200~{\rm cm}^{-3}$. There may thus be locations where the charge balance favors \csixtyminus, and we thus also searched for spectroscopic evidence of \csixtyminus. We compared our observations to new room-temperature laboratory measurements of gas-phase \csixtyminus obtained using infrared multiple-photon electron detachment (IRMPED) spectroscopy at the FELIX free-electron laser facility (Finazzi et al., in prep). These measurements cover the 400--1800~\wno (5.5--25~\micron) range at $\sim$0.5\% resolution and represent a significant improvement over previous matrix-isolation spectra. The two strongest \csixtyminus bands occur at 566~cm$^{-1}$ (17.67~\micron) and 1367~cm$^{-1}$ (7.32~\micron), corresponding to radial cage and tangential vibrational modes respectively, with weaker bands at 660~cm$^{-1}$ (15.15~\micron), 1196~cm$^{-1}$ (8.36~\micron), 1264~cm$^{-1}$ (7.91~\micron), and 1509~cm$^{-1}$ (6.62~\micron).

Figure~\ref{Fig:C60anion} compares the two strongest bands from the FELIX \csixtyminus spectrum to our Tc~1 observations. The laboratory band at 566~cm$^{-1}$ (17.66 \um) aligns well with an emission band in the red wing of the 17.4 \um \csixty band; we included this component in our spectral decomposition as a band centered at 17.75~\micron. However, at this frequency we also expect a \cseventy band, and given the strength of the other \cseventy bands, we expect most (if not all) of this emission to be due to \cseventy as well (Section~\ref{Sec:C70_analysis}; Appendix \ref{App:fitting}). While it is possible that \csixtyminus contributes to the emission in this band, it is not possible to isolate it from the \cseventy feature. 

The strongest \csixtyminus band, at 1367~cm$^{-1}$ (7.32~\micron) is located in the red wing of the strong 7.0~\micron\ neutral \csixty band, and we do not see a recognizable feature at this wavelength. If \csixtyminus is present in our data, it could at best be represented by the 7.2 \um component from our decomposition of this band; however, as noted above, this component is most likely due to neutral \csixty. The other, weaker \csixtyminus bands mentioned above coincide with weak features in the JWST data, although the features at 6.6 and 7.95 \um are better explained as \csixty combination bands (Section~\ref{Sec:C60_spectral_characteristics}). Thus, while there are some tentative correspondences, we can certainly not claim a definitive detection of \csixtyminus emission; it could contribute to the observed emission however. 

\subsection{A search for other fullerenes}

\begin{figure*}[t]
    \centering
    \resizebox{\hsize}{!}{
    \includegraphics[width=\textheight, clip, trim= 4cm 0cm 4cm 0cm]{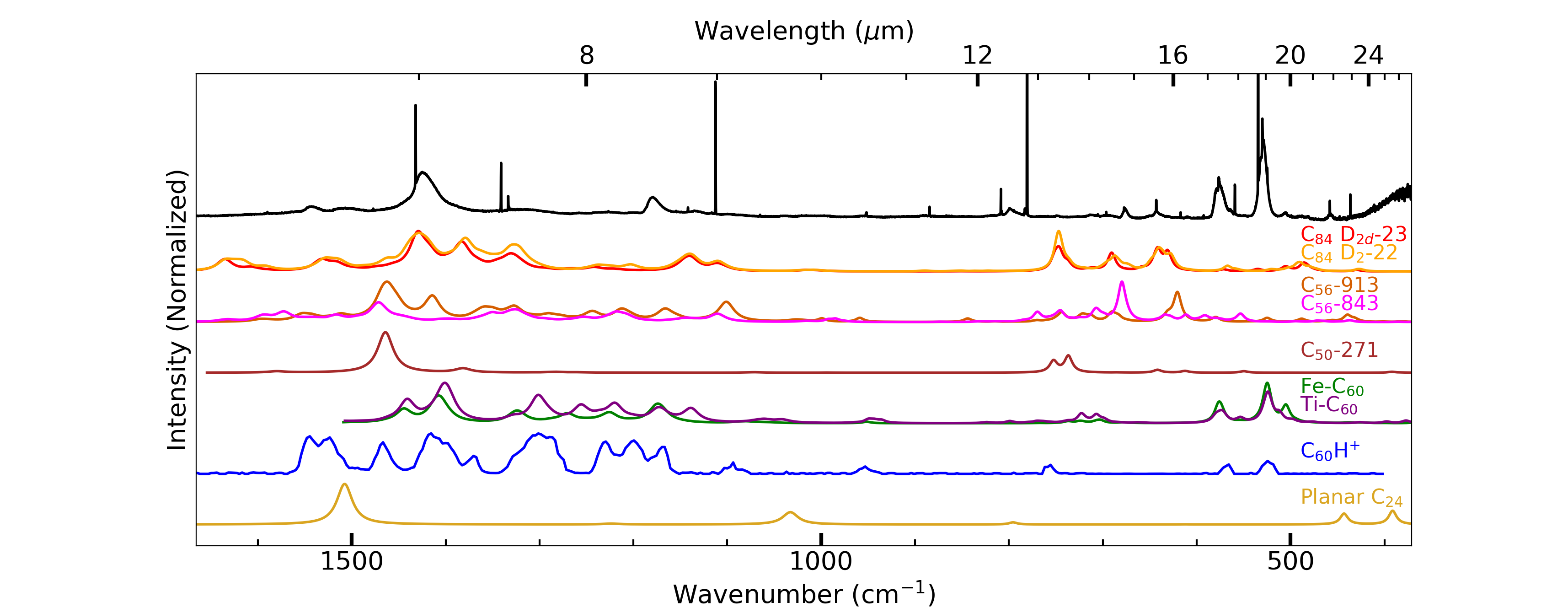}}
    \caption{Model IR spectra of related carbon species plotted against the continuum-subtracted spectrum of Tc~1 (black). See the text for a description of how the spectra were calculated. All spectra are normalized to their peak value.}
    \label{fig:other_fullerene}
\end{figure*}

To obtain a more complete inventory of large carbonaceous species in the spectrum of Tc~1, we also searched for the spectroscopic signatures of other neutral fullerenes. Given the strength of the \cseventy features relative to the \csixty features, we first considered the next large stable fullerene, C$_{84}$ and in particular its two most-stable isomers of ($D_{2d}-23$ and $D_{2}-22$). We compared our data to model spectra calculated using positions and intensities from \citet{Bettinger2000} calculated at the B3LYP/6-31G* level of theory. The $D_{2d}-23$ isomer has 93 IR-active transitions while the $D_{2}-22$ isomer has 183; the strongest transitions for both occur in the 7--8 \um and 13--17 \um ranges \citep{Dennis2000, Bettinger2000}. To turn the calculated stick spectra (frequencies and intensities) into an estimated emission spectrum for a molecule undergoing IR fluorescence, we multiply the intensities by a 750~K blackbody \citep{2008ARA&A..46..289T}. We then convolved the spectrum with a Lorentzian with line widths of 20 cm$^{-1}$ and 10 cm$^{-1}$ for wavelengths less than and greater than 10 \um, respectively. These line widths were chosen to roughly match those of the \csixty features for direct comparison to our data.

As noted in \cite{Bettinger2000}, the original calculations will require an unknown scaling factor for the vibrational transition frequencies that is not taken into consideration. This will cause a slight shift of all bands to longer wavelengths, meaning we would not expect to have a perfect match between the simulated and observed spectra. To better estimate the positions here, we assume a scale factor of 0.978 as done for other fullerenes in \citet{Candian:cages}. Given that C$_{84}$ has many more IR-active modes than \csixty, the same energy (per absorbed UV photon) would be distributed over many more modes. If C$_{84}$ would have the same abundance as \csixty, its bands would be an order of magnitude weaker. However, we don't see even weak features that correspond with either the short-wavelength features or the long-wavelength features. The set of C$_{84}$ features in the 6--9 \um range does correspond to the location of the broad plateau in the JWST data (\dustpaper), but if we were to assume that C$_{84}$ was present and contributing emission to the plateau, we would also expect to see clear features at longer wavelengths (e.g. at 13.5 \um). We do not see any evidence of these longer wavelength features between 13--17 \um in our data. We therefore conclude a non-detection of C$_{84}$ in Tc~1.

\bigskip

Laboratory studies have shown that UV irradiation of large PAHs can result in the formation of fullerenes, and that intense radiation can also further fragment 
\csixty \citep{Zhen:formation_lab}. Fragmentation of \csixty follows a path through ejection of C$_2$ units, and not all smaller cages are equally stable. Laboratory experiments and theoretical calculations show increased stability for fullerene cages with so-called ``magic numbers" of carbon atoms. Indeed, fullerenes with 56, 50, or 44 carbon atoms are more stable than cages of other sizes  \citep[smaller than \csixty; ][]{Zhen:formation_lab, Candian:cages}. We thus also searched for the presence of the fullerene species C$_{56}$ and C$_{50}$, the two next smallest, most-stable fullerenes after \csixty. We calculated an approximate emission spectrum for the two most stable isomers of C$_{56}$ and the most stable of C$_{50}$ in a similar manner to our analysis for C$_{84}$, taking positions and intensities from \cite{Candian:cages}, multiplying them by a 750 K blackbody, and convolving the created stick spectrum with a Lorentzian.

The two C$_{56}$ isomers have their strongest features at $\sim$7.0 and $\sim$16.0 \um with weak features between $\sim$7--10 and $\sim$12--20 \um. While the C$_{56}-843$ isomer has a strong feature at $\sim$14.8, this would be completely blended with the \cseventy feature at the same position. The C$_{50}$ isomer has its strongest feature near the neutral \csixty band at 7 \um and some additional weaker features at $\sim$13.5 \um. In our data, we see no isolated features that directly correspond to the positions of either of these additional fullerene species.

As noted in \cite{Candian:cages}, it is difficult to make any substantive claims of detection for these smaller cages based off individual features alone. If present, the abundance of any of these cages will be significantly lower than that of \csixty or \cseventy in astrophysical environments. Furthermore, the intrinsic strengths of the IR-active modes of these species are also lower, and the emitted energy is typically distributed over more IR active modes. Together, this would result in considerably weaker emission from either C$_{56}$ or C$_{50}$ in our JWST data, if present. 

\bigskip
To summarize, we do not see any clear evidence for the presence of any other neutral fullerene species in our data other than \csixty or \cseventy. At the same time, we cannot rule out that weak emission from these species may be present, and together contribute to the emission plateaus underneath the \csixty and \cseventy features. 

\subsection{Fullerene compounds}

\citet{2023ApJS..269...26B} and \citet{2023ApJ...952...13H} suggest that metal-complexed fullerenes (``metallofullerenes") could contribute significantly to the observed \csixty emission seen in space. Metallofullerenes are complexes formed between \csixty and a metal atom. They can be endohedral (atom located inside the fullerene cage) or exohedral (metal bonded to the outside of the cage).

Endohedral metallofullerenes form as readily as empty cages in condensing carbon vapor \citep{Dunk:metallofullerenes}, and could thus be present wherever fullerenes are found. Bottom-up formation experiments show that alkali metals such as Na and K most readily form M@\csixty \citep{Dunk2014_EMF}; of these, Na is by far the most abundant. Na@\csixty is best described as Na$^+$@C$_{60}^-$, and its IR spectrum closely resembles that of \csixtyminus \citep{Dunk:metallofullerenes}. Since we find no clear evidence for \csixtyminus (Section~\ref{sec:c60minus}), Na@\csixty is unlikely to contribute significantly to our spectra. Other endohedral metallofullerenes show distinct features, many in the 6--9~\um\ range, that can be stronger than the bands of pristine \csixty \citep{2023ApJS..269...26B}. Apart from the bands we assigned to \csixty and \cseventy, we see no such discrete features in our data, although a mix of such species could contribute to the 6--9~\um\ plateau (\dustpaper). Finally, molecular endohedral fullerenes are spectroscopically very similar to the corresponding empty cages: the gas-phase IR spectra of H$_2$@\csixtyplus and H$_2$O@\csixtyplus closely resemble that of \csixtyplus, with only small blue shifts \citep{Strelnikov2018_H2C60, ONeill2026_H2OC60}. We would thus expect neutral species such as H$_2$@\csixty or H$_2$O@\csixty to be essentially indistinguishable from \csixty in our data. Exohedral metallofullerenes, on the other hand, can form by reactions of metal atoms or ions with existing fullerenes, regardless of how the fullerenes themselves formed; we therefore discuss these in more detail below.

We searched for the presence of metallofullerenes in our data by taking the two species from \cite{2023ApJS..269...26B} with the highest binding energies Fe-\csixty and Ti-\csixty and creating simulated spectra in the same way as for the other fullerenes (see Figure~\ref{fig:other_fullerene}). The details for how the original stick spectra were calculated can be found in \citet{2023ApJS..269...26B}. While both species exhibit features in the vicinity of the pure \csixty features at 7.0, 8.5, 17.4, and 18.9 \um and some pure \cseventy features, each metallofullerene also exhibits distinct features isolated from other fullerene bands with isolated features near 7.69, 7.96, and 8.20~\um. We do see weak emission in our data at similar wavelengths 7.95, 8.05, and 8.15~\um as shown in Figure \ref{Fig:overviewspectrum}. To test whether these bands in our data are due to metallofullerenes, we use their theoretical relative strengths compared to that of the stronger metallofullerene bands at 7.11 and 7.14~\um (for Fe-\csixty and Ti-\csixty, respectively) and estimate the expected strength of such bands in our data. In doing so, we find that these stronger bands would be twice as strong as the bands $\sim$8~\um. At such strengths and similar line widths, we would expect to see distinct features on top of the \csixty 7.0~\um band corresponding to the metallofullerenes. We do not see any such features in our data. The absence of the strong metallofullerene bands near 7.1~\um thus rules out that metallofullerenes are a major component of the astronomical spectra. Moreover, the weak features at 7.95, 8.05, and 8.15~\um\ in our data are better explained as \csixty combination bands (Section~\ref{Sec:C60_spectral_characteristics}).

Another fullerene compound that has been suggested as a possibly abundant interstellar species is protonated \csixty, \csixtyHplus \citep{Kroto1992}. When \csixty gains an extra proton and becomes \csixtyHplus, its symmetry is broken, and many more vibrational modes become IR-active resulting in a much richer vibrational spectrum \citep{Palotas:C60H+}. \citet{Palotas:C60H+} obtained a gas-phase IR absorption of \csixtyHplus, and we compare it directly to our data (Figure~\ref{fig:other_fullerene}). 

\csixtyHplus has strong IR features throughout the 6--9 \um region with some weaker features at 13.2, 17.7, and 19.1 \um. While there are some coincidences between the laboratory data and the features we see in the 6--9 \um region, there is no overall match between the laboratory data and our JWST data in this range, as was already discussed by \citet{Palotas:C60H+}. We see no evidence of an isolated 13.2 \um feature in our data. The features at the longer wavelengths, 17.7 and 19.1 \um, would be blended with the \cseventy and \csixty features (respectively) at similar wavelengths.

If we were to assume that the isolated 13.2 \um \csixtyHplus feature was hidden within the noise of our data, the possible scaled contribution from \csixtyHplus to the 17.4 and 18.9 \um \csixty bands would be $<$1\%, as these \csixtyHplus features are of similar strengths. The \csixtyHplus bands between 6--9 \um are $\sim$4--5 times stronger than that at 13.2 \um, and would be more likely to contribute emission if the molecule were present. But again, there is not a perfect match between our data and such features. Thus, in the best case, there could only be weak and extremely blended \csixtyHplus emission in our data. It is important to note however that the symmetry breaking implies that the same amount of energy is distributed over many more modes that are intrinsically weaker than the \csixty bands; thus, even if \csixtyHplus were present at the same abundance as \csixty, we would expect its features to be at least an order of magnitude weaker. 

\subsection{A search for ``planar \texorpdfstring{C$_{24}$}{C24}"}\label{sec:c24}

Finally, we searched for spectroscopic evidence for ``planar C$_{24}$", a species that was suggested to be the carrier of the 6.6 \um band in the spectra of fullerene containing objects \citep{Garcia-Hernandez:MC, 2020FNCN...28..637S}. We should note that the term ``planar C$_{24}$" is ambiguous, as there are many isomers with 24 carbon atoms in a planar structure. For example, the NASA Ames PAH IR Spectral Database lists 12 such isomers. \citet{Garcia-Hernandez:MC} specifically considered the $D_{6h}$ symmetric 7-ring cluster that could be described as fully dehydrogenated coronene, and we only consider that species here as well. Following \citet{Manna2016:C20:C24}, we refer to this species as \textit{dehydrocoronene}. 

We first downloaded the frequencies and IR intensities for dehydrocorone from the NASA AMES PAH IR Spectral Database \citep{PAHdb, PAHdbv2, Bauschlicher_2018, Mattioda_2020, Ricca_2025} and created a simulated fluorescence spectrum as described above. Dehydrocoronene has its strongest feature at 6.63 \um with weaker features at 9.68 \um, 22.57 \um, and 25.56 \um \citep{charlie:dehydrogenated}; in our simulated fluorescence model, these features each carry about 30\% of the intensity of the 6.6 \um band. Given the intensity of the 6.6 \um feature in the Tc~1 spectrum, we would therefore expect to see these other features clearly in our data as well. There is at best a very weak feature at $\sim$9.6 \um---far too weak given the relative band strengths---and there is no hint of a feature near 22.5 or 25.5 \um. We must therefore conclude that there is no dehydrocoronene in the spectrum of Tc~1, and that the 6.6 \um feature is likely a \csixty combination band (see Section \ref{Sec:C60_spectral_characteristics}).

\section{The spatial distribution of the fullerene emission} \label{sect:maps}
\begin{figure*}
      \centering

\resizebox{\hsize}{!}{
    \includegraphics[trim={0.6cm 1.5cm 0cm 1.5cm},clip]{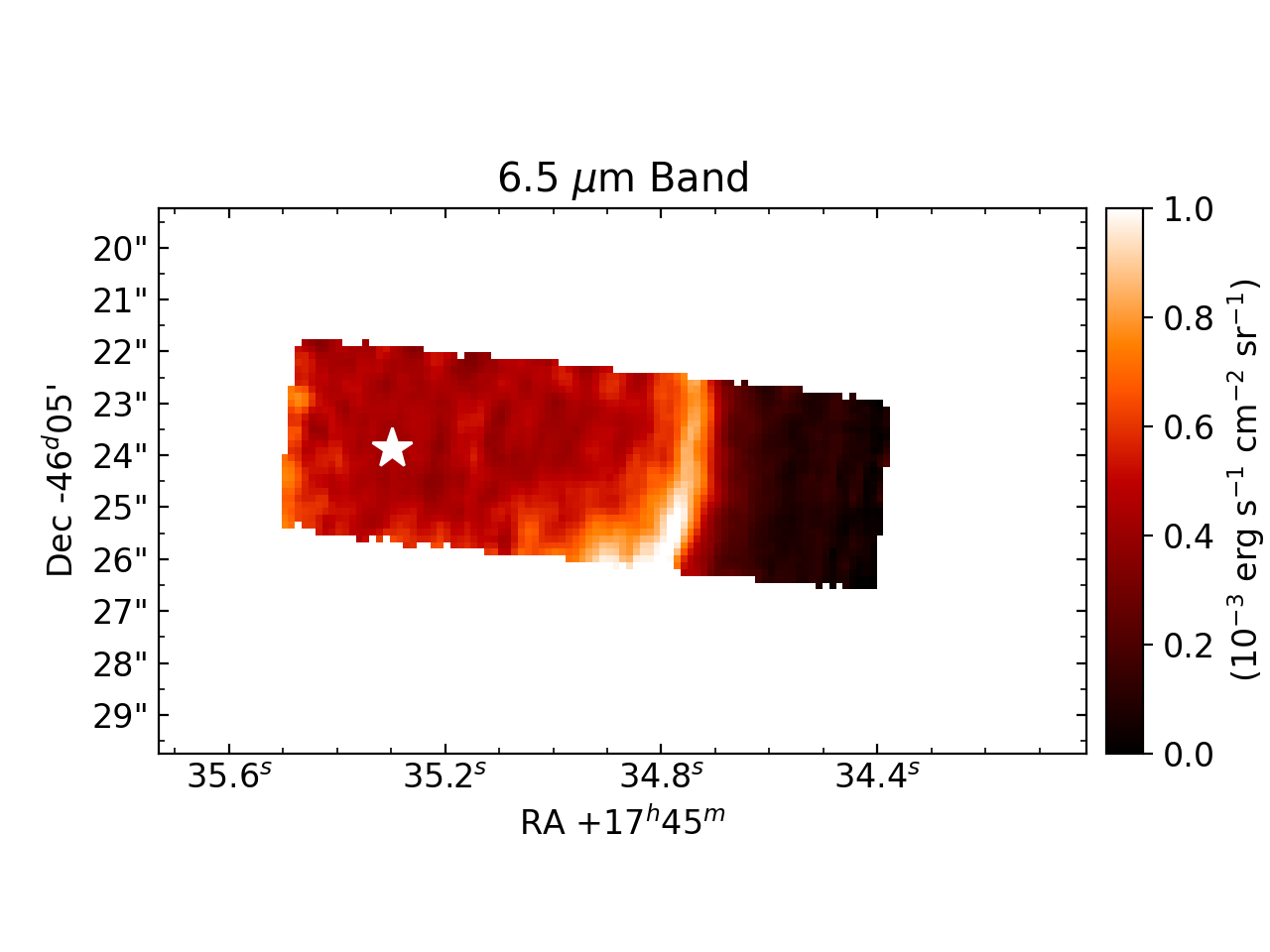}
    \includegraphics[trim={0.6cm 1.5cm 0cm 1.5cm},clip]{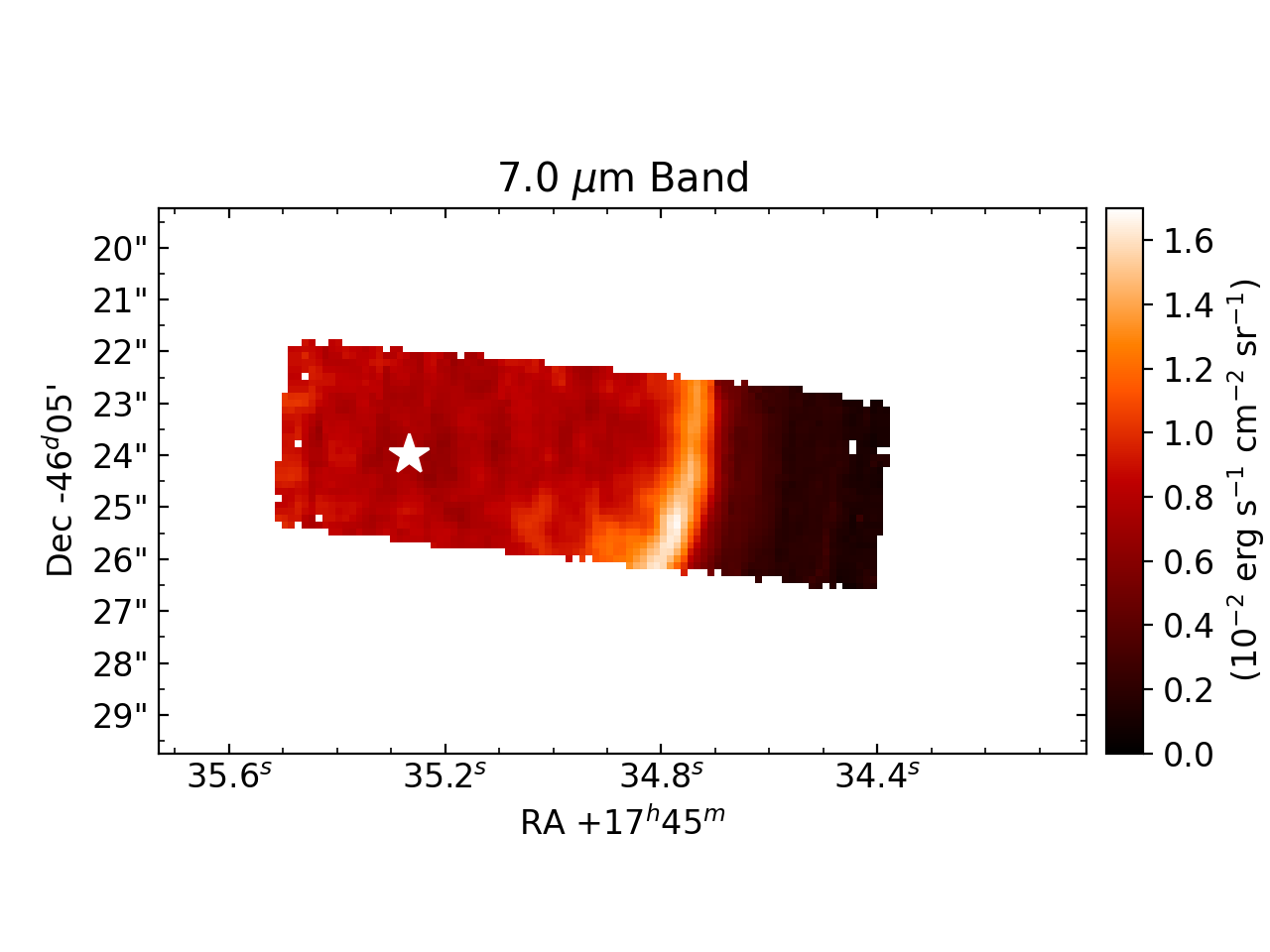}
    \includegraphics[trim={0.6cm 1.5cm 0cm 1.5cm},clip]{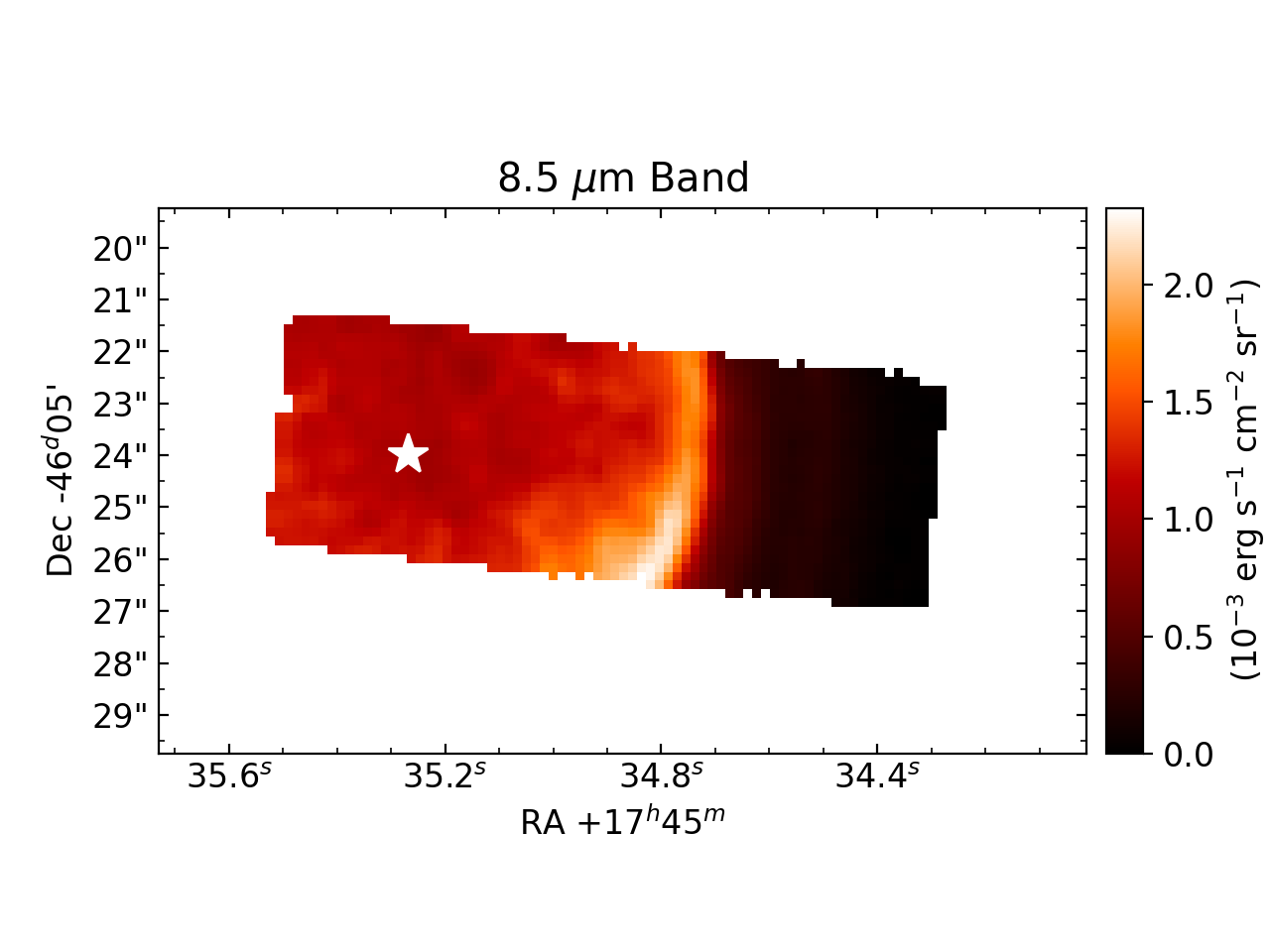}
    }
\resizebox{\hsize}{!}{
    \hspace{.15\linewidth}
    \includegraphics[trim={0.6cm 1.5cm 0cm 1.5cm},clip,width=0.3\linewidth]{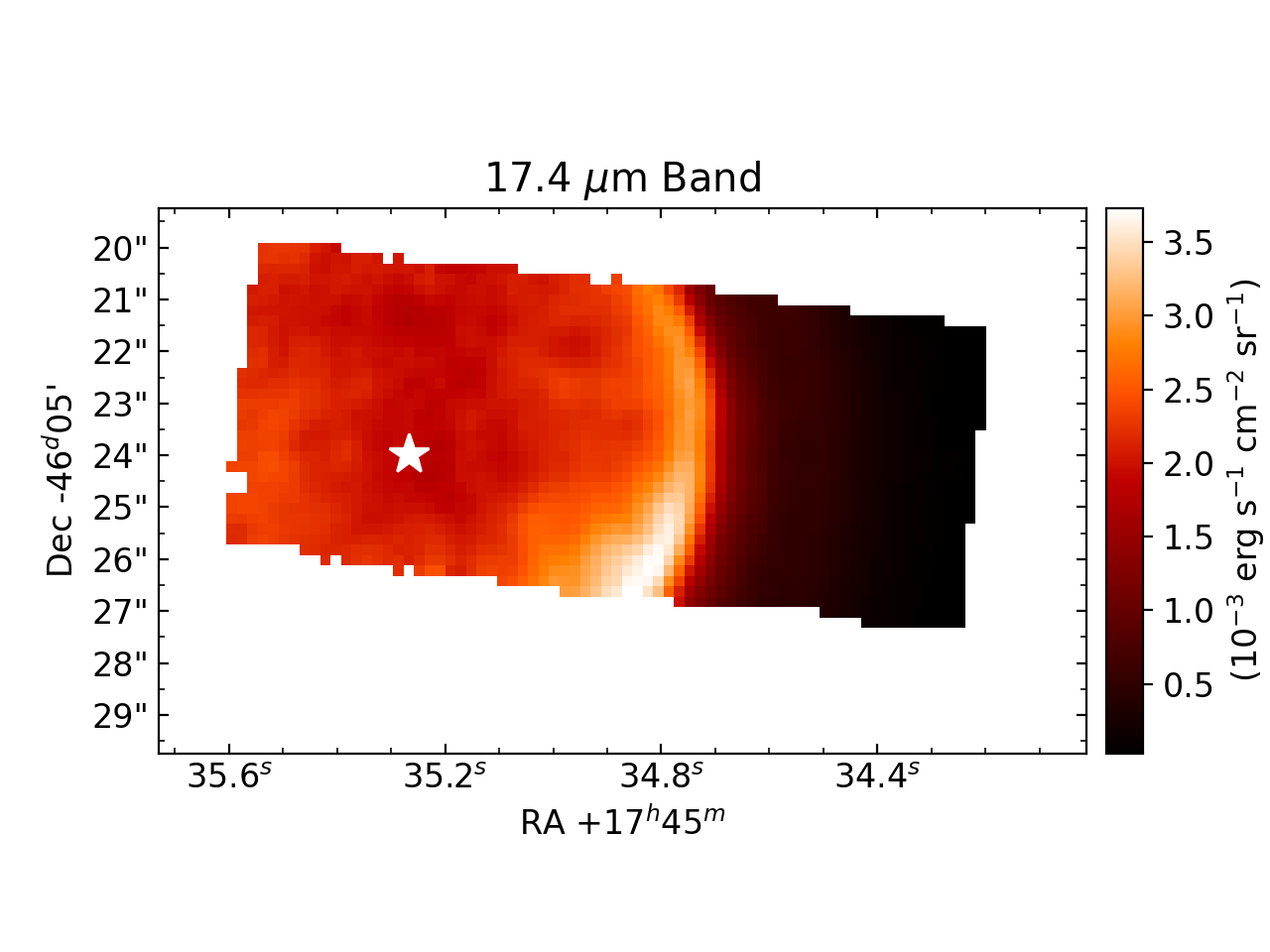}
    \includegraphics[trim={0.6cm 1.5cm 0cm 1.5cm},clip,width=0.3\linewidth]{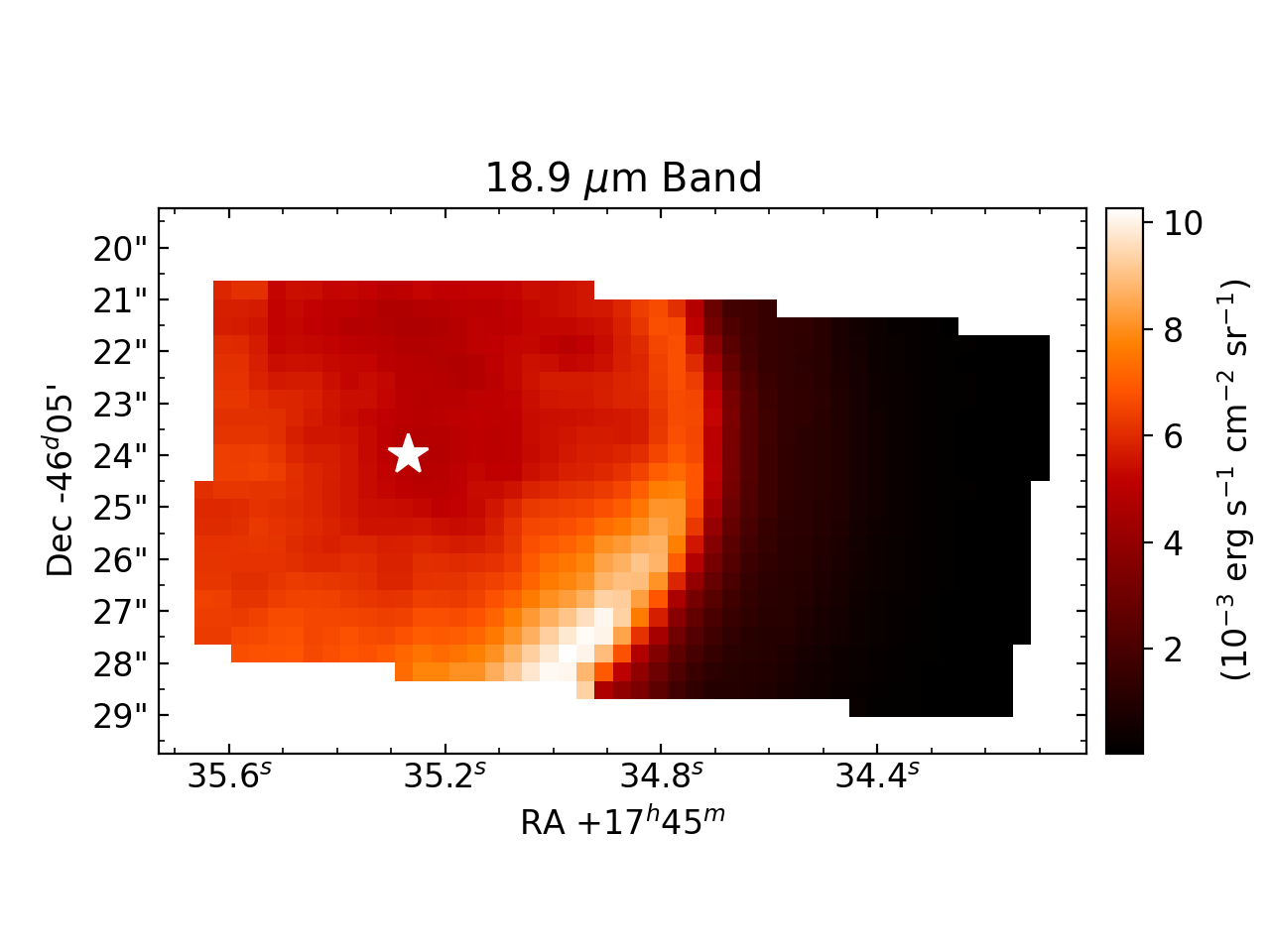}
    \hspace{.15\linewidth}
    }    
    \caption{Integrated surface brightness maps of the \csixty combination band at 6.5 \um and the five blended \csixty/\cseventy bands (see Section \ref{sec:discussion}). The location of the central star is denoted by a white star.}
    \label{fig:c60_individual_maps}
\end{figure*}

\begin{figure*}
      \centering
\resizebox{\hsize}{!}{
    \includegraphics[trim={0.6cm 1.5cm 0cm 1.5cm},clip]{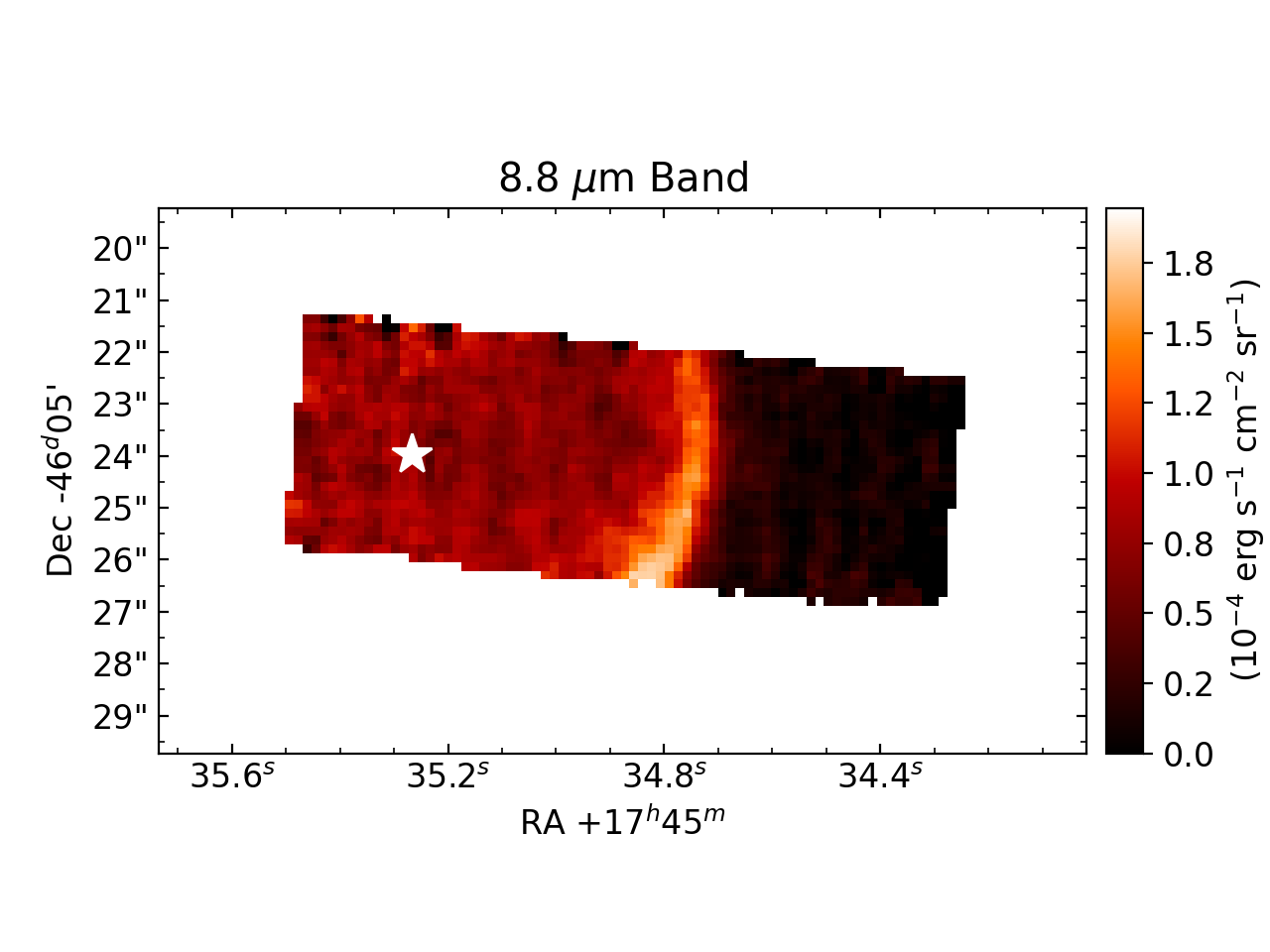}
    \includegraphics[trim={0.6cm 1.5cm 0cm 1.5cm},clip]{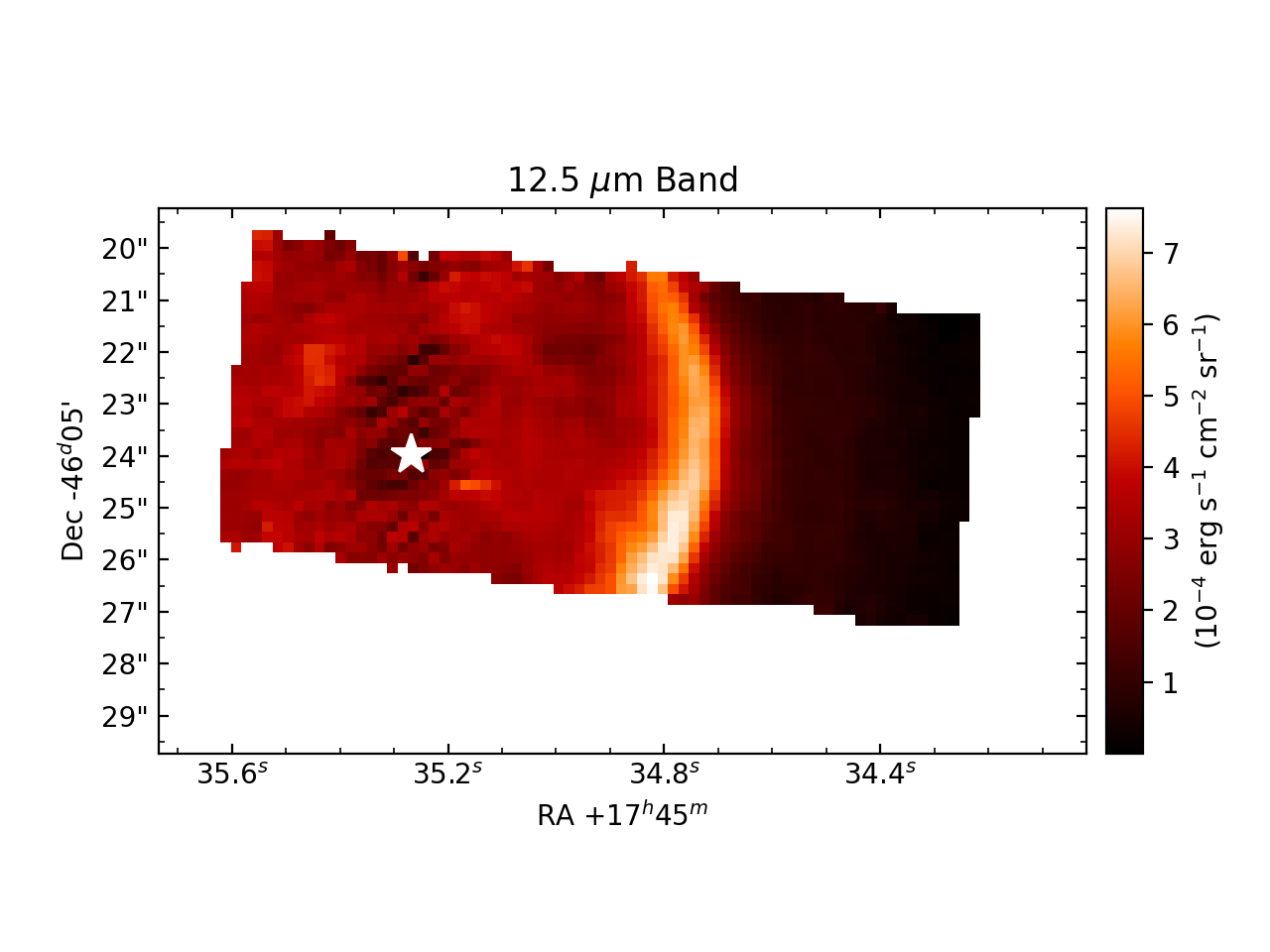}
    \includegraphics[trim={0.6cm 1.5cm 0cm 1.5cm},clip]{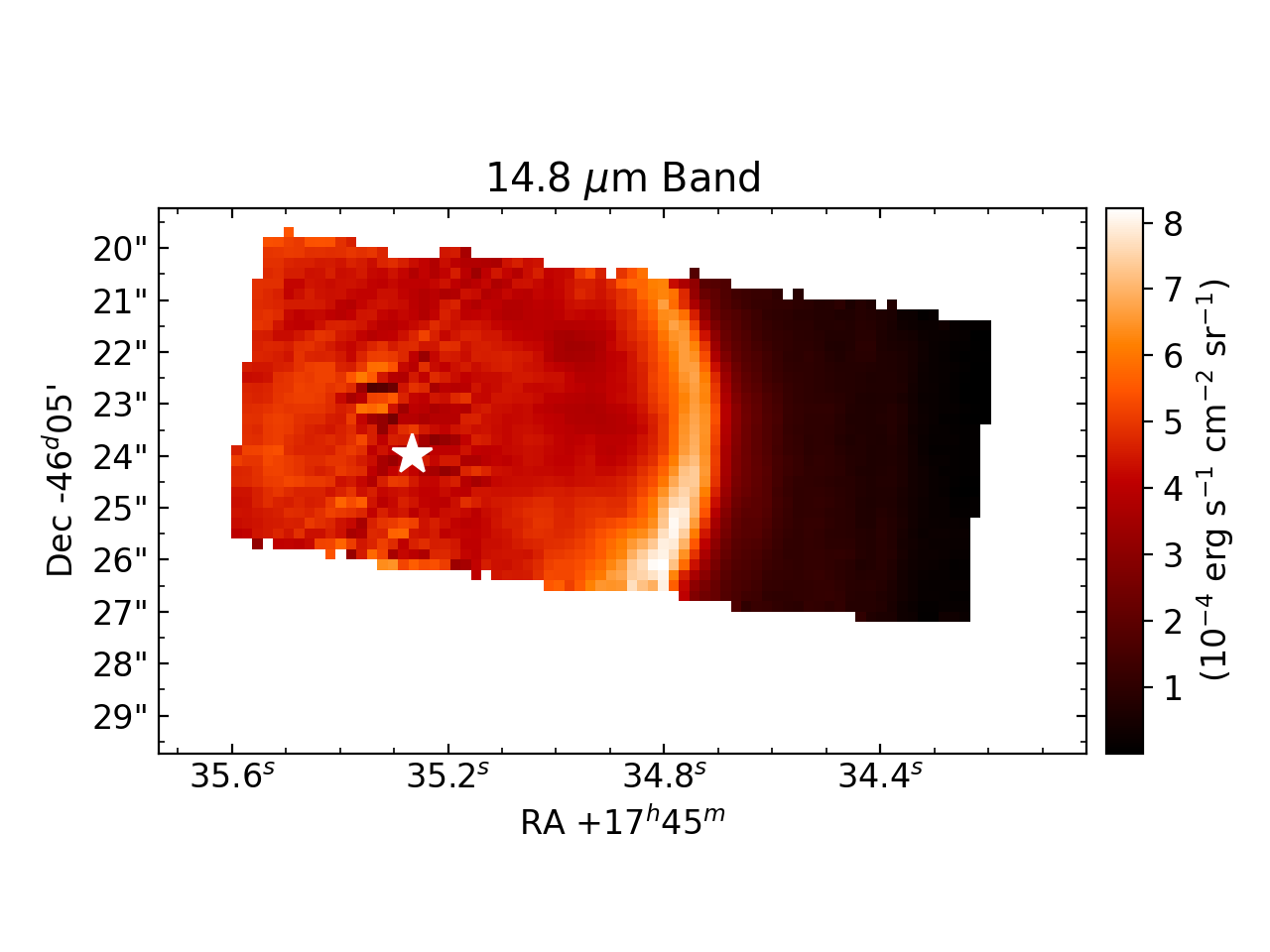}
    }
\resizebox{\hsize}{!}{
    \includegraphics[trim={0.6cm 1.5cm 0cm 1.5cm},clip]{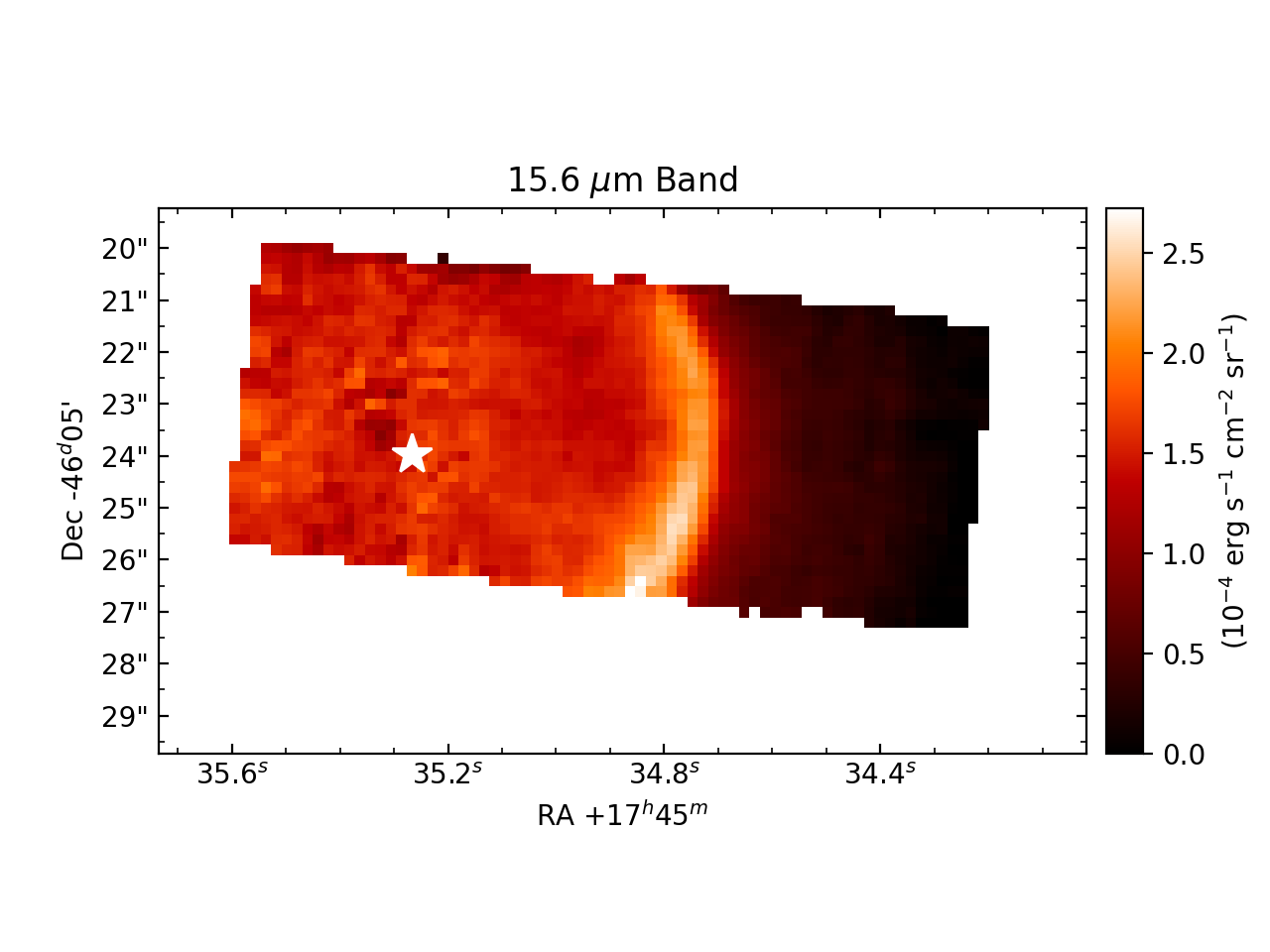}
    \includegraphics[trim={0.6cm 1.5cm 0cm 1.5cm},clip]{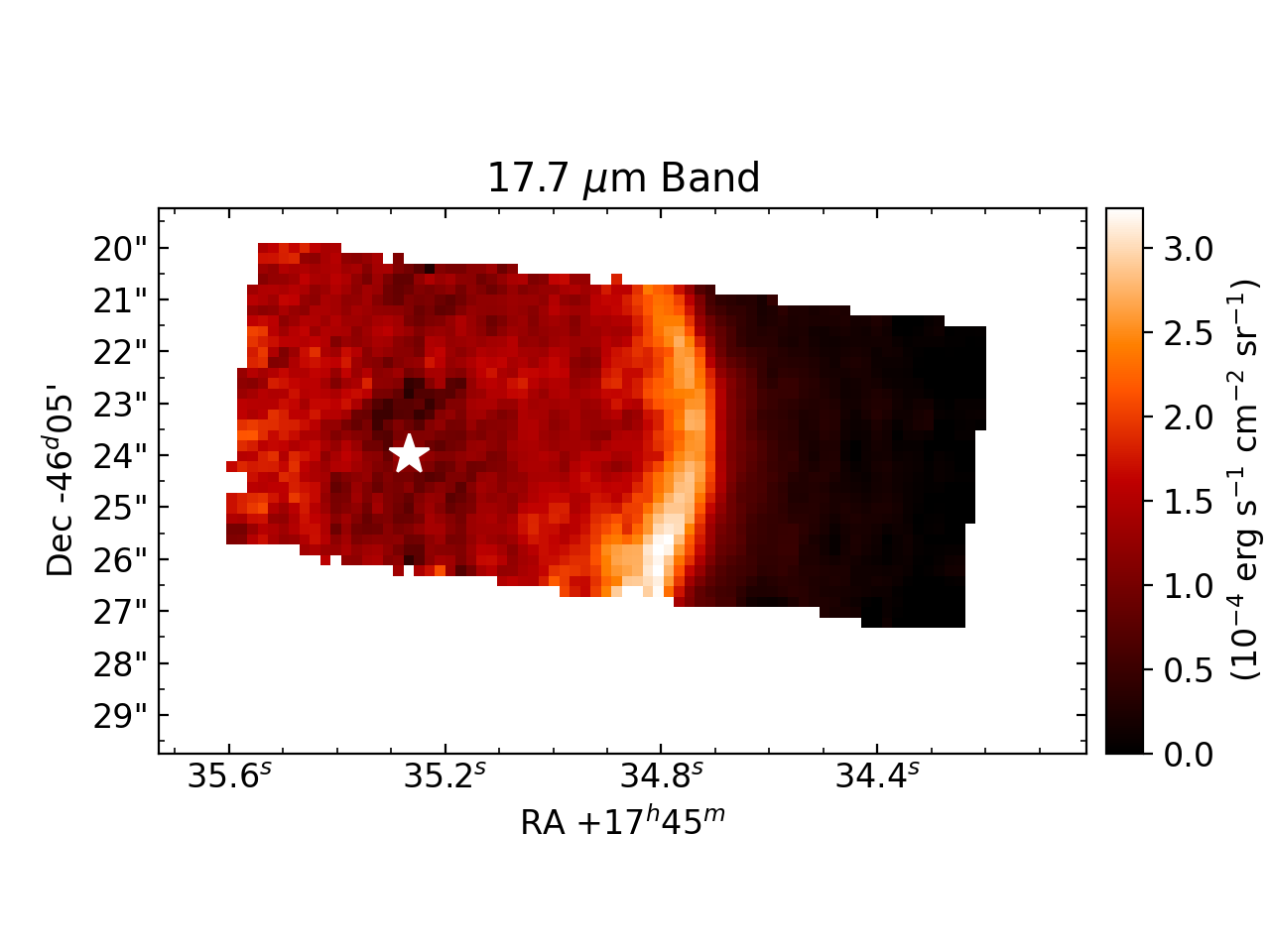}
    \includegraphics[trim={0.6cm 1.5cm 0cm 1.5cm},clip]{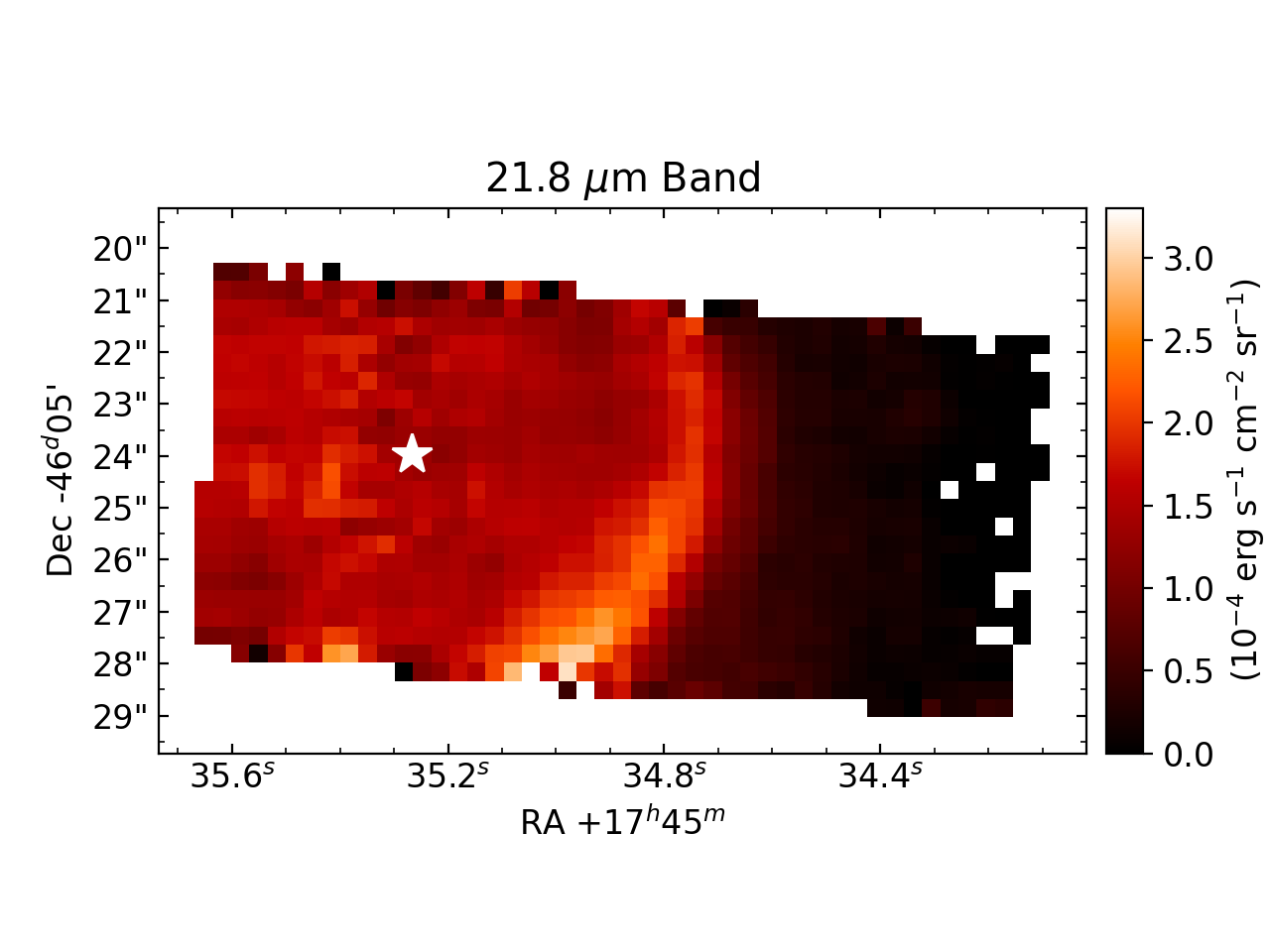}
    } 
    \caption{Integrated surface brightness maps of each of the six \cseventy features. The location of the central star is denoted by a white star.}
    \label{fig:c70_individual_maps}
    
\end{figure*}

We now turn our attention to the spatial distribution of the fullerene emission. 

\subsection{Maps of the fullerene features}

We first created maps to show the spatial distribution of each of the fullerene features. To do so, we determined the integrated surface brightness of the features at each pixel, using the results of our spectral decomposition (see Section~\ref{Sec:C60_spectral_characteristics} and Appendix~\ref{App:fitting}) to exclude any blended atomic lines and continuum contributions from the spectra. To create spatial maps at the highest possible resolution, we used the original (i.e., non-regridded) data cubes for this exercise. We masked pixels at the edges of the field of view due to their higher noise levels. 

Figure~\ref{fig:c60_individual_maps} shows the integrated surface brightness maps for the features corresponding to the fundamental \csixty modes as well as the 6.5 \um \csixty combination band. All maps show a similar morphology: a roughly spherical distribution, with the peak of the emission for all bands occurring in an asymmetric ring-like structure near the edge of the central core region. The southern part of this ring appears slightly brighter than the northern side. Figure \ref{fig:c70_individual_maps} shows that the six isolated \cseventy bands follow the same spatial distribution. We note though that the emission from these bands is considerably weaker than that of the \csixty fundamental bands, resulting in a lower signal-to-noise ratio and varying fit quality.

\subsection{Radial structure of the fullerene emission}

The spatial maps above suggest that the \csixty and \cseventy emission originates from a geometrically thin spherical shell; the ring appearance is then the consequence of limb brightening. To further evaluate this, it is insightful to study the radial profiles. 

\begin{figure}
\centering
\resizebox{\hsize}{!}{
    \includegraphics[clip, trim= 0cm 0cm 1cm 1cm]{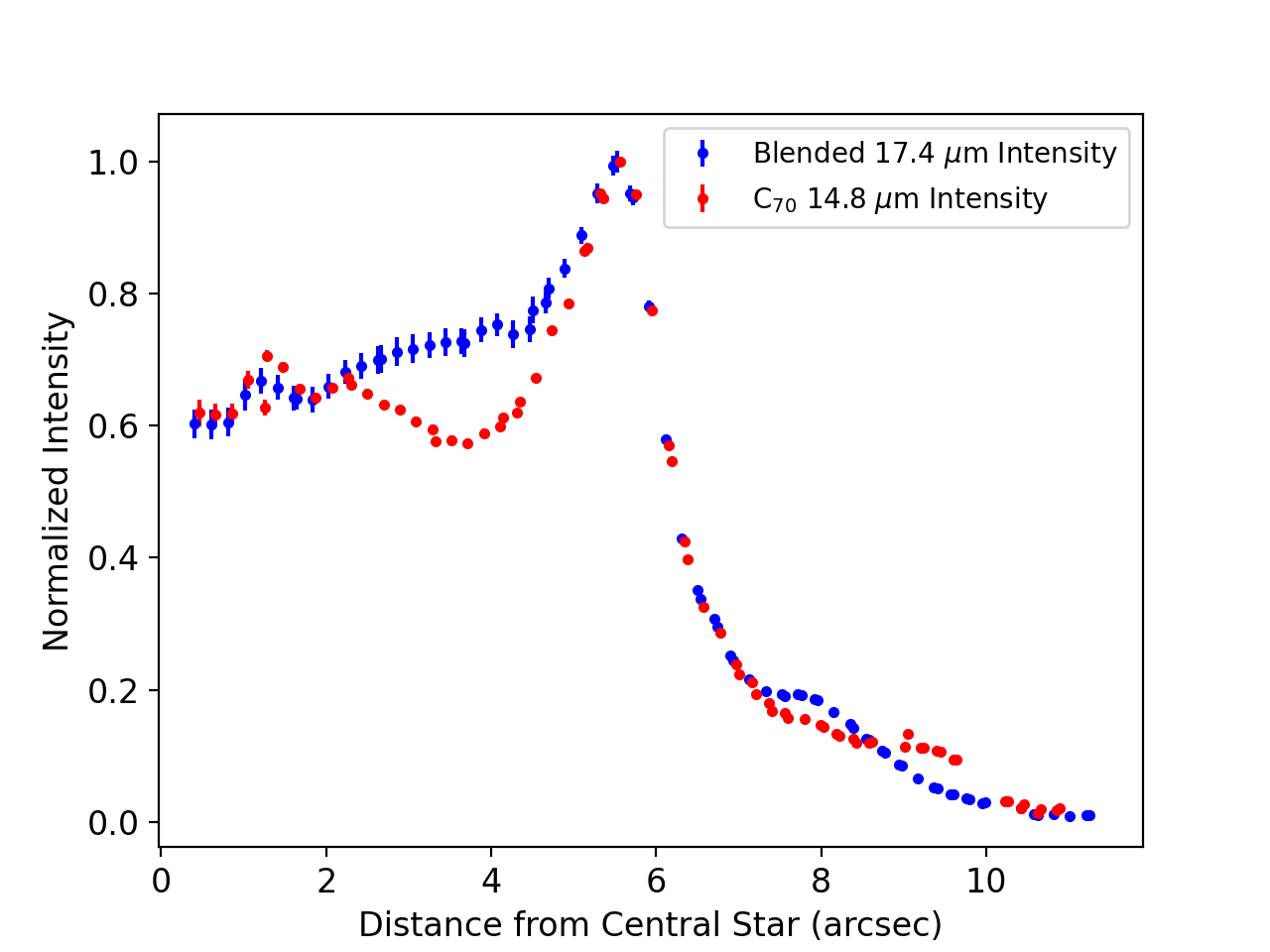}}
\caption{Normalized radial intensity profiles of the blended 17.4 \um band and the 14.8 \um isolated \cseventy band as a function of distance from the central star. Individual points describe an averaged value over pixels along the 275\degree~azimuth (toward the northern edge of the ring with less visible substructure).}
\label{fig:radial_profile}
\end{figure}

A comparison of the radial profiles of the 17.4~\um \csixty/\cseventy band and the 14.8~\um isolated \cseventy band (Figure~\ref{fig:radial_profile}) shows that profiles peak at the same radial distance from the star ($\sim$5.3\arcsec), in the primary ring. A second bump in the 17.4 \um profile occurs at a distance of $\sim$7–9\arcsec\ from the star; for the isolated \cseventy band, there is no second bump at that location, but instead a bit further out, $\sim$8–10\arcsec. 

\begin{figure}
    \centering
    \resizebox{\hsize}{!}{
    \includegraphics{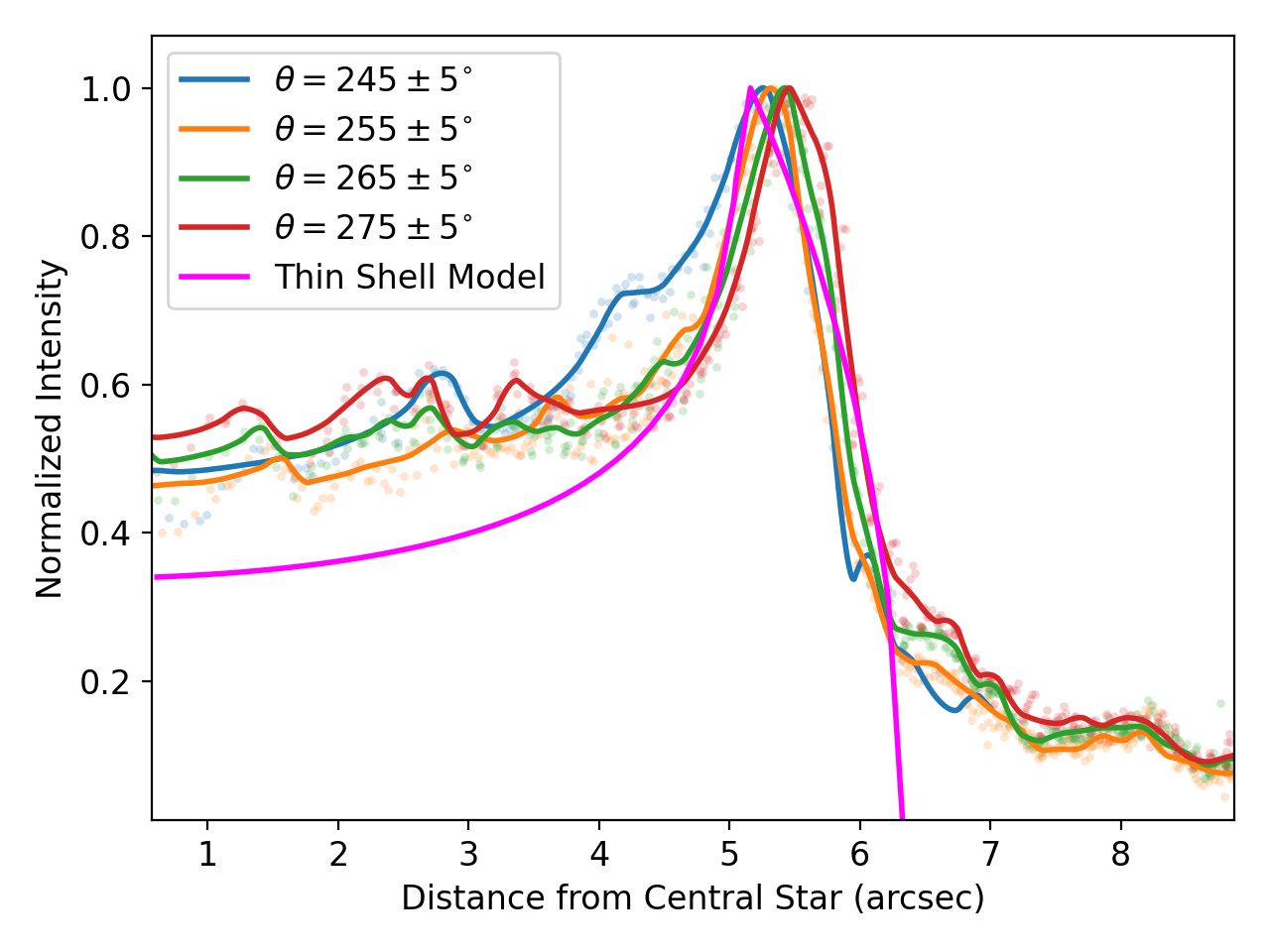}}
    \caption{Integrated surface brightness of the 7.0 \um fullerene feature as a function of distance from the central star. The first four lines denote the fit models describing the intensity at different position angles relative to the central star. 245\degree ~and 275\degree ~denotes the top and bottom of the ring in the spatial maps, respectively. The magenta line describes a fit model that would correspond to all fullerene intensity originating from a thin spherical shell surrounding the central core region.}
    \label{fig:seven_scatter}
    
\end{figure}

We further studied the radial and azimuthal distribution of the fullerenes using the 7.0~\um band. The dots in Figure~\ref{fig:seven_scatter} show the surface brightness at four different position angles (245\degree, 255\degree, 265\degree, and 275\degree), averaged over $\pm5\degree$ in azimuth. The magenta curve represents the expected radial profile for a hollow spherical shell that is optically thin; we find that the best fit is a model with an inner radius of $\sim$5.0\arcsec\ and an outer radius of $\sim$6.4\arcsec. While the observed radial distribution looks similar to that of a hollow spherical shell, the observations show excess emission both inside and outside this shell. This suggests that the shell is not completely hollow, and embedded in halo material that exhibits fullerene emission as well. The other curves in Figure~\ref{fig:seven_scatter} represent models for spherical shells as well, but this time with an internal and external density distribution derived from the averaged surface brightness measurements at the four different position angles. The derived profiles show clear density variations (and thus also variations in the projected intensity) both inside and outside the shell, and reach their peak emission at different distances from the central star---indicative of a slight asymmetry of the fullerene emission. All derived radial profiles once more show the secondary ring as a bump at $\sim$8\arcsec.

\section{Fullerene band ratios}
\label{Sec:bandratios}

\begin{table*}
  \centering
  
    \caption{\label{tab:ratios}Observed Integrated Surface Brightness \csixty Ratios}

    \begin{tabular}{c|c|c|c|c}
        \hline
        Aperture & $6.5 ~/ ~18.9$ & $7.0 ~/ ~18.9$ & $8.5 ~/ ~18.9$ & $17.4 ~/ ~18.9$\\
        \hline
        Star    & 0.099$\pm$0.004 & 1.58$\pm$0.09 & 0.227$\pm$0.003 & 0.415$\pm$0.005\\
        Core Region & 0.097$\pm$0.002 & 1.53$\pm$0.08 & 0.220$\pm$0.002 & 0.421$\pm$0.003\\
        Fullerene Ring    & 0.118$\pm$0.002 & 1.89$\pm$0.05 & 0.256$\pm$0.002 & 0.437$\pm$0.003\\
        Extended Halo    & 0.126$\pm$0.010 &  1.77$\pm$0.07 & 0.224$\pm$0.004 & 0.433$\pm$0.008\\
        \ArII Peak & 0.094$\pm$0.002 & 1.35$\pm$0.08 & 0.214$\pm$0.002 & 0.426$\pm$0.003\\
        \hline
        Previous Literature$^a$ & --- & 1.31$\pm$0.28 & 0.29$\pm$0.02 & 0.47$\pm$0.06\\
    \end{tabular}
    \begin{tablenotes}
\item $^a$ Taken from Table 1 of \cite{Bernard-Salas2012}
\end{tablenotes}

\end{table*}

\begin{table*}
  \centering
  
    \caption{\label{tab:c70_ratios}Observed Integrated Surface Brightness \cseventy Ratios}

    \begin{tabular}{c|c|c|c|c|c}
        \hline
        Aperture & $8.8 ~/ ~21.8$ & $12.5 ~/ ~21.8$ & $14.8 ~/ ~21.8$ & $15.6 ~/ ~21.8$ & $17.7 ~/ ~21.8$\\
        \hline
        Star           & 0.63$\pm$0.08 & 2.9$\pm$0.2    & 2.5$\pm$0.3 & 0.80$\pm$0.09 & 0.50$\pm$0.04\\
        Core Region    & 0.49$\pm$0.03 & 1.80$\pm$0.09 & 3.1$\pm$0.3 & 0.83$\pm$0.04 & 0.51$\pm$0.03\\
        Fullerene Ring & 0.76$\pm$0.02 & 2.52$\pm$0.07 & 3.7$\pm$0.2 & 0.97$\pm$0.03 & 0.53$\pm$0.02\\
        Extended Halo  & 0.15$\pm$0.05 & 2.26$\pm$0.3  & 2.7$\pm$0.5 & 1.3$\pm$0.2   & 0.28$\pm$0.05\\
        \ArII Peak     & 0.50$\pm$0.03 & 1.74$\pm$0.10 & 3.7$\pm$0.3 & 0.87$\pm$0.05 & 0.60$\pm$0.04\\
    \end{tabular}

\end{table*}

\begin{figure*}
     \centering
\resizebox{\hsize}{!}{
   \includegraphics[trim={0.6cm 1.5cm 0cm 1.5cm},clip]{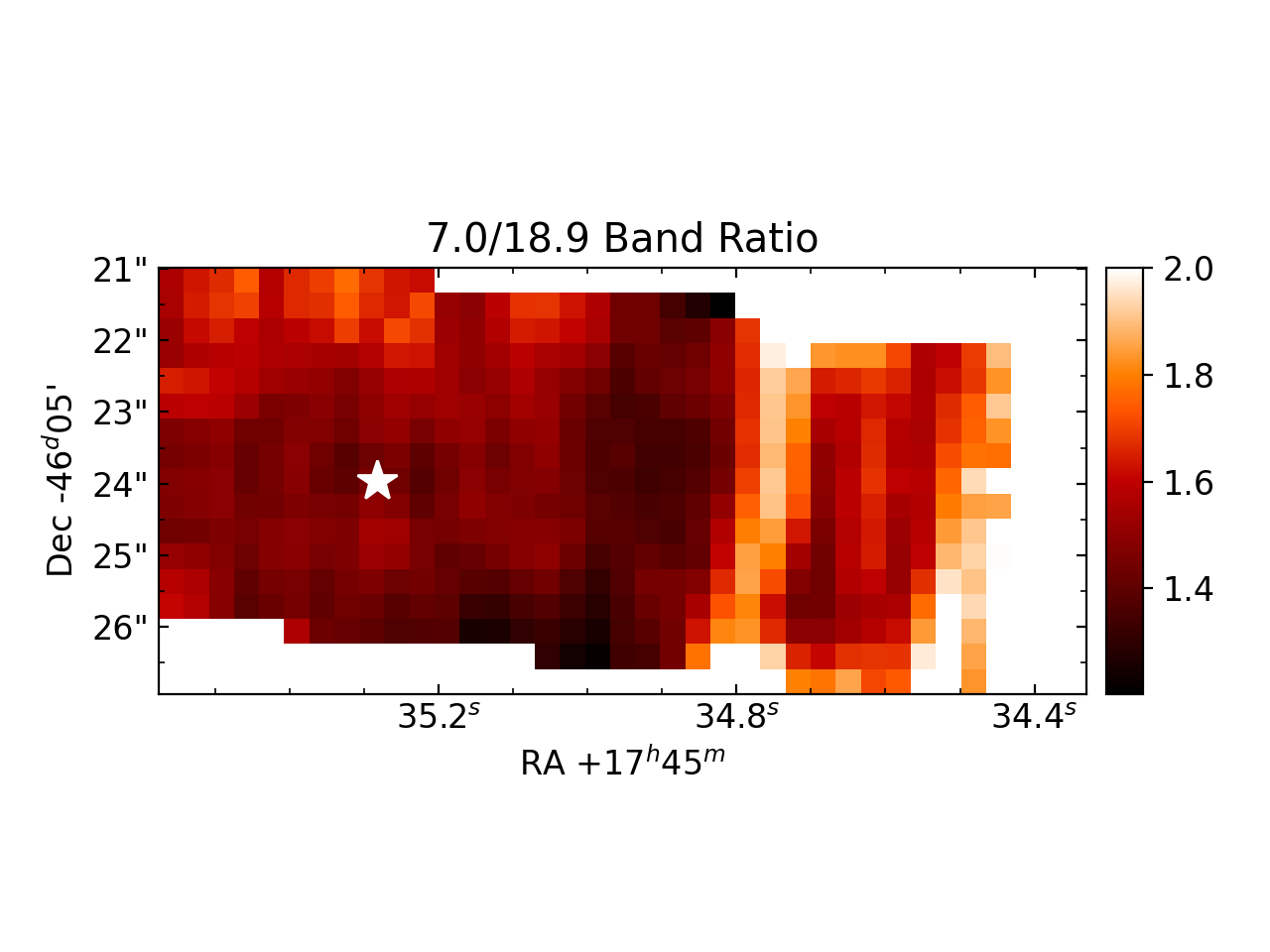}
   \includegraphics[trim={0.6cm 1.5cm 0cm 1.5cm},clip]{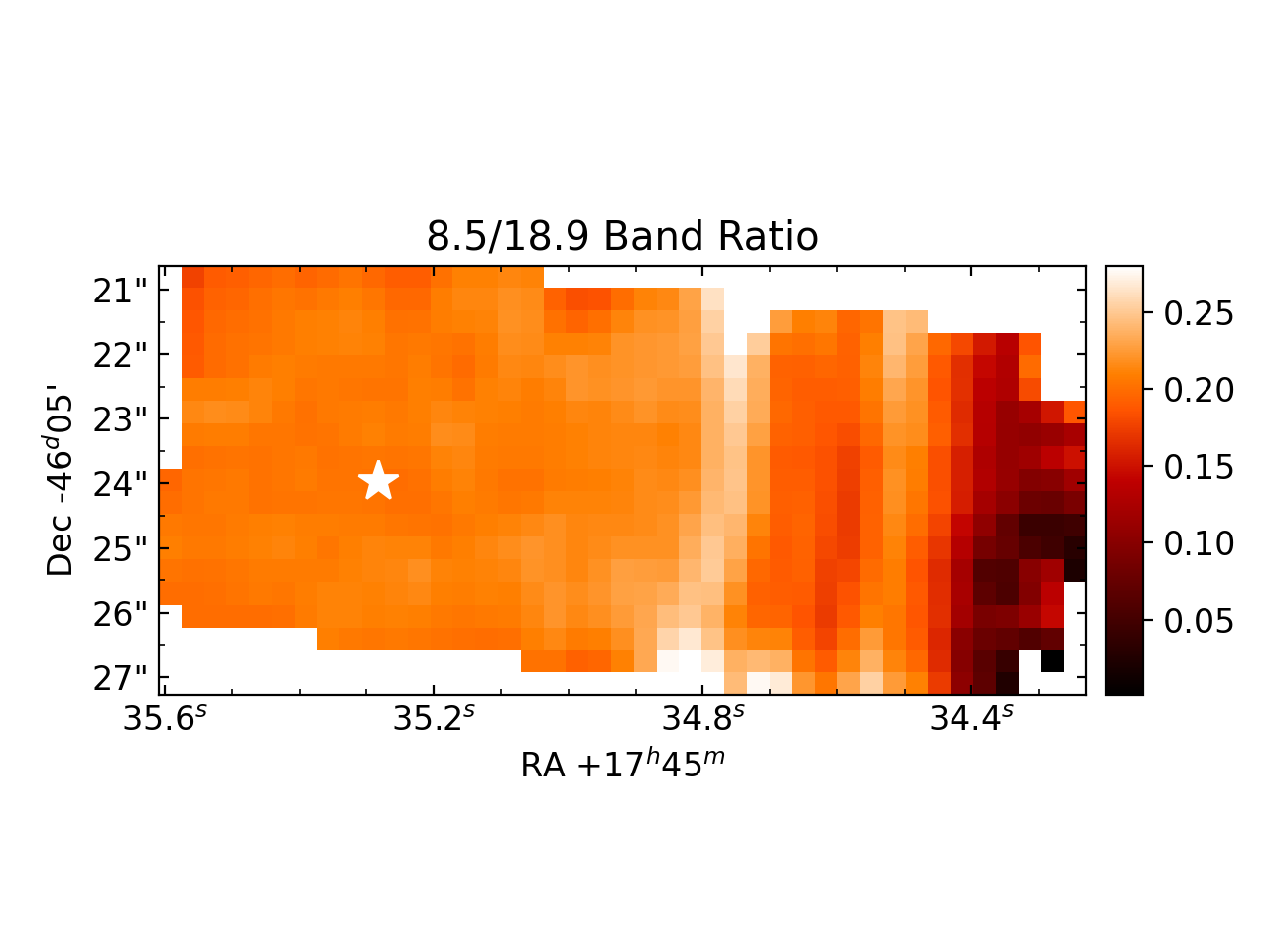}
   \includegraphics[trim={0.6cm 1.5cm 0cm 1.5cm},clip]{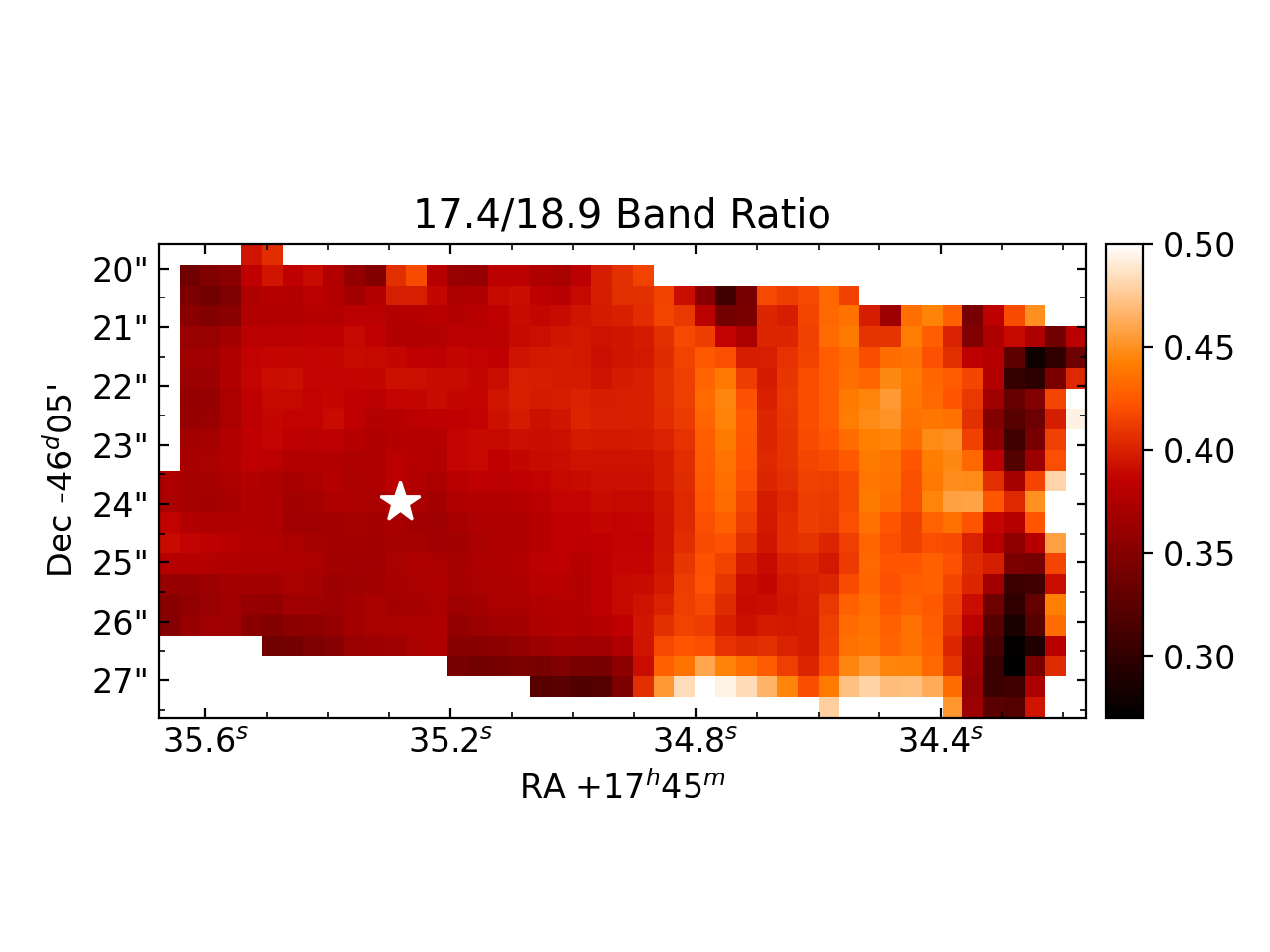}
    }
    \caption{Integrated surface brightness maps showing the observational band ratios of the 7.0, 8.5, and 17.4 \um features against the 18.9 \um feature. Each map's spatial pixels correspond to the regridding of the individual subchannel cubes onto the longest wavelength channel's spatial coordinates. The location of the central star is denoted by a white star.}
    \label{fig:ratio}
    
\end{figure*}

The relative intensities of the \csixty and \cseventy bands provide important constraints on the excitation mechanism, as different models predict different band ratios \citep{Bernard-Salas2012,Brieva:C60ratios}. Previous observations were hampered by the low spectral resolution of Spitzer, which did not allow to cleanly separate the contribution of forbidden emission lines from some of the \csixty bands. In Section~\ref{Sec:spectral_characteristics}, we decomposed the spectra at each pixel into contributions from continuum, atomic lines and fullerene emission; we can thus determine the clean fullerene emission for the first time, determine accurate ratios, and study if the ratios change spatially. Band ratios for \csixty are typically normalized to the emission in the 18.9 \um band; the resulting ratios for the five apertures are listed in Table~\ref{tab:ratios}. Spatial maps of the band ratios are shown in Figure \ref{fig:ratio}.

All of our measured band ratios, apart from the 7.0/18.9 ratio, are in rough agreement with previously reported values for Tc 1 \citep{Bernard-Salas2012}. Our 7.0/18.9 ratio is 1.4--1.9 across our field of view, higher than the value of 1.31$\pm$0.28 reported by \cite{Bernard-Salas2012}. This is perhaps the best example of the consequence of the higher spectral resolution of our MIRI observations, as it allows us to cleanly separate the fullerene emission from the contaminating \ArII line at 6.99 \um. \cite{Bernard-Salas2012} estimated the \ArII line to contribute approximately 15\% to the observed flux at 7 \um.  We do indeed find such a ratio between the \ArII emission to the 7.0 \um fullerene emission, but only in the ``\ArII Peak" aperture that samples the low-ionization structure found within the core (see \overviewpaper). Interestingly, our ratio value within this aperture is very similar to that of \cite{Bernard-Salas2012} at 1.35. Outside of this structure, the \ArII emission contributes only $\sim$4\% of the 7.0 \um fullerene emission.

Figure~\ref{fig:ratio} shows how these ratios vary across the nebula. Just as in the apertures, all ratios are elevated within the fullerene ring, and all three maps also show increased ratios in the extended halo compared to the core. The 8.5/18.9 ratio is fairly constant throughout the core, peaks in the ring, drops just outside it, and increases again in a secondary ring structure at $\sim$8\arcsec.

Interpreting these band ratios is complicated by several factors. One source of uncertainty is that different laboratory measurements yield different intrinsic relative band strengths for \csixty, complicating any direct comparison between observed and predicted ratios \citep{Bernard-Salas2012}. Moreover, there is an additional crucial consideration when analyzing these bands and their ratios: the contribution of \cseventy. \cseventy has strong features at 6.99 \um, 7.07 \um, 17.30 \um, and 18.73 \um (along with a weaker feature at 8.50 \um) that are heavily blended with the \csixty bands and will contribute to the observed emission \citep{Stratmann1998}. As such, the band ratios reported above will differ from the actual band ratios of \csixty. In the case of the 7.0 \um band particularly, the reported value above will drastically over-predict the actual ratio.

There is some difficulty in accurately estimating the \cseventy contributions to the \csixty bands which stems from a large disagreement between laboratory and theoretical measurements of the intrinsic strengths of the \cseventy bands \citep[see e.g.][]{vonCzarnowski1995321, Stratmann1998, Schettino2002, Sun:Kertesz:2002, Iglesias-Groth2011, Candian:cages}. While all agree that the \cseventy feature at 6.99 \um is the strongest, the relative intensities of all other bands change drastically depending on the study. As such, we only make a rough comparison here. Following the procedure described in J. Cami et al. 2026a, submitted, we estimate that the relative contribution from the \cseventy bands at 17.30 and 18.73 \um will be $\sim$2.5 times greater than the related isolated \cseventy features at 17.7 and 21.8 \um (respectively). Subtracting this contribution from our total 17.4 and 18.9 \um values and calculating a new band ratio for the ``Fullerene Ring" aperture results in a new 17.4/18.9 \csixty ratio of $\sim$0.35$\pm$0.01---slightly lower than the initially-calculated value of 0.44 and bringing the value further away from that reported in \cite{Bernard-Salas2012}. This greater difference is expected, however, as the previously-reported value would also have unaccounted-for contributions from \cseventy. As the 8.5 \um band has minimal \cseventy contamination and the \cseventy contribution to the overall 18.9 \um band is $\lesssim$~10\%, the 8.5/18.9 ratio is the best tracer of changes in the \csixty emission alone. The implications of these fullerene band ratios are discussed in Section~\ref{sec:discussion}.

\section{Discussion}\label{sec:discussion}

\subsection{Fullerene Abundances}\label{sec:abundances}

The abundance of large carbonaceous molecules such as PAHs and fullerenes can be estimated by comparing the total energy they emit at infrared wavelengths to the total far-infrared dust continuum emission \citep{Tielens:book}. The underlying assumption is that all carbonaceous species---whether in dust grains, PAHs, or fullerenes---compete for the same far-ultraviolet (FUV) photons. At steady state, the energy absorbed in the FUV is re-emitted in the infrared, so the fraction of IR emission contributed by a given species reflects the fraction of FUV photons it absorbs, which in turn depends on its abundance and FUV absorption cross section. 

Following \citet{Tielens:book}, the fraction of carbon locked in a species is then given by
\begin{equation}
    f_C = 0.23 \left(\frac{7 \times 10^{-18} \text{ cm}^2}{\sigma_{\text{FUV}}}\right) \frac{f_{\text{IR}}}{1 - f_{\text{IR}}},
\end{equation}

\noindent where 
\begin{equation}
    f_{IR} = \frac{I_{\rm s}}{I_{\rm s} + I_{FIR}}
\end{equation}

\noindent is the ratio of the emission from that species $I_{\rm s}$ to the total far-infrared emission $I_{\rm FIR}$, and $\sigma_{\rm FUV}$ is its intrinsic FUV absorption cross section per carbon atom. The numerical factor of 0.23 incorporates the ratio of FUV to visual dust absorption, the dust albedo in the FUV, and the elemental carbon abundance \citep[see][for details]{Tielens:book}. We adopt $\sigma_{\rm FUV}=7\times10^{-18}$~cm$^2$ per C atom; this value is appropriate for PAHs, but should also be a good estimate for fullerenes \citep{BerneTielens2012}.

We then need to determine $f_{\rm IR}$. The total FIR emission for Tc~1 is $I_{\rm FIR} = 2.65\times10^{-12}$ W/m$^2$; we determined this value by integrating a composite spectrum combining Spitzer-IRS and Herschel observations. This represents the emission for the entire nebula, albeit with some uncertainties due to the different fields of view by the instruments involved.  

While our MIRI/MRS observations provide accurate fullerene intensities, they only cover a small segment of the nebula. To obtain an estimate of the total fullerene emission over the entire nebulae, we employed a patchwork technique. For the inner region (within 1.6\arcsec\ of the central star), we have complete spatial coverage; summing the fullerene emission in this region ($I_{\rm circle}$) is thus a direct measurement of the total fullerene emission in this region. For larger distances from the central star, we defined the largest wedge that fits within MIRI's rectangular FOV, and found that this wedge spans position angles 254\degree--275\degree. We summed the total fullerene emission in this wedge ($I_{\rm wedge}$), and assuming azimuthal symmetry, we can then estimate the total fullerene emission for the entire nebula as 
\begin{equation}
I_{\text{Total}} = I_{\text{circle}} + \frac{360\degree}{\Delta\theta} I_{\text{wedge}}.
\end{equation}
where $\Delta\theta=21$\degree is the angular width of the wedge. This estimate for the total fullerene emission ignores contributions further out in the halo than our MRS observation covers. While the fullerene emission greatly decreases beyond the \csixty ring (Figure~\ref{fig:radial_profile}), it is not zero, and the halo extends out quite far. Our estimate for the total fullerene emission is thus a lower limit to the total fullerene flux. 

With these assumptions and approximations, we find that the combined fullerene emission accounts for $\sim$2.6\% of the total IR emission, corresponding to $f_C \sim 5.9 \times 10^{-3}$, i.e. approximately 0.6\% of all carbon in Tc~1 is locked in fullerenes. This is in agreement with the 0.1--0.6\% range for \csixty found by \cite{Sellgren2010} and the 0.3--0.9\% range for \csixtyplus found by \cite{Foing:C60_1}, but it is much higher than similar estimates in various PDRs \citep{BerneTielens2012,Castellanos2014}.

\subsection{Constraints on the fullerene excitation mechanism}\label{sec:strength}

The excitation of the fullerene bands has mostly been discussed in terms of two
scenarios, each with clear problems. In the fluorescence scenario, gas-phase
fullerenes absorb UV photons and cool through infrared emission; however, this does
not account for the variations in band ratios observed between fullerene-rich
planetary nebulae \citep{Bernard-Salas2012}, nor for the weak 7.0~\um\ band in the
reflection nebula NGC~2023 \citep{Sellgren2010}. The alternative, thermal emission
from fullerenes attached to dust grains \citep{Cami2010}, fails to reproduce the
observed band ratios as well \citep{Bernard-Salas2012}. Shocks have also been
invoked, mostly in the context of fullerene formation \citep{Cami:APN_formation}.
Our observations allow us to discriminate between these possibilities. The resolved
rovibrational structure in the 17.4 and 18.9~\um\ bands establishes that the
emitting \csixty\ molecules are in the gas phase (J. Cami et al. 2026a, submitted),
ruling out the thermal scenario. Shocks are unlikely to play a significant role
either: diagnostic diagrams show that the gas in Tc~1 is photoionized and exclude
fast shocks, and no emission line shows a localized enhancement at the location of
the fullerene ring (\overviewpaper). Photo-excitation thus remains
the most likely mechanism. Our results do not resolve its problems, but they do
offer clues as to the physical processes at play.

First, the environment differs from that assumed in most models: the fullerenes in
Tc~1 reside in fully ionized gas, directly exposed to the stellar radiation field,
rather than in a photodissociation region. Second, the band ratios vary across the
nebula, and these variations do not follow the radiation field. Although the FUV
field declines by a factor of $\sim$40 from the horseshoe to the fullerene ring
(\overviewpaper), all ratios are elevated in the ring and in the
halo compared to the core (Section~\ref{Sec:bandratios}). The 8.5/18.9 ratio, which
is least affected by \cseventy, peaks in the ring, drops at the position of the
low-extinction annulus \citep{Walsh_MUSE2026}, and rises again in a secondary ring
at $\sim$8\arcsec. In the single-photon regime, fluorescent band ratios depend on
the energy of the absorbed photons rather than on the intensity of the radiation
field; the spatial structure in the ratios thus indicates that something in the
excitation changes across the nebula, although it is not yet clear what changes,
or why.

A quantitative interpretation of the band ratios is, however, still hampered by
two factors. Most \csixty\ bands are blended with \cseventy\ bands, whose intrinsic
strengths differ strongly between laboratory and theoretical studies
\citep[see e.g.][]{Stratmann1998, Iglesias-Groth2011, Candian:cages}, so that the
\csixty\ ratios corrected for this contribution (Section~\ref{Sec:bandratios})
remain uncertain. Moreover, current fluorescence models were developed for
photodissociation regions and do not include processes that may be important in
ionized gas, such as interactions with free electrons. These processes are explored
in more quantitative detail by Cami et al. (in prep.).

\subsection{The spatial distribution and origin of the fullerenes}\label{sec:formation}

The spatial distribution of the fullerene emission offers additional constraints on
the origin of the fullerenes in Tc~1. Several formation pathways have been proposed,
including formation in hydrogen-poor environments
\citep{Kroto1985, Kroto1988, deVries1993, Wang1995}, photochemical processing of PAHs
\citep{Cami2011, BerneTielens2012}, and the destruction of hydrogenated amorphous
carbon grains \citep{Scott1997, Garcia-Hernandez:PN}; we refer to \citet{Cami2011}
for a discussion of their difficulties.

Our maps show that \csixty\ and \cseventy\ follow the same spatial distribution
(Section~\ref{sect:maps}). Their emission peaks in a thin, asymmetric shell at
5--6\arcsec, at the outer edge of the core, but is also detected inside the core and
in the halo, with a hint of a secondary shell at $\sim$8\arcsec. Moreover, the
spectra of the ring and the halo are very similar, differing mainly in intensity
(\overviewpaper), with only modest differences in the band ratios
(Section~\ref{Sec:bandratios}).

The presence of fullerenes in the halo has important implications for their origin.
The halo results from an earlier mass-loss episode than the core (\overviewpaper), and its fullerenes are located well beyond the ring, in
low-density ionized gas. They must therefore have formed independently of the ring,
and have survived in this environment as well. The ring is thus not a formation site.
It does, however, mark a location where the conditions favor much stronger fullerene
emission. Given the similarity of the ring and halo spectra, this more likely
reflects differences in density and excitation conditions
(Section~\ref{sec:strength}), possibly combined with the destruction or conversion
of fullerenes closer to the star (\overviewpaper), rather than
differences in formation.

We therefore speculate that the fullerenes formed in the ejected material itself, at
the time of ejection. \overviewpaper show that the core --- and
possibly the halo --- resulted from discrete, massive ejections whose momentum cannot
have been supplied by radiation pressure on dust, and suggest a common-envelope
ejection as a possible origin. Material ejected in such an event cools adiabatically
as it expands, which makes efficient dust formation in the ejecta likely
\citep{Iaconi2020}. Laboratory experiments show that the condensation of carbon vapor
at high temperatures ($\gtrsim$3500~K) favors the formation of fullerenes and
fullerene-like grains, whereas PAHs dominate the condensates at lower temperatures
($\lesssim$1700~K; \citealt{Jager2009}; see also \citealt{Dunk:metallofullerenes}).
The rapid condensation of very carbon-rich ejecta from high temperatures would thus
naturally produce fullerenes rather than PAHs, consistent with the weakness of the
PAH features in Tc~1 (S. Van Schuylenbergh et al. 2026a, submitted). Fullerenes have
indeed also been found in the circumstellar environments of binary post-AGB stars
\citep{Gielen:C60p-AGB}. If fullerene-rich planetary nebulae preferentially result
from such binary ejections, this could also explain why fullerenes are detected in
only a small fraction of planetary nebulae (\overviewpaper).

\section{Summary \& Conclusions}

We have presented a detailed analysis of the fullerene emission in the planetary nebula Tc~1 using JWST MIRI/MRS observations. The high spectral and spatial resolution of these data provide new insights into the spectral characteristics, spatial distribution, abundances, and excitation of \csixty and \cseventy in this benchmark source. Our main findings are as follows:

\begin{enumerate}

\item We detect all four infrared-active fundamental bands of \csixty (at 7.0, 8.5, 17.4, and 18.9~\micron) as well as the 6.5~\micron\ combination band. We also detect six \cseventy bands at 8.8, 12.5, 14.8, 15.6, 17.7, and 21.8~\micron. The band profiles are well characterized by combinations of Gaussian and Lorentzian components, with the 17.4 and 18.9~\micron\ \csixty bands showing resolved rovibrational structure indicative of gas-phase molecules. The 12.5, 14.8, 17.7, and 21.8~\micron\ \cseventy bands also show similarities to this rovibrational structure.

\item We detect additional features at 5.6, 6.6, 7.95, 8.05, 8.15, 9.1, 9.6, 10.0, 10.5, 14.0, and 14.4 \um. The positions of these bands agree with both gas-phase laboratory \csixty spectra and the anharmonic calculations of \csixty used for identifying the near-IR \csixty combination bands. Although these bands are very weak, all show some spatial correlation with the fullerene ring. We tentatively assign all of these features to \csixty combination bands.

\item We searched for fullerene cations but find no clear evidence for \csixtyplus or \cseventyplus in our data. The strongest expected cation features in the 6--9~\micron\ region are not seen, although weak, broadened cation emission blended with the neutral fullerene bands cannot be excluded.

\item We find no direct spectroscopic evidence of \csixtyminus. While we do detect both a broad feature at 17.75~\um and other smaller features throughout our spectrum that all coincide with the laboratory bands of \csixtyminus (Finazzi et al., in prep.), we attribute the band at 17.75~\um (the strongest \csixtyminus band) to \cseventy. If there were to be emission from \csixtyminus in our data, it would be either broad or severely blended with the neutral fullerene bands, making it impossible to claim a definitive detection.

\item Our MIRI observations allow us to cleanly separate the \csixty bands from blended atomic emission lines for the first time. Contrary to expectations, the 7.0/18.9~\micron\ band ratio is higher ($\sim$1.4--1.9) than previously reported ($\sim$1.3), even after accounting for the removal of [\ion{Ar}{2}] contamination. The 8.5/18.9 and 17.4/18.9 ratios are consistent with previous measurements. However, these band ratios will be directly affected by contamination from \cseventy. The band ratios also vary across the nebula: all ratios are elevated in the fullerene ring and in the halo compared to the core, and the 8.5/18.9 ratio, which is least affected by \cseventy, shows additional structure outside the ring.

\item The fullerene emission is concentrated in a thin asymmetric shell at a radius of approximately 5--6\arcsec ~from the central star. Both \csixty and \cseventy show similar spatial distributions, peaking in this ring structure with weaker emission extending into the interior and a hint of a secondary shell at larger radii.

\item The combined fullerene emission accounts for approximately 2.6\% of the total infrared emission from Tc~1, corresponding to roughly 0.6\% of the total carbon budget locked in fullerenes.

\item The resolved rovibrational structure in the 17.4 and 18.9~\micron\ bands rules out thermal emission from grain-attached fullerenes, and the photoionized nature of the gas makes shock excitation unlikely. Photo-excitation thus remains the most likely mechanism, but our results do not resolve its long-standing problems. The spatial variations in the band ratios, which do not follow the radiation field, indicate that the excitation conditions change across the nebula. A quantitative interpretation is hampered by the contribution of \cseventy\ to most \csixty\ bands and by processes specific to ionized gas that are not included in current fluorescence models; these are explored by Cami et al. (in prep.).

\item The fullerene emission peaks in a thin shell at the outer edge of the core, but is also detected in the core and in the halo, with very similar spectra. Since the halo results from an earlier mass-loss episode, the fullerenes formed independently of the ring and survived in the low-density ionized gas. The ring is thus not a formation site, but a location where density and excitation conditions favor stronger emission. We speculate that the fullerenes formed in the ejected material itself, possibly during a common-envelope ejection, where the rapid, high-temperature condensation of carbon-rich material favors fullerenes over PAHs. If fullerene-rich planetary nebulae preferentially result from such events, this could explain why fullerenes are found in only a small fraction of them.

\end{enumerate}

These JWST observations demonstrate that Tc~1 remains an exceptional laboratory for studying fullerenes in space. The rovibrational structure, detailed spatial mapping, and updated band ratios presented here provide critical constraints for models of fullerene excitation and formation in astrophysical environments. Future work combining these results with improved laboratory data and theoretical models will be essential for a complete understanding of the life cycle of fullerenes in planetary nebulae.

\begin{acknowledgments}

This work is based on observations made with the NASA/ESA/CSA James Webb Space Telescope. All of the data presented in this article were obtained from the Mikulski Archive for Space Telescopes (MAST) at the Space Telescope Science Institute. The data of this specific observing program can be accessed via \dataset[doi:10.17909/9p7c-fn82]{https://archive.stsci.edu/doi/resolve/resolve.html?doi=10.17909/9p7c-fn82}. This research is based upon work from COST Action NanoSpace, CA21126, supported by COST (European Cooperation in Science and Technology). M.M.G., J.C., S.V.S., E.P., C.B., and D.V.D.P. acknowledge support from the University of Western Ontario, the Canadian Space Agency (CSA) [24JWGO3A13], and the Natural Sciences and Engineering Research Council of Canada. V.J.E thanks Chapman University for support. D.A.G.H. and A.M. acknowledge the support from the State Research Agency (AEI) of the Ministry of Science, Innovation and Universities (MICIU) of the Government of Spain, and the European Regional Development Fund (ERDF), under grant PID2023--147325NB--100/AEI/10.13039/501100011033. Astrochemistry in The Netherlands is supported by the NWO Dutch Astrochemistry Network (grant no. ASTRO.JWST.001). R.S.’s contribution to the research described here was carried out at the Jet Propulsion Laboratory, California Institute of Technology, under a contract with NASA (80NM0018D0004). K.J. acknowledges the support of the Swedish National Space Agency. Support for participation in JWST-GO-04706 by G.C.S., H.L.D, and R.S.\ was provided through grants from the STScI under NASA contract NAS5-03127. P.B.C. is supported by NIST. L.F. acknowledges the Max Planck-Radboud University Center for Infrared Free Electron Laser Spectroscopy for support.

Portions of the text in this paper were drafted and edited with the assistance of
Claude, a large language model developed by Anthropic
(\url{https://claude.ai}). The
tool was used under the direction of the authors, who verified all content and take
full responsibility for the analysis, results and conclusions presented here.
\end{acknowledgments}

\vspace{5mm}

\software{\textsc{astropy} \citep{2013A&A...558A..33A, 2018AJ....156..123A, 2022ApJ...935..167A},  
          \textsc{scipy} \citep{SciPy2020}, 
          \textsc{lmfit} \citep{newville_2025_15014437},
          \textsc{photutils.aperture} \citep{photutils_Bradley2024}
          }

\appendix

\section{Fitting Procedure}\label{App:fitting}

To generally characterize the profile shapes of all fullerene bands, and to describe any potential spatial variations more easily, we decomposed them into combinations of Gaussian and Lorentzian components using the fitting software package \textsc{lmfit} \citep{newville_2025_15014437}. Given that the 17.4 and 18.9~\um bands show resolved rovibrational structure with intrinsic Lorentzian line profiles (see J. Cami et al. 2026a, submitted), we expect Lorentzian components to be appropriate for the fullerene emission in general. However, for some bands, we find that adding Gaussian components provide a better fit. When unresolved atomic lines are superposed on the bands, we include a narrow Gaussian for these lines as well; in some cases however, we masked the narrow emission line if we could not obtain satisfactory fits otherwise. All fits were weighted by the data uncertainties and included a linear component to account for the local continuum. 

\input{ringparameters_table}

\begin{figure*}[t]
    \centering
        \centering
    \resizebox{\hsize}{!}{
    \includegraphics{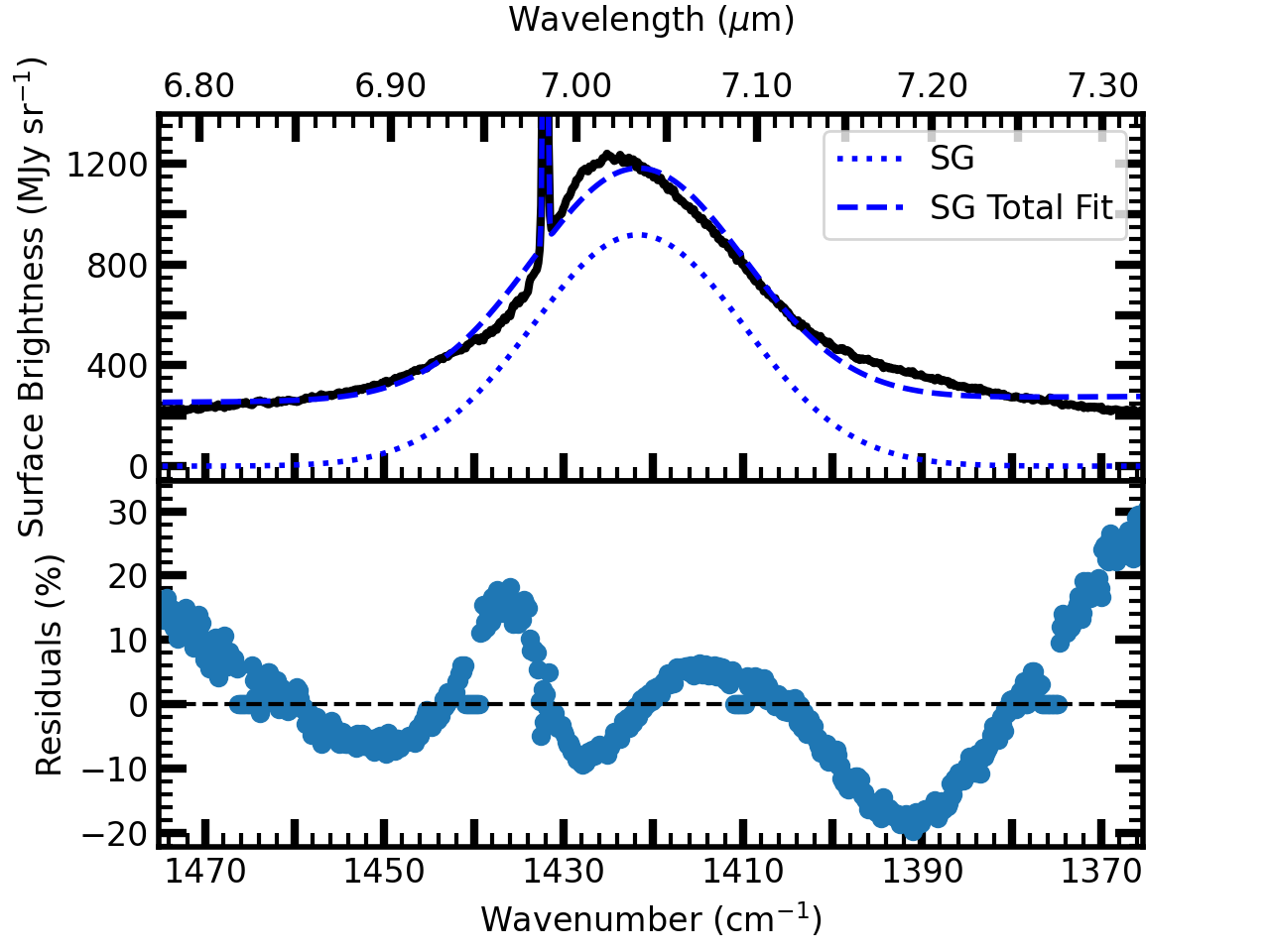}
    \includegraphics{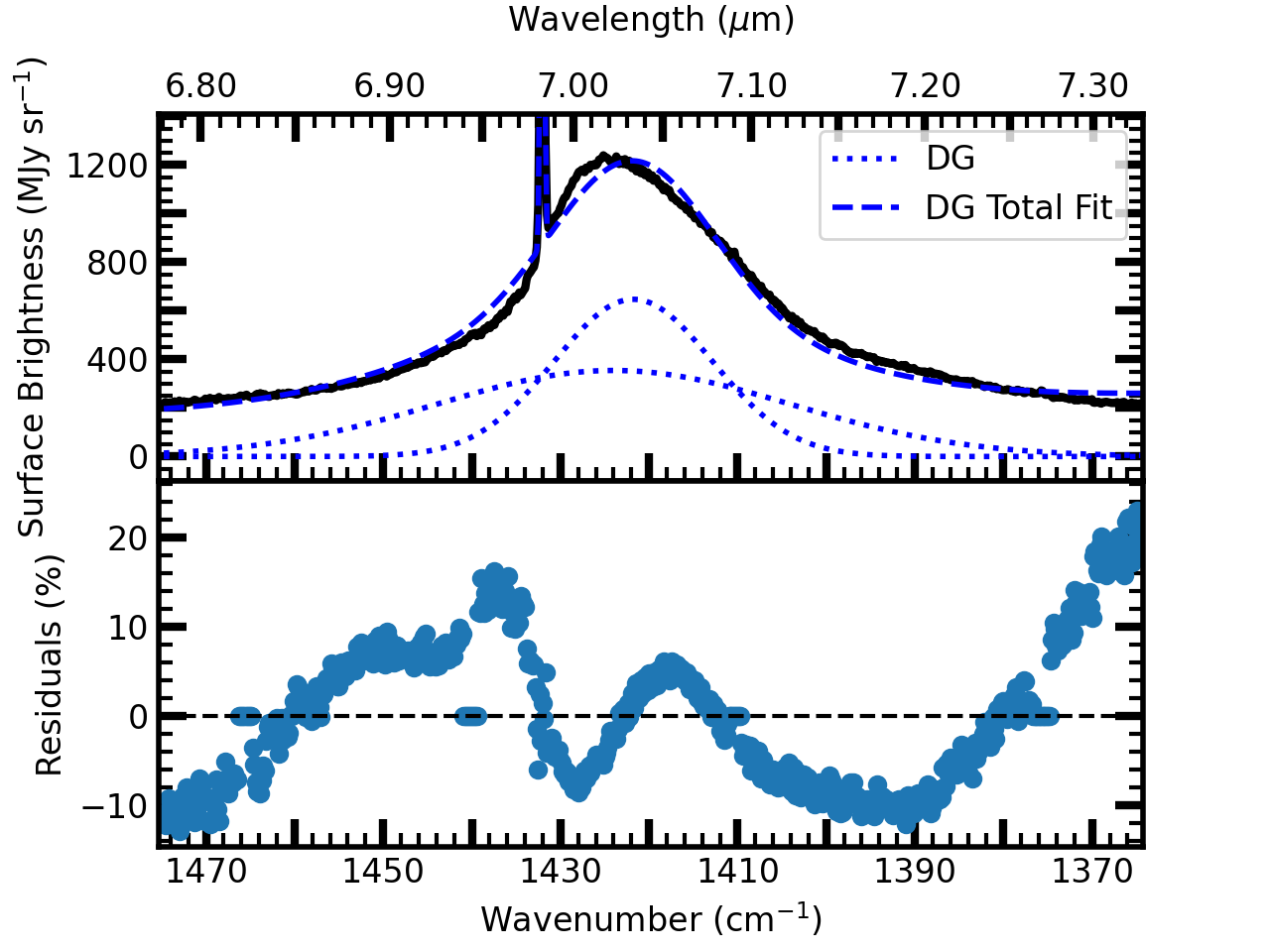}}
   \resizebox{\hsize}{!}{
    \includegraphics{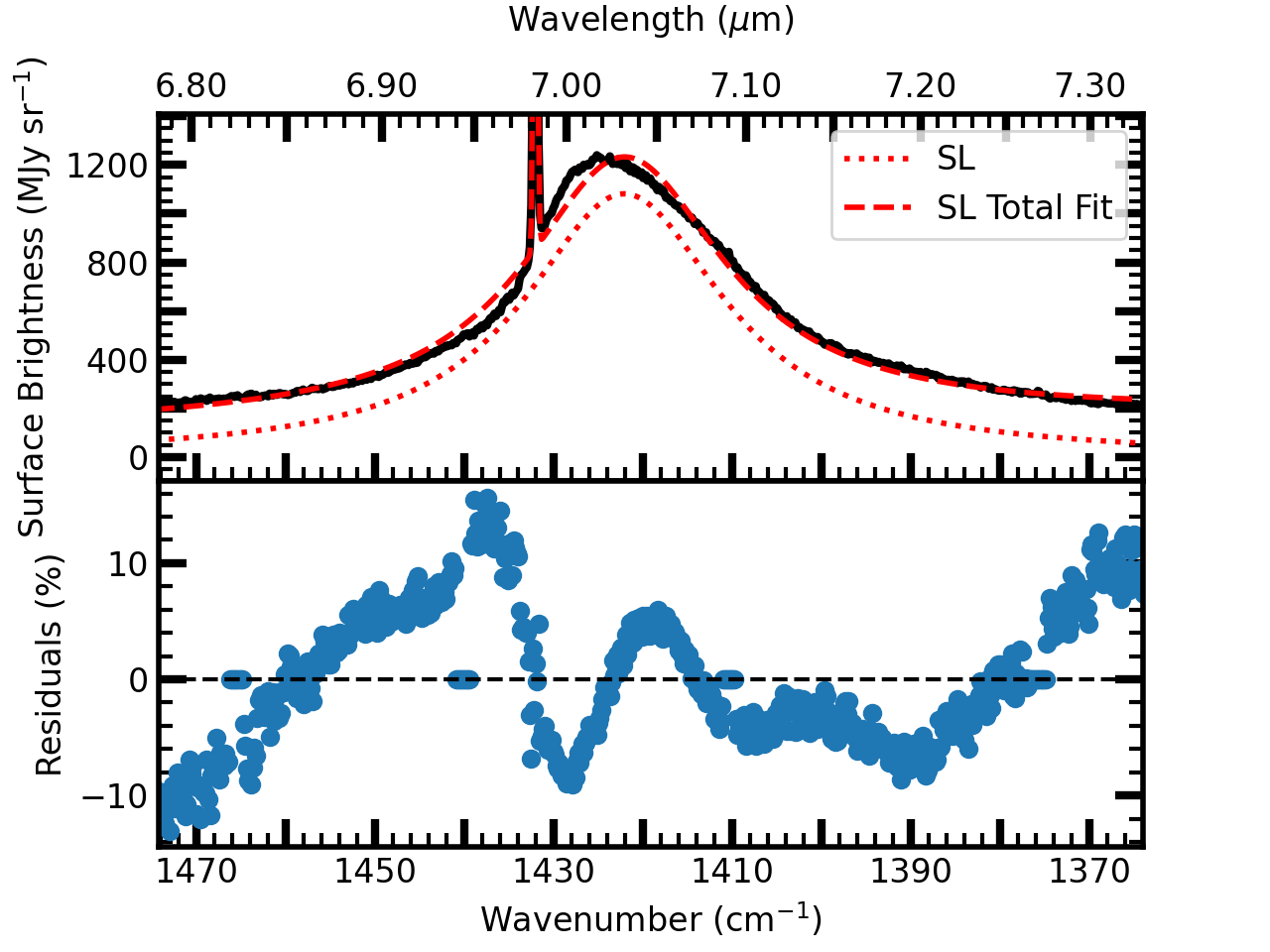}
    \includegraphics{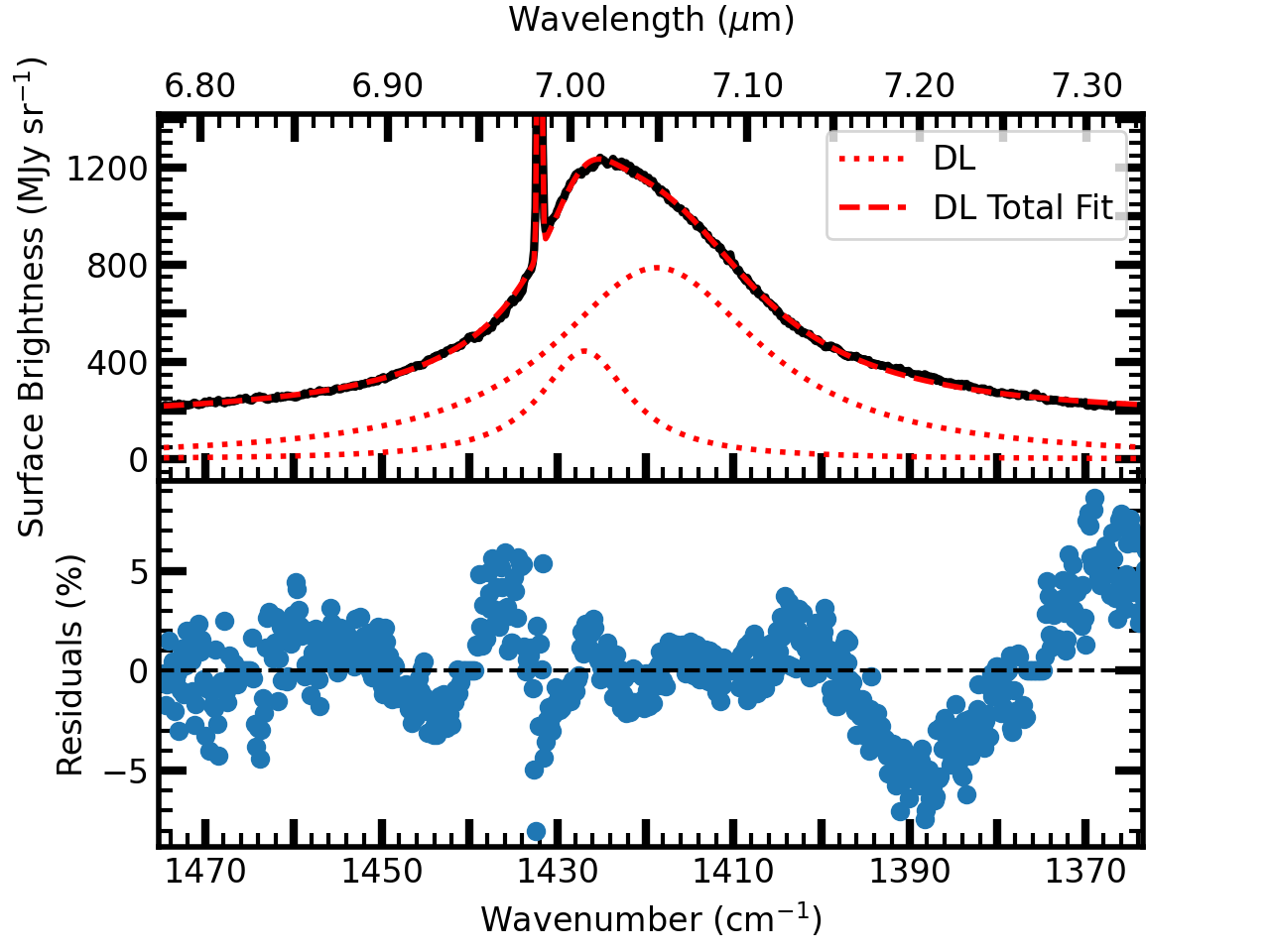}}
    
    \caption{Comparing different combinations of profiles to fit the 7.0 \um fullerene feature. Top: Single and double Gaussians (SG and DG, blue). Bottom: Single and double Lorentzian (SL and DL, red). Dotted curves represent individual components; dashed lines the composite fit. All fits also include a narrow Gaussian component for the unresolved \ArII emission line and a linear continuum (not shown). The bottom half of each panel shows the residuals displayed as percentage of the original, non-continuum subtracted spectrum.}
    \label{fig:decomp}
    
\end{figure*}

We illustrate our approach with the 7.0~\um band. Figure \ref{fig:decomp} shows a close-up of the 7.0 \um fullerene band in the ``Fullerene Ring" aperture. The feature is blended with an unresolved \ArII emission line at 6.99 \um, but the high spectral resolution allows us to cleanly separate the two. The 7.0 \um feature has a slightly asymmetrical shape with broad, extended wings, making it hard to define where the feature starts and ends. This furthermore depends on the choice of continuum; we refer to \dustpaper for a more detailed discussion on the continuum and its effects on the features in the 6--9 \um range. Figure~\ref{fig:decomp} shows that a single Gaussian cannot capture the asymmetry of the band. Two Gaussians provide a slightly better fit, but even then the broad wings are not well reproduced. A single Lorentzian on the other hand reproduces the wings well, but not the asymmetry; two Lorentzians reproduce the overall fullerene profile very well, and just leaves two weak bumps in the wings of the main band at 6.9 and 7.2 \um. Those were best reproduced with a Gaussian profile (see Figure~\ref{fig:c60_fits}). Reproducing the overall observed band profile thus requires two Lorentzian components, two weaker Gaussians, and another narrow Gaussian for the \ArII line (see Table~\ref{tab:ringparameters}). This general procedure was used to fit every feature discussed in this paper.

The two Gaussian components of the 7.0 \um band each contain about 1\% of the total power emitted in the 7 \um band, making it difficult to create clear spatial maps of these features to test spatial similarities of the emission. However, it is clear that the 7.2 \um feature shows similar spatial structure to the rest of the fullerene bands while the 6.9 \um feature does not; the 7.2 \um Gaussian component is thus treated as a ``fullerene" component and its emission included in the 7.0 \um spatial map shown in Figure \ref{fig:c60_individual_maps}. The 6.9 \um spatial map does not show the ring and is extremely noisy; it is therefore excluded from further analysis.

The 8.5 \um band contains a fairly broad Gaussian component centered at 8.6 \um (see Figure~\ref{fig:c60_fits}). This component shows strong spatial variations (see Appendix~\ref{sect:pah_86}), and most likely contains a combination of fullerene and weak PAH emission (S. Van Schuylenbergh et al. 2026a, submitted). We therefore fixed the line center and width to the values listed by \citet{Peeters:NGC2023}. Additionally, the profiles of the 17.4 and 18.9 \um bands show clear ro-vibrational structure and evidence for contributions due to $^{13}$C substituted fullerenes (see J. Cami et al. 2026a, submitted). The observed profiles are thus much more complex than our fit components can describe. Our aim here is not to interpret the profiles in terms of rotational contours, but to more generally describe the profile shape. 

\section{Fullerene band profile variation across Tc~1 }\label{App:profile}

\begin{figure}
\resizebox{\hsize}{!}{
\includegraphics[clip, trim= 0cm 0cm 1cm 0cm]{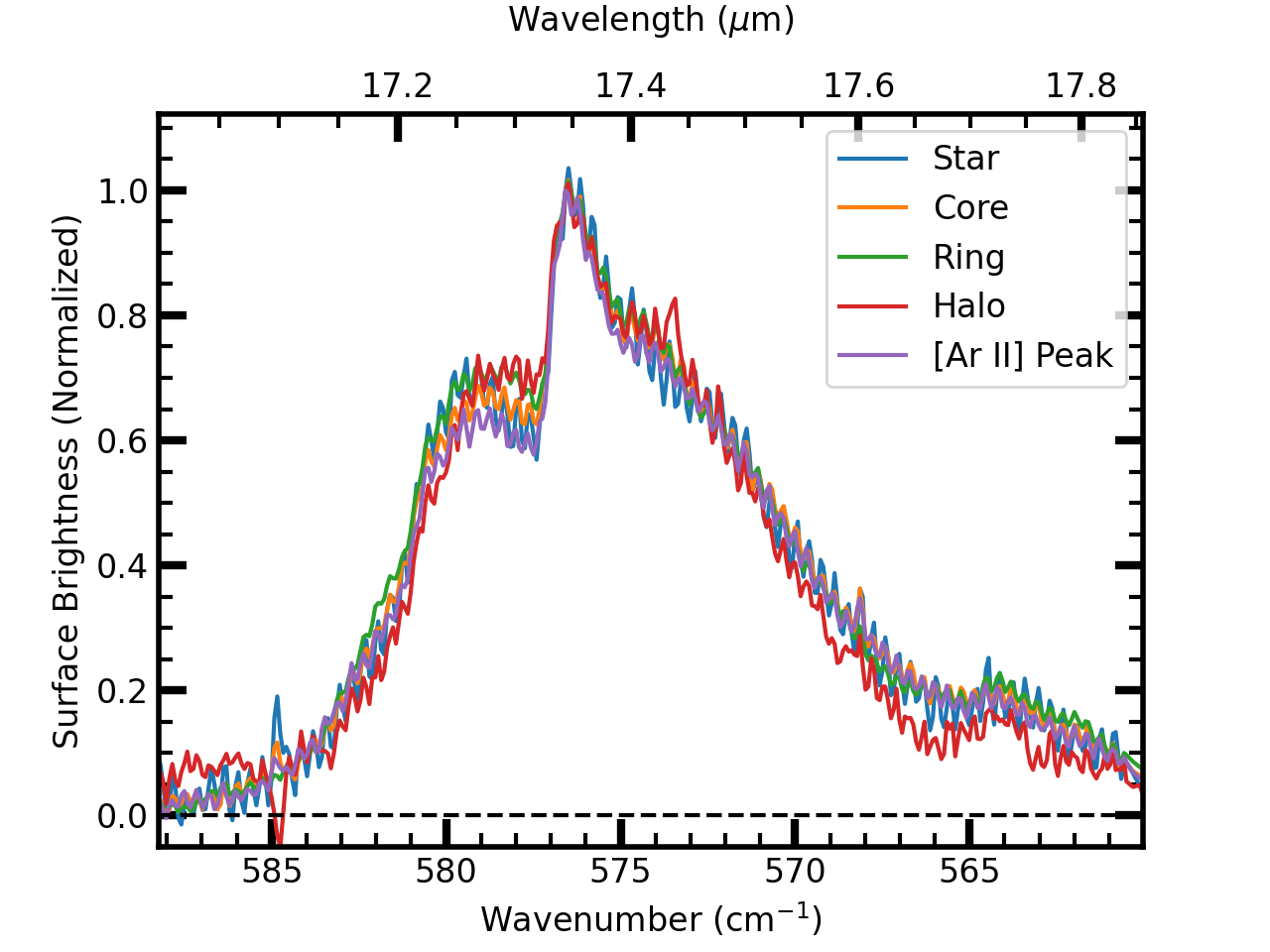}
\includegraphics[clip, trim= 0cm 0cm 1cm 0cm]{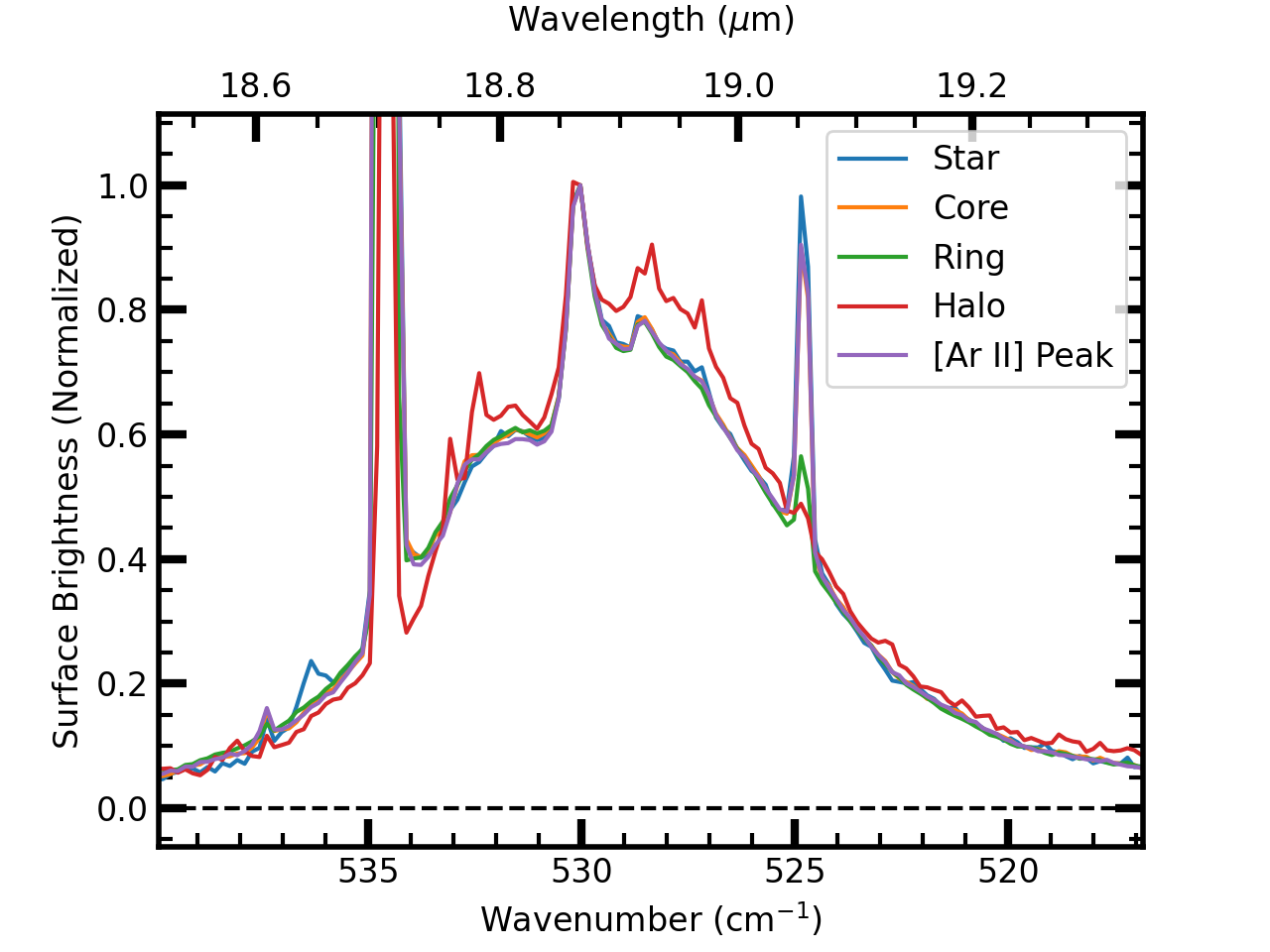}}
\caption{\label{Fig:spatial_vary} Profile variations of the 17.4 and 18.9 \um fullerene bands across the five apertures.}
\end{figure}

To further test how the band profiles vary across our five different apertures, we normalized each of the four main \csixty features in the different spectra and overlaid them; the 17.4 and 18.9 \um bands are shown in Figure \ref{Fig:spatial_vary}. For the 17.4 \um band, there is slightly increased emission from the blue wing of the feature ($\sim10\%$ relative to band peak) in the ``Fullerene Ring" and ``Extended Halo" apertures. For the 18.9 \um band, the ``Extended Halo" aperture has increased relative emission in both wings of the feature along with additional substructures across the band.

As discussed in J. Cami et al. 2026a, submitted, both bands contain contributions from \cseventy and $^{13}$C substitution. This additional emission will directly affect the broad wings of these features, and different relative amounts of either could have effects on the profile shape across different apertures. For the analysis in this paper, we conclude that the general profiles are overall very similar across the different regions of Tc~1. Future, more-detailed modeling of these bands that include contributions from ro-vibrational structure, $^{13}$C substitution, and \cseventy contribution will better determine how exactly the emission varies spatially.

\section{Fitting the additional \texorpdfstring{\csixty}{C60} mid-IR combination bands}\label{sec:append_combination}

\begin{figure}
      \centering
\resizebox{\hsize}{!}{
    \includegraphics[trim={0.6cm 1.5cm 0cm 1.5cm},clip]{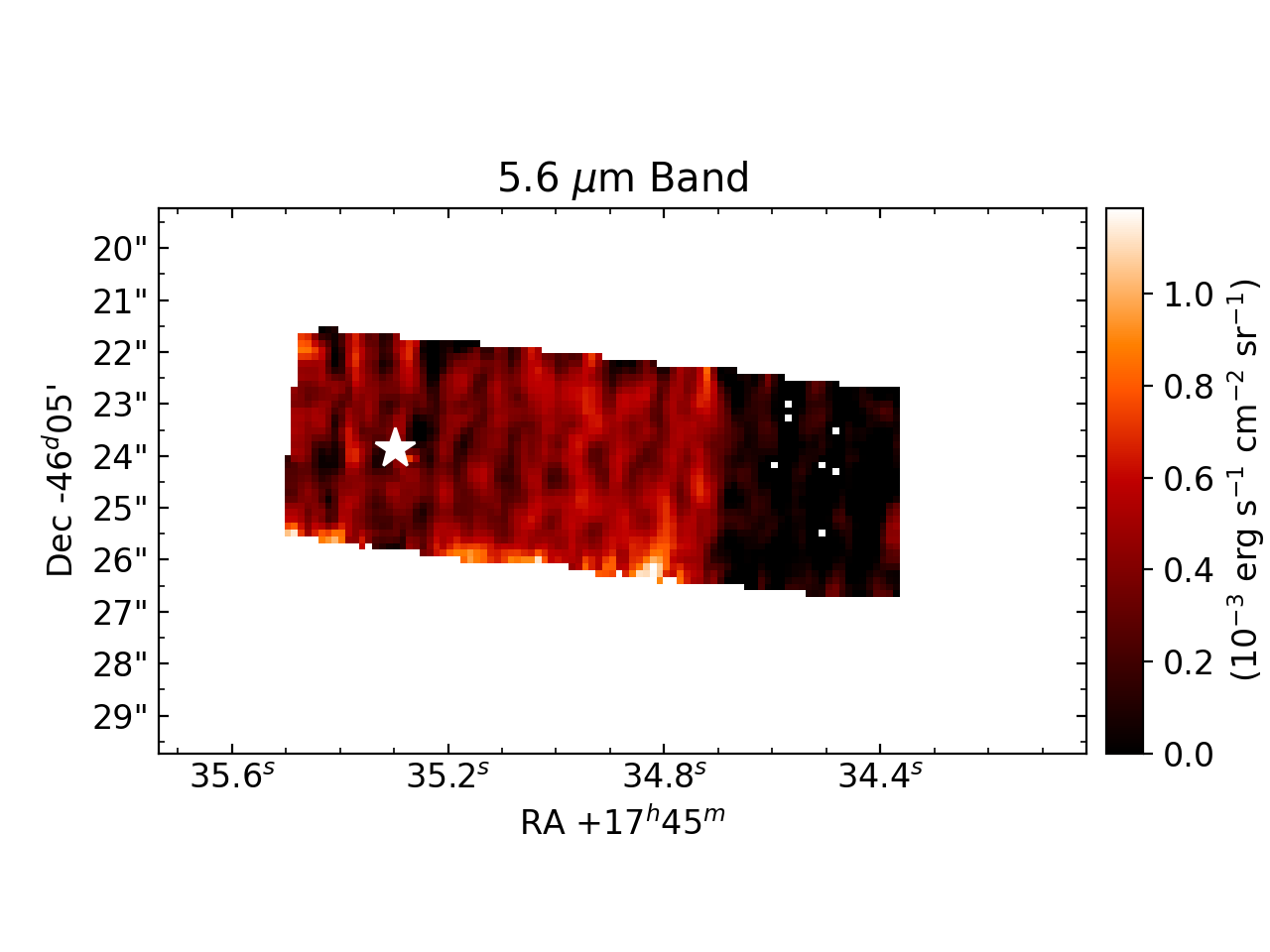}
    \includegraphics[trim={0.6cm 1.5cm 0cm 1.5cm},clip]{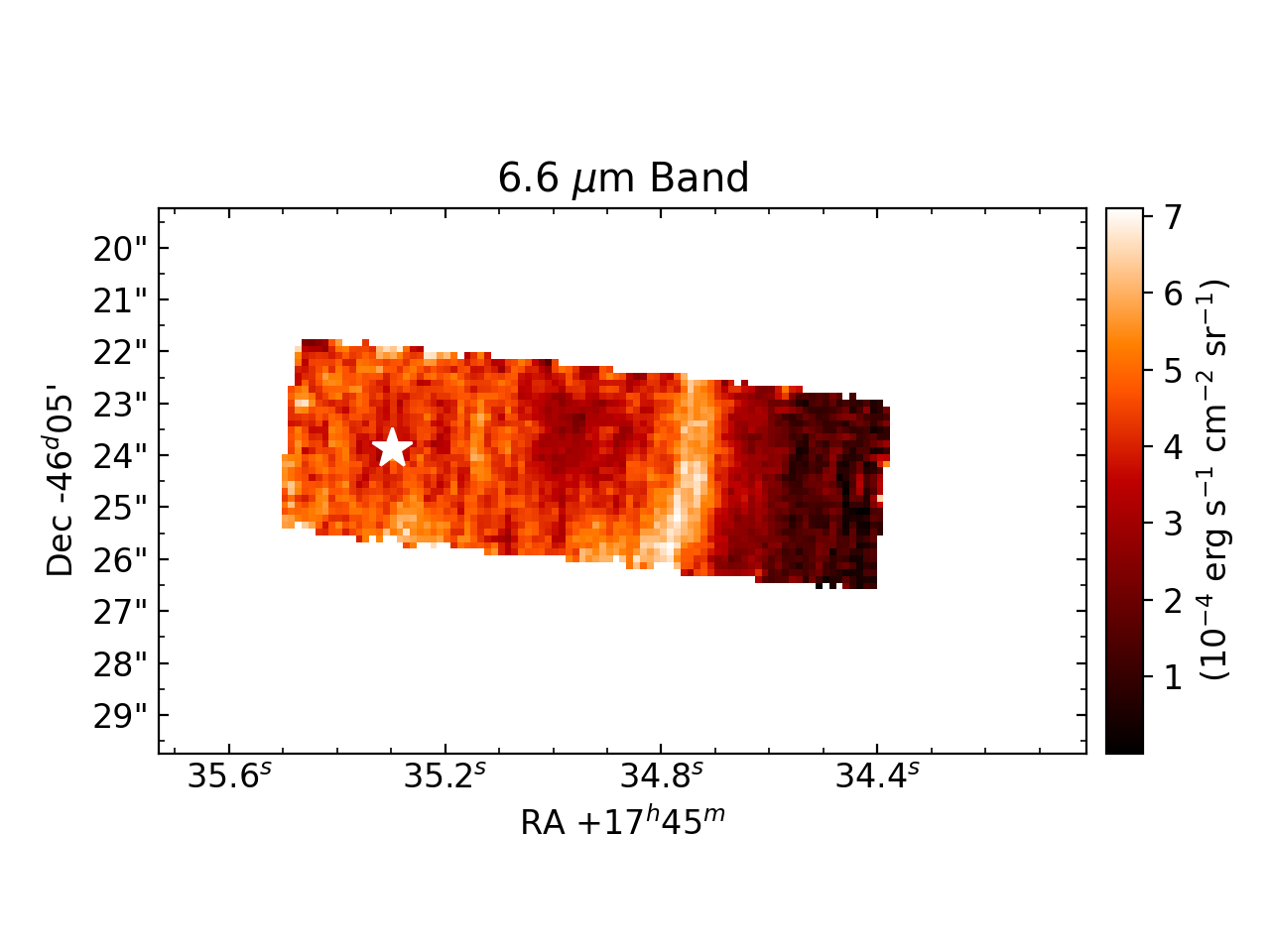}
    \includegraphics[trim={0.6cm 1.5cm 0cm 1.5cm},clip]{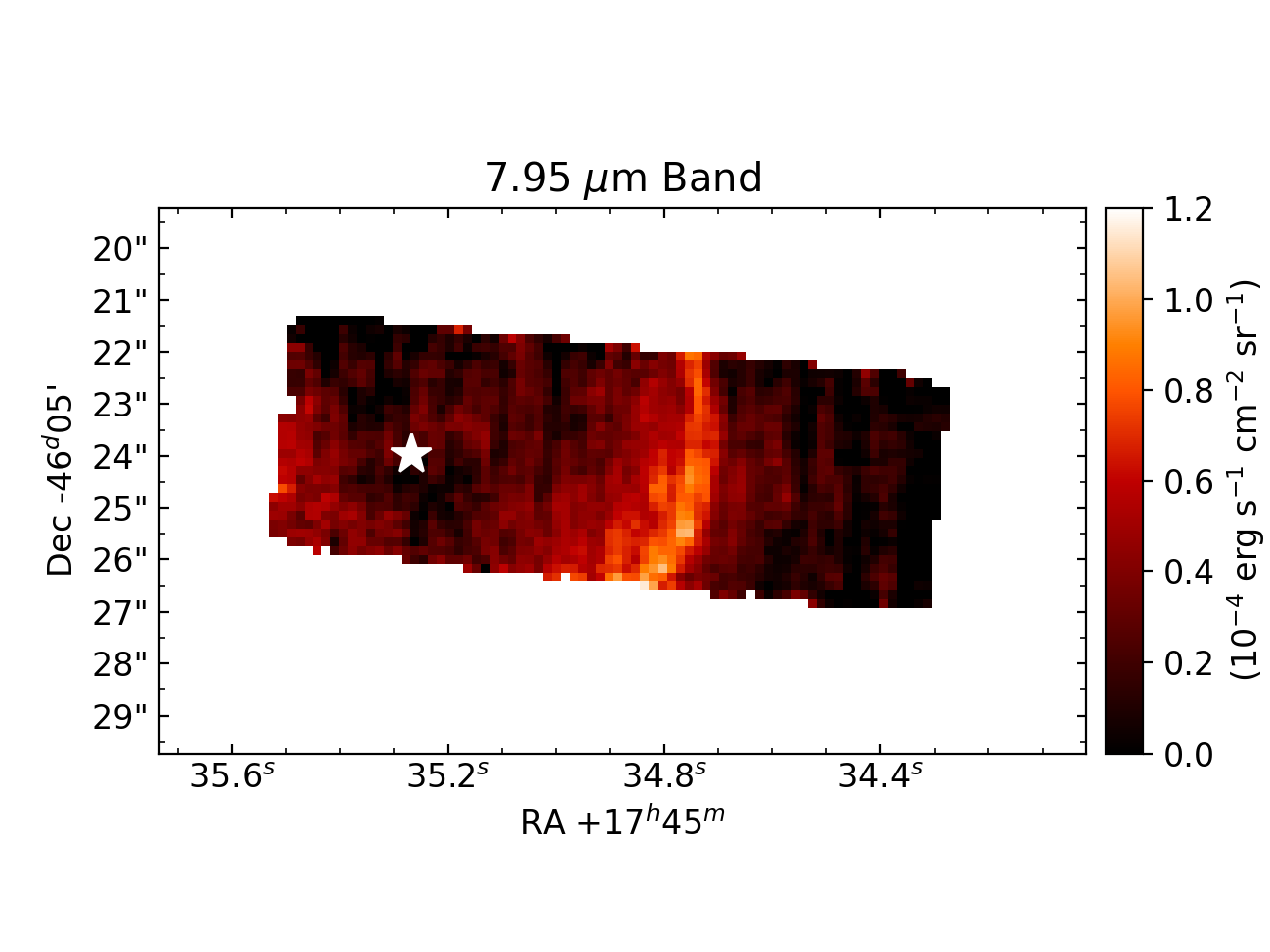}
    }
\resizebox{\hsize}{!}{
    \includegraphics[trim={0.6cm 1.5cm 0cm 1.5cm},clip]{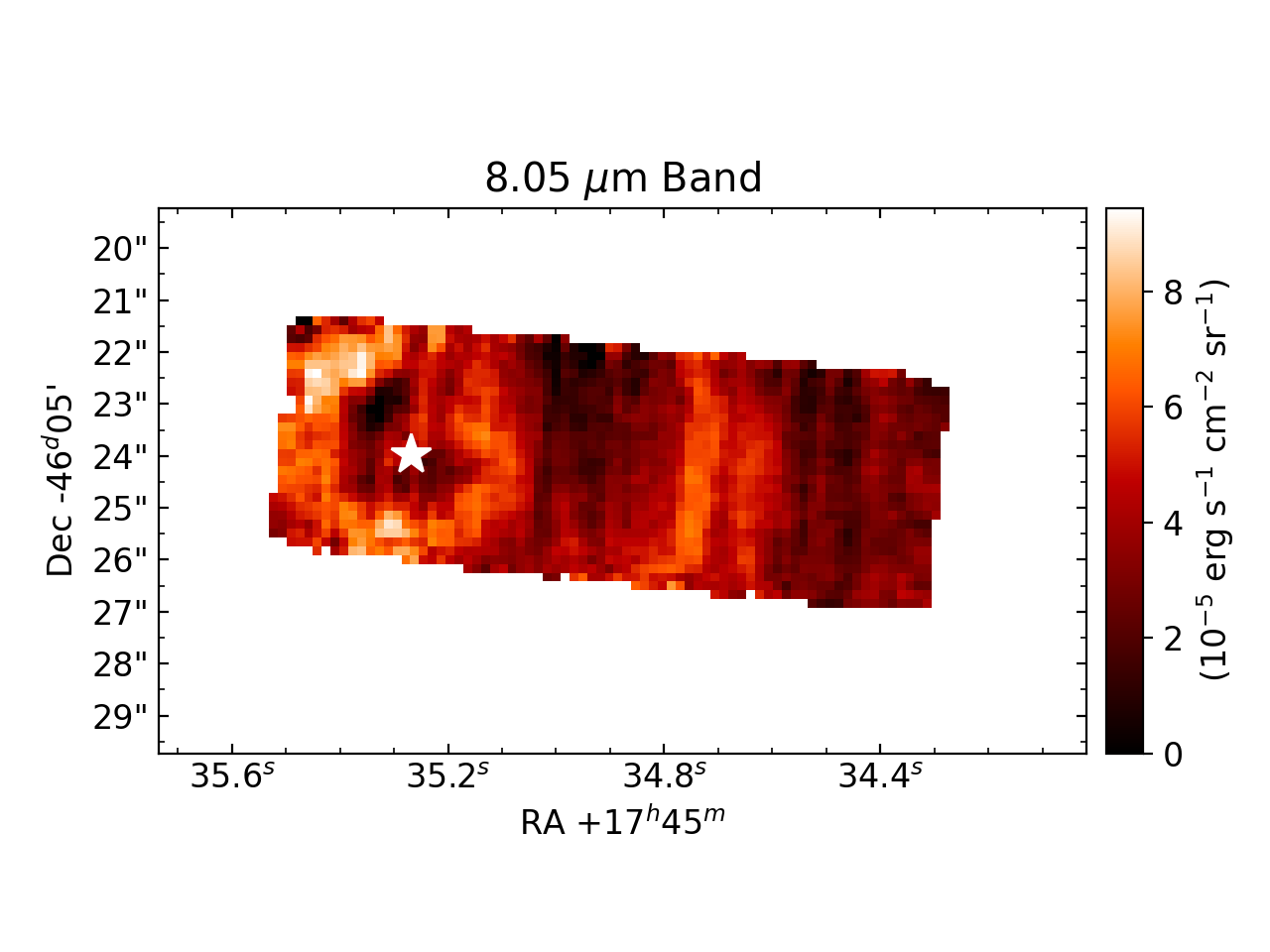}
    \includegraphics[trim={0.6cm 1.5cm 0cm 1.5cm},clip]{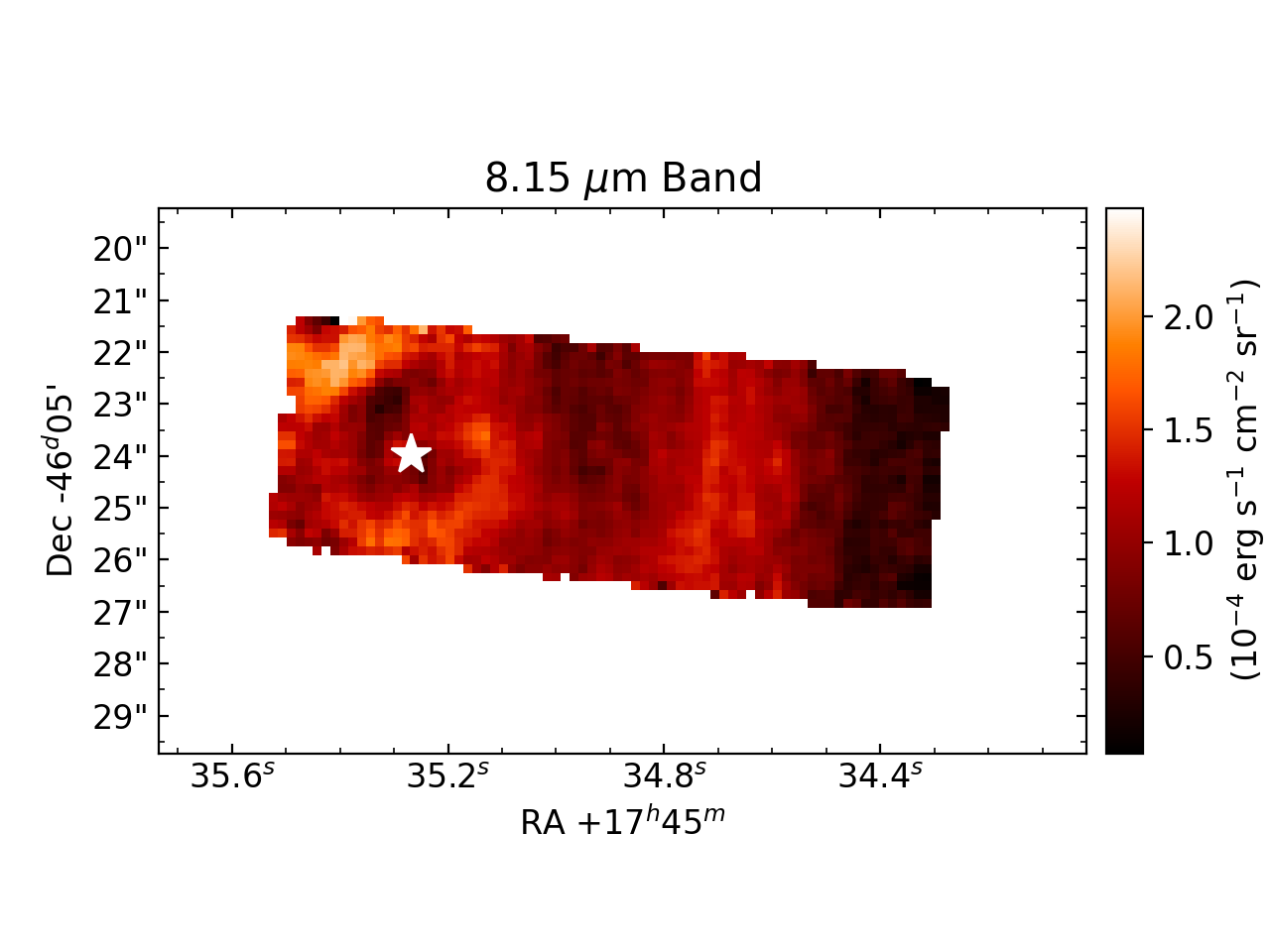}
    \includegraphics[trim={0.6cm 1.5cm 0cm 1.5cm},clip]{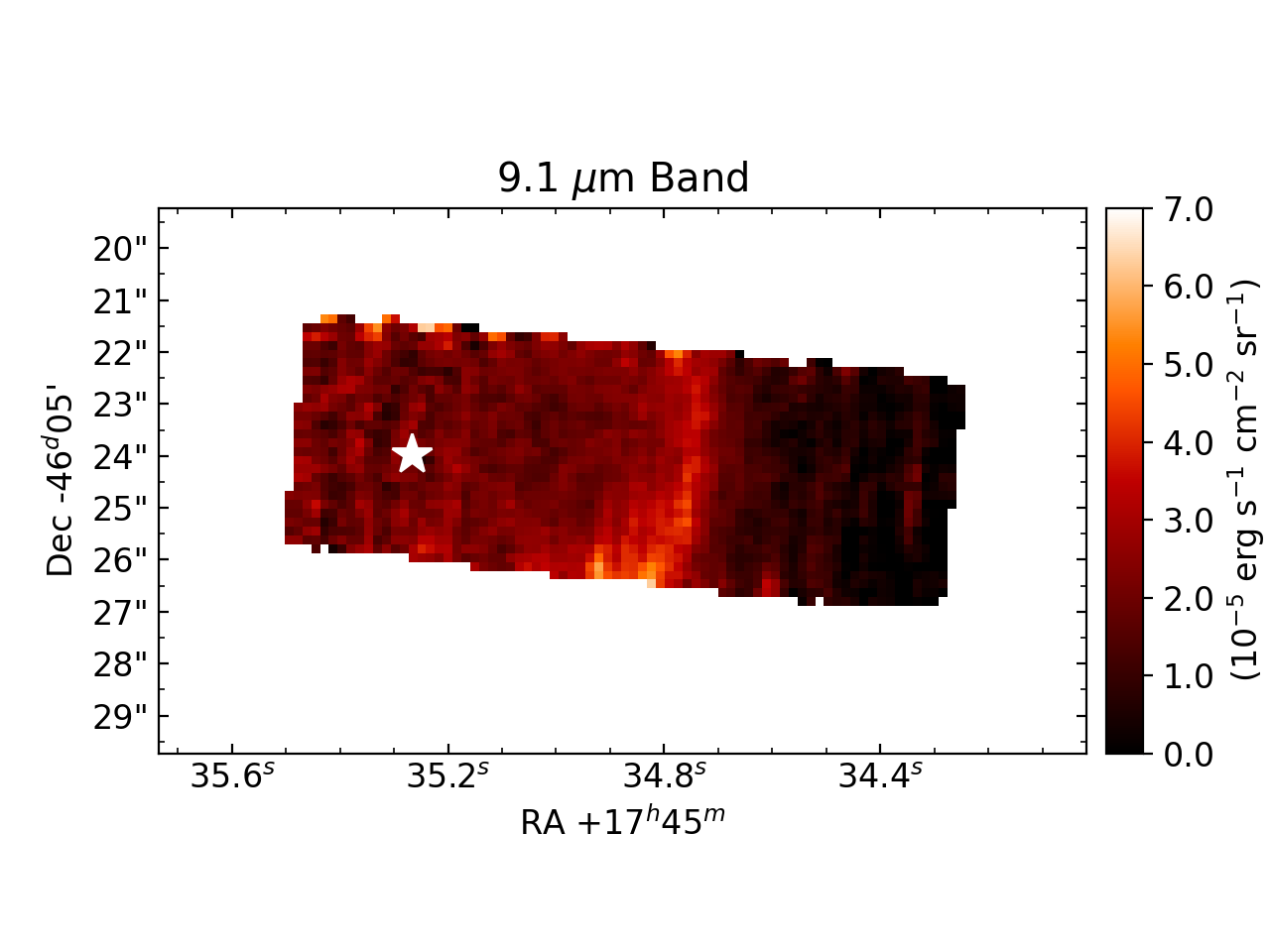}
    } 
\resizebox{\hsize}{!}{
    \includegraphics[trim={0.6cm 1.5cm 0cm 1.5cm},clip]{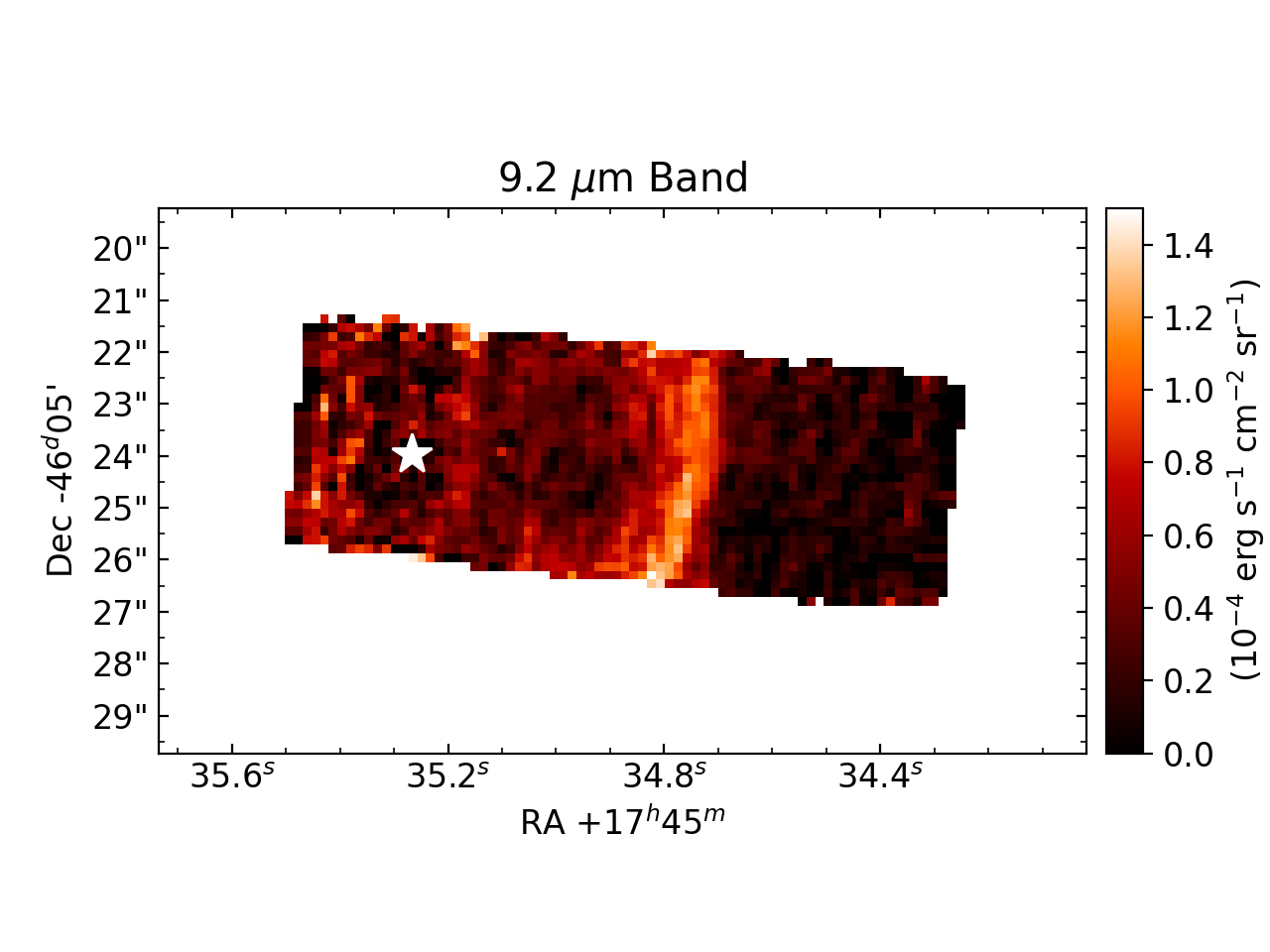}
    \includegraphics[trim={0.6cm 1.5cm 0cm 1.5cm},clip]{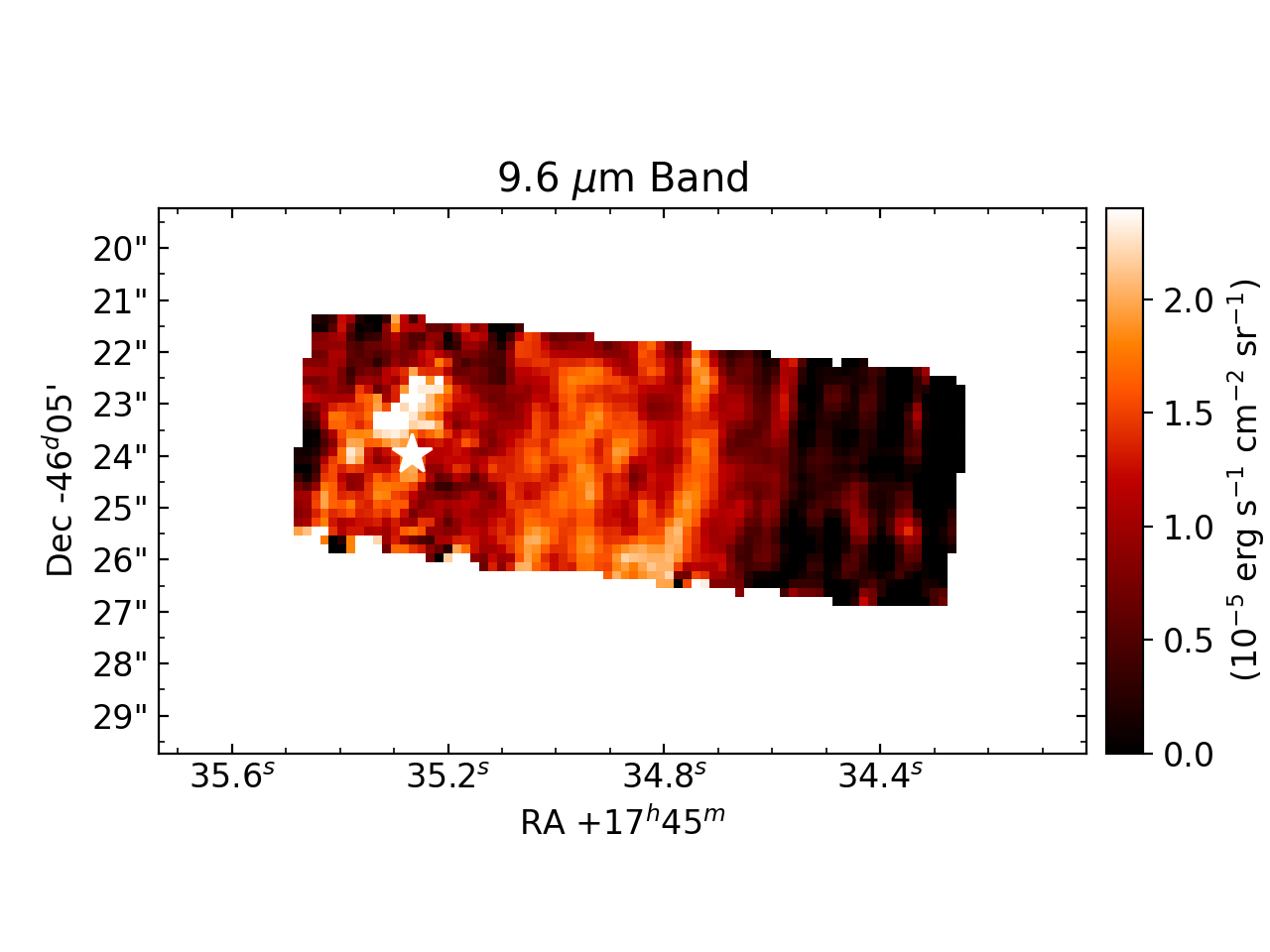}
    \includegraphics[trim={0.6cm 1.5cm 0cm 1.5cm},clip]{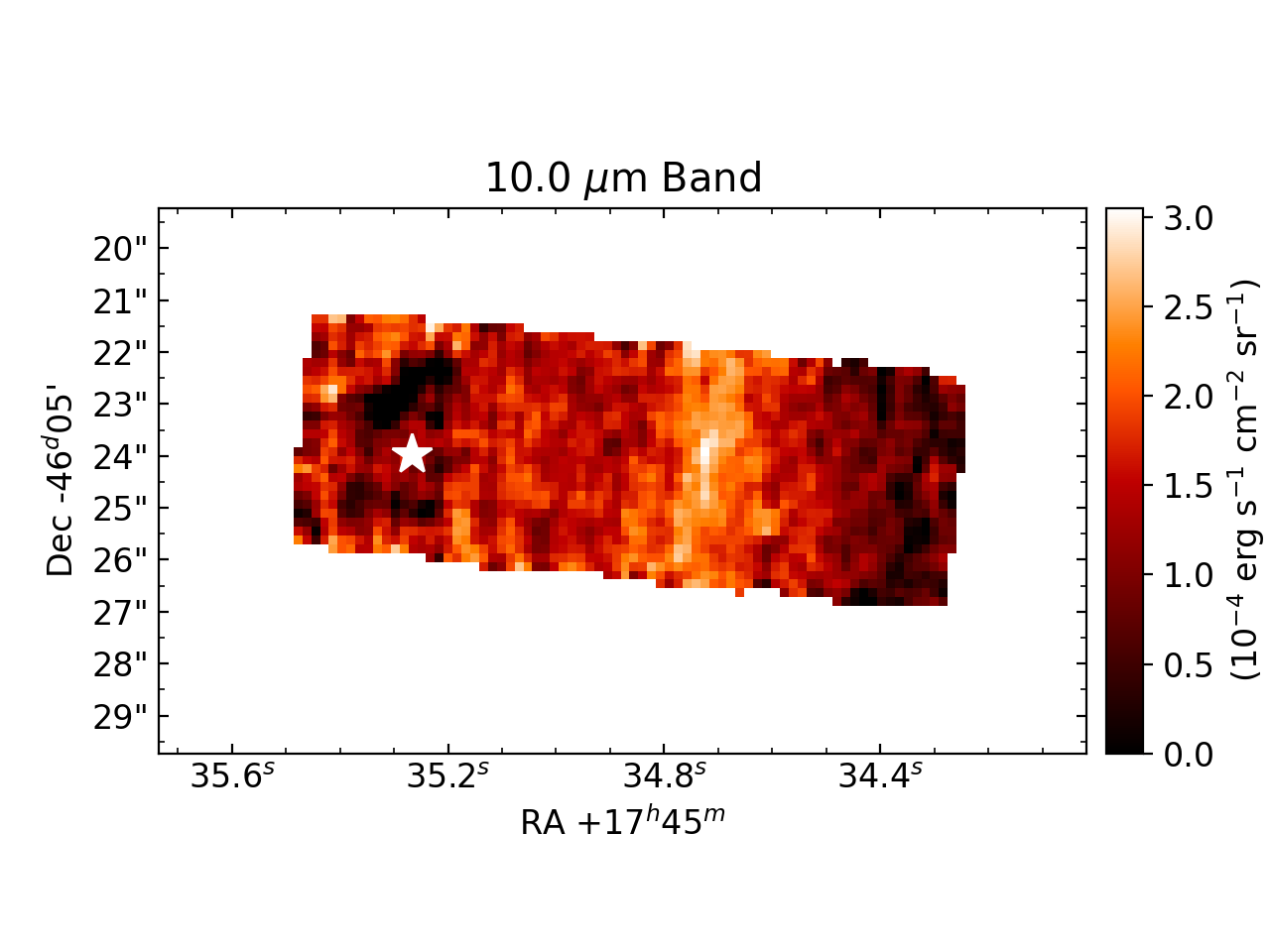}
    } 
\resizebox{\hsize}{!}{
    \includegraphics[trim={0.6cm 1.5cm 0cm 1.5cm},clip]{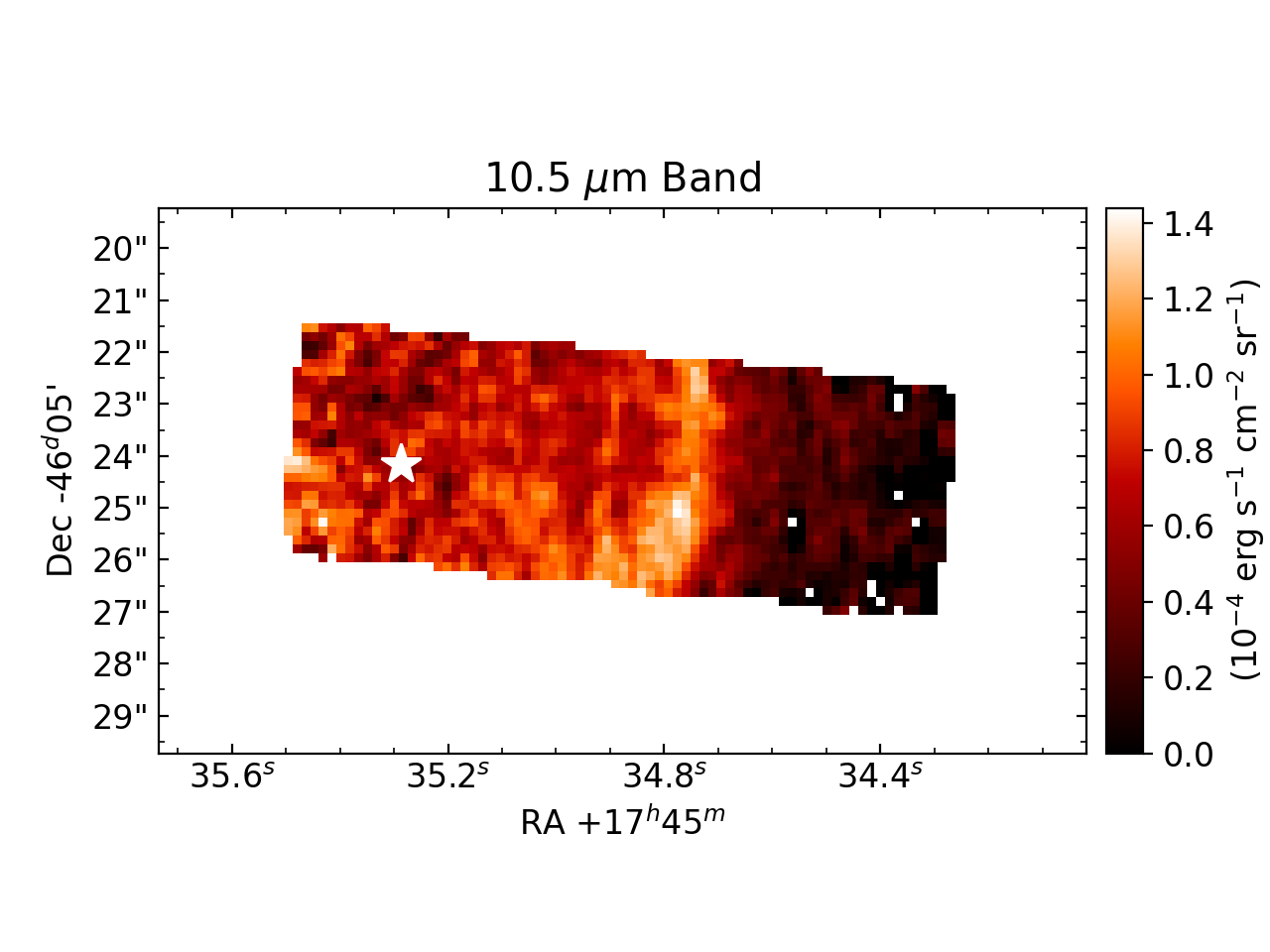}
    \includegraphics[trim={0.6cm 1.5cm 0cm 1.5cm},clip]{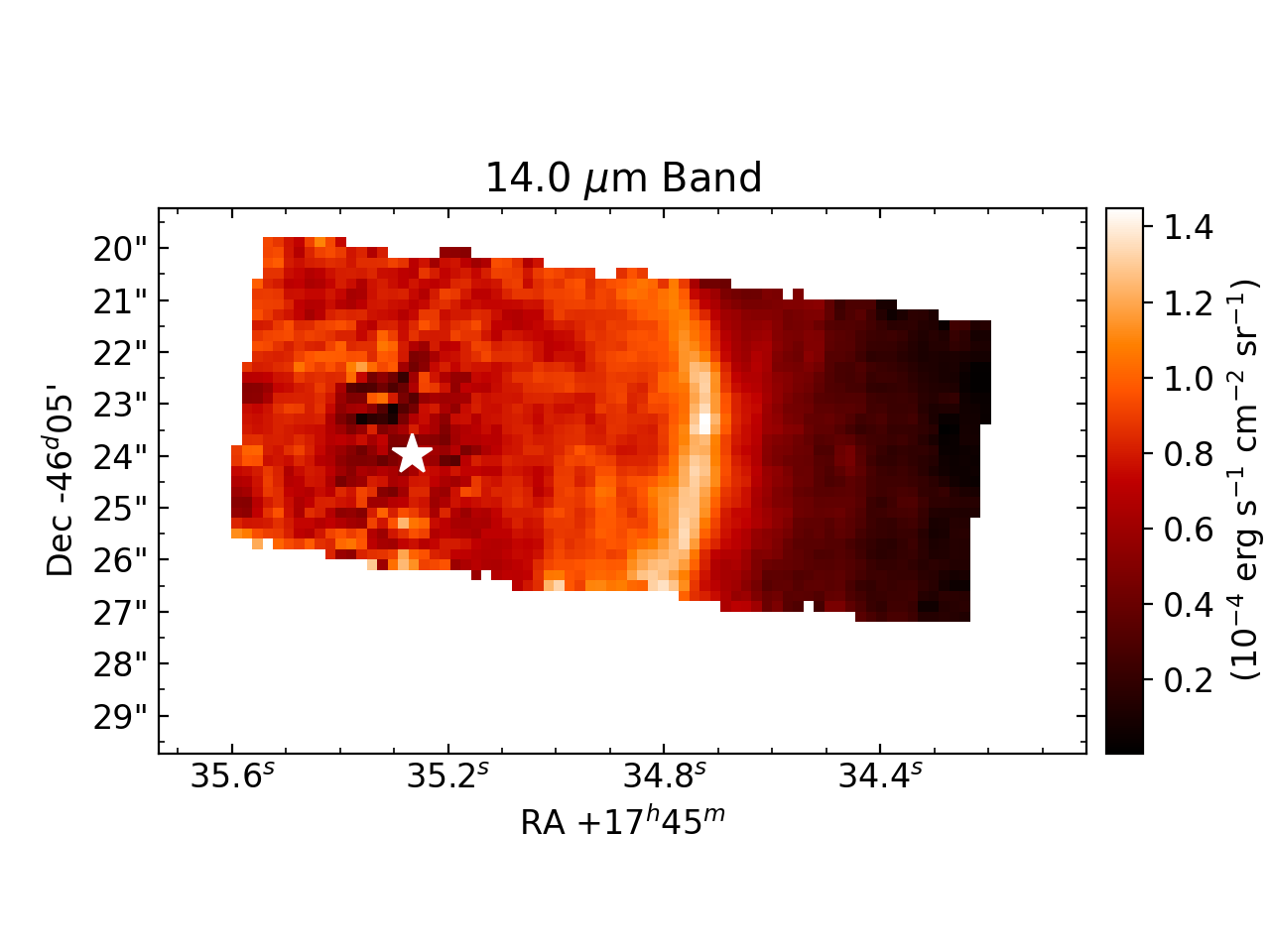}
    \includegraphics[trim={0.6cm 1.5cm 0cm 1.5cm},clip]{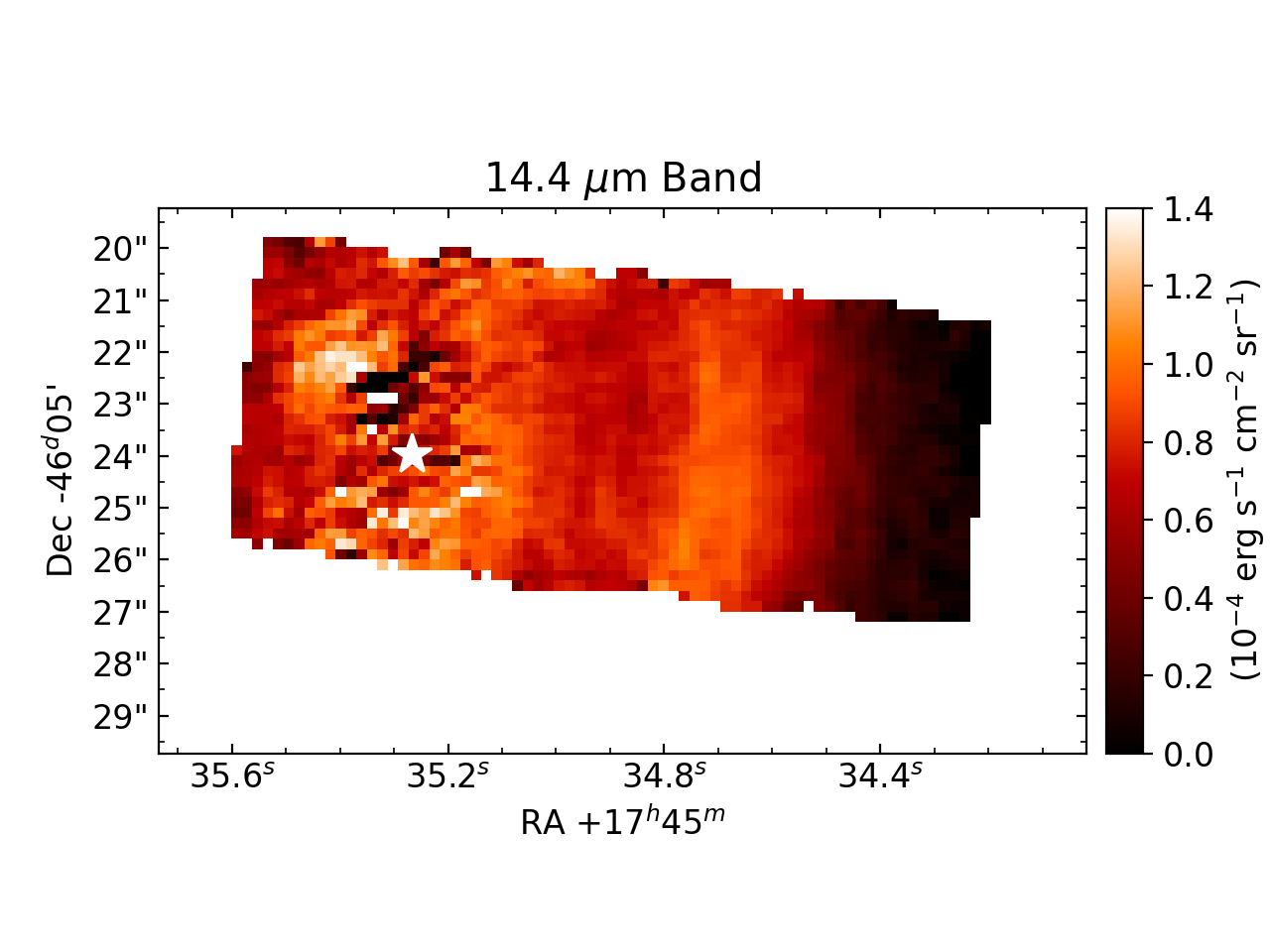}
    } 
    \caption{Integrated surface brightness maps of each of the features tentatively due to \csixty combination bands. The location of the central star is denoted by a white star. The white/black pixels near the central star and past the fullerene ring in some maps indicate pixels affected by diffraction and fringing artifacts.}
    \label{fig:combination_maps}
    
\end{figure}

We fit a single Gaussian component and a linear local continuum to each of the weak features at 5.6, 7.95, 8.05, 8.15, 9.1, 9.6, 10.0, 10.5, 14.0, and 14.4 \um (the 6.6 \um feature is fit alongside the 6.5 \um feature, see Figure \ref{fig:c60_fits}). This approach is slightly simpler than the fitting process used for the fundamental bands due to the difference in strength: many of these bands are so weak that they disappear within the noise in spectra taken from outside the fullerene ring. Surface brightness spatial maps are found in Figure \ref{fig:combination_maps}. All maps show the fullerene ring to some degree. We note that the fits of many of these features are severely affected by noise, causing artifacts to appear along with spatial features not correlated with fullerenes (such as the horseshoe). Due to the weak nature of these features, it is difficult to further separate individual spatial components and completely isolate the ring to make any definitive assignments.

\section{The 8.6 \um feature}\label{sect:pah_86}

\begin{figure}
    \centering
    \resizebox{\hsize}{!}{
    \includegraphics[trim={0.5cm 1.5cm 0cm 1.5cm},clip]{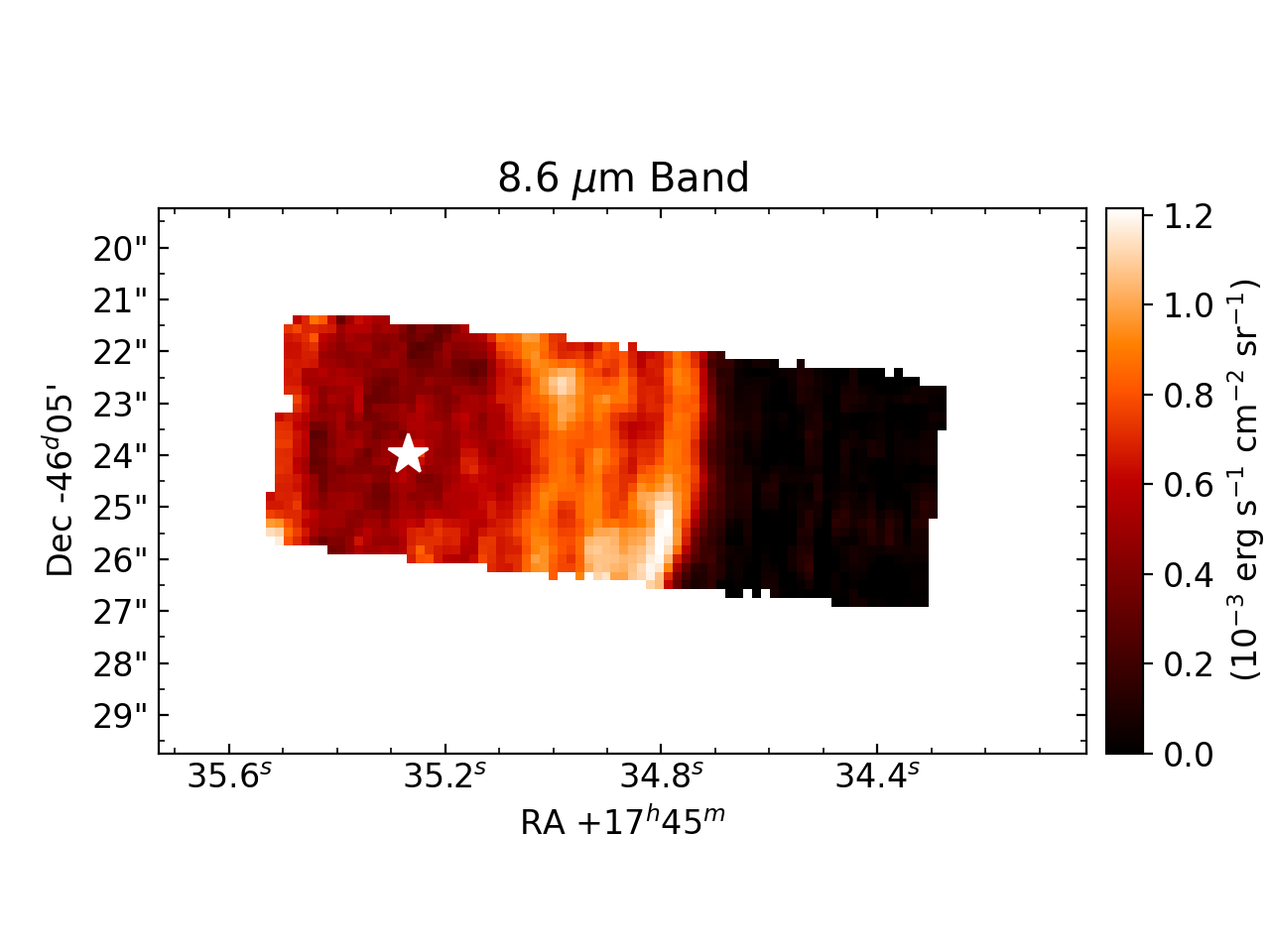}}
    \caption{Integrated surface brightness map of the 8.6 \um feature.}
    \label{fig:86map}
    
\end{figure}

The 8.6~\um feature, generally attributed to PAH emission, has never previously been detected in Tc~1 due to severe blending with the 8.5 \um \csixty band. Our fits allow us to partially separate these contributions, and Figure~\ref{fig:86map} shows the spatial distribution of this 8.6~\um component.
Unlike the fullerene bands, the 8.6~\um emission is not confined to the ring. While there is enhanced emission near the southern part of the ring, there is also significant emission within the core region that is co-spatial with low-ionization structure seen in some atomic emission lines (see \overviewpaper). This suggests that the 8.6 \um component in our fits includes contributions from both the PAH feature and the extended wing of the 8.5~\um \csixty band. Due to severe blending and the broad nature of the features, it is difficult to separate these contributions further.

\bibliography{morgan_ALL}{}
\bibliographystyle{aasjournalv7.1}

\end{document}

%% file: fullerene_measurements_table.tex
\begin{table*}[t]
\caption{\label{tab:features}Fullerene Feature Measurements}
    \begin{tabular}{c|c|ccccc} 
        \hline
        \hline
        Species & Wavelength  & \multicolumn{5}{c}{Integrated Surface Brightness per Aperture}\\
        & {[$\mu$m]} & \multicolumn{5}{c}{$10^{-3}$ [erg/s/cm$^{-2}$/sr]}\\
        & & Star & Core Region & Fullerene Ring & Extended Halo & \ArII Peak \\
        \hline
        \multirow{2}{*}{\textbf{\csixty}}
        & 6.5  & 0.46$\pm$0.02 & 0.51$\pm$0.01 & 0.80$\pm$0.01 & 0.13$\pm$0.01 & 0.55$\pm$0.01 \\
        & 8.5   & 1.052$\pm$0.008 & 1.154$\pm$0.005 & 1.741$\pm$0.007 & 0.229$\pm$0.003 & 1.243$\pm$0.006 \\
        \hline
        \multirow{6}{*}{\textbf{\cseventy}}
        & 8.8  & 0.076$\pm$0.007  & 0.070$\pm$0.002  & 0.147$\pm$0.002  & 0.004$\pm$0.001  & 0.066$\pm$0.002 \\
        & 12.5  & 0.42$\pm$0.01   &  0.404$\pm$0.006 &  0.663$\pm$0.004 &  0.124$\pm$0.004 &  0.378$\pm$0.007 \\
        & 14.8  & 0.30$\pm$0.03   &  0.45$\pm$0.03   & 0.70$\pm$0.03    & 0.07$\pm$0.01  & 0.50$\pm$0.03 \\
        & 15.6  & 0.109$\pm$0.009 & 0.132$\pm$0.003  & 0.205$\pm$0.004  & 0.035$\pm$0.001  & 0.127$\pm$0.003 \\
        & 17.7  & 0.184$\pm$0.009 & 0.221$\pm$0.006  & 0.304$\pm$0.009  & 0.023$\pm$0.002  & 0.242$\pm$0.007 \\
        & 21.8  & 0.122$\pm$0.009 & 0.143$\pm$0.007  & 0.192$\pm$0.005  & 0.027$\pm$0.003  & 0.134$\pm$0.007 \\
        \hline
        \multirow{3}{*}{\textbf{\csixty$+$~\cseventy}}
        & 7.0  & 7.3$\pm$0.4    & 8.0$\pm$0.4     & 12.9$\pm$0.3  & 1.81$\pm$0.07   & 7.8$\pm$0.4 \\
        & 17.4  & 1.92$\pm$0.01 & 2.207$\pm$0.009 & 2.98$\pm$0.01 & 0.444$\pm$0.004 & 2.47$\pm$0.01 \\
        & 18.9  & 4.63$\pm$0.05 & 5.24$\pm$0.03   & 6.81$\pm$0.04 & 1.02$\pm$0.02   & 5.81$\pm$0.04 \\
        \hline
        \hline
    \end{tabular}
\end{table*}

%% file: ringparameters_table.tex
\begin{table}
\centering
\caption{\label{tab:ringparameters}Best-fit parameters for the ``Fullerene Ring" aperture. }
    \begin{tabular}{ccc} 
        \hline
        \hline
        Fit  &  Center & Width \\
        Component$^a$ & [cm$^{-1}$] & [cm$^{-1}$] \\
        \hline
        \multicolumn{3}{c}{\textbf{6.5 \um, 6.6 \um}} \\
        %\hline
        Full. G  & 1542.17$\pm$0.05 & 7.21$\pm$0.05 \\
        Full. G  & 1506.2$\pm$0.2 & 12.5$\pm$0.2 \\
        \hline
        \multicolumn{3}{c}{\textbf{7.0 \um}} \\
        %\hline
        Full. L & 1426.95$\pm$0.07 & 6.8$\pm$0.2 \\
        Full. L    & 1417.8$\pm$0.3 & 13.2$\pm$0.3 \\
        Full. G & 1390.2$\pm$0.9 & 10.0$\pm$1.0 \\
        6.9 \um G    & 1444.7$\pm$0.4 & 3.1$\pm$0.5 \\
        \hline
        \multicolumn{3}{c}{\textbf{8.5 \um}} \\
        %\hline
        Full. G    & 1181.5$\pm$0.2 & 3.8$\pm$0.2 \\
        Full. G & 1175.6$\pm$1.0 & 5.4$\pm$0.5 \\
        8.6 \um G$^b$ & 1165.5 & 18.35 \\
        \hline
        \multicolumn{3}{c}{\textbf{8.8 \um}} \\
        %\hline
        Full. G  & 1133.00$\pm$0.07 & 4.9$\pm$0.1 \\
        \hline
        \multicolumn{3}{c}{\textbf{12.5 \um}} \\
        %\hline
        Full. L  & 802.18$\pm$0.05 & 0.8$\pm$0.1 \\
        Full. L  & 799.87$\pm$0.03 & 1.0$\pm$0.1 \\
        Full. L  & 798.2$\pm$0.1 & 3.50$\pm$0.09 \\
        Full. G  & 791.6$\pm$0.4 & 5.0$\pm$0.3 \\
        \hline
        \multicolumn{3}{c}{\textbf{14.8 \um}} \\
        %\hline
        Full. L  & 678.91$\pm$0.05 & 0.4$\pm$0.1 \\
        Full. L  & 677.24$\pm$0.02 & 0.45$\pm$0.05 \\
        Full. L  & 675.83$\pm$0.08 & 3.0$\pm$0.1 \\
        \hline
        \multicolumn{3}{c}{\textbf{15.6 \um}} \\
        %\hline
        Full. G  & 642.51$\pm$0.07 & 2.5$\pm$0.1 \\
        Full. G & 637.2$\pm$0.2 & 1.3$\pm$0.2 \\
        \hline
        \multicolumn{3}{c}{\textbf{17.4 \um}} \\
        %\hline
        Full. G   & 579.98$\pm$0.03 & 1.27$\pm$0.05 \\
        Full. G    & 576.32$\pm$0.02 & 0.49$\pm$0.02 \\
        Full. G & 575.04$\pm$0.05 & 4.53$\pm$0.05 \\
        Full. G    & 563.33$\pm$0.08 & 2.6$\pm$0.1 \\
        \hline
        \multicolumn{3}{c}{\textbf{18.9 \um}} \\
        %\hline
        Full. L   & 532.34$\pm$0.07 & 2.32$\pm$0.07 \\
        Full. G    & 530.057$\pm$0.006 & 0.239$\pm$0.008 \\
        Full. L & 528.84$\pm$0.08 & 2.1$\pm$0.2 \\
        Full. L & 526.5$\pm$0.3 & 3.3$\pm$0.2 \\
        \hline
        \multicolumn{3}{c}{\textbf{21.8 \um}} \\
        %\hline
        Full. G   & 457.47$\pm$0.07 & 2.52$\pm$0.09 \\
        \hline
        \hline
    \end{tabular}
    \begin{tablenotes}
\item $^a$ G and L indicate a Gaussian profile and Lorentzian profile, respectively. Components not labeled with ``Full." are blended with the overall feature but are not associated with fullerenes.
\item $^b$ These values were fixed corresponding to values found in \cite{2017ApJ...836..198P}.
\end{tablenotes}
\end{table}